\documentclass[twocolumn,english,showpacs,aps,pre,secnumarabic]{revtex4}
\usepackage[T1]{fontenc}
\usepackage[latin1]{inputenc}
\usepackage{textcomp}
\usepackage{amsmath}
\usepackage{graphicx}
\usepackage{amssymb}

\makeatletter

\newcommand{\noun}[1]{\textsc{#1}}
\providecommand{\tabularnewline}{\\}

\usepackage{float}
\usepackage{graphicx}
\usepackage{graphicx}
\usepackage{amsfonts}
\usepackage{hyperref}

\usepackage{babel}
\makeatother

\begin{document}

\title{Analytic application of the mean crossover function to the description
of the isothermal compressibility of xenon}

\author{Yves Garrabos, Carole Lecoutre, Fabien Palencia }

\affiliation{Equipe du Supercritique pour l'Environnement, les Matériaux et l'Espace
- Institut de Chimie de la Matière Condensée de Bordeaux - UPR 9048,
Centre National de la Recherche Scientifique - Université Bordeaux
I - 87, avenue du Docteur Schweitzer, F 33608 PESSAC Cedex, France}

\date{24 July 2007}

\begin{abstract}
We use the mean crossover functions {[}Garrabos and Bervillier, Phys
Rev. E \textbf{74}, 021113 (2006)] estimated from the bounded results
of the Massive Renormalization scheme applied to the $\Phi_{d}^{4}\left(n\right)$
model in three dimensions ($d=3$) and scalar order parameter ($n=1$)
{[}Bagnuls and Bervillier, Phys. Rev. E \textbf{65}, 066132 (2002)],
to represent the singular behavior of the isothermal compressibility
of xenon along the critical isochore in the homogeneous preasymptotic
domain. The validity range and the Ising nature of the crossover description
are discussed in terms of a single scale factor whose value can be
analytically estimated \emph{beyond} the Ising-like preasymptotic
domain.
\end{abstract}


\maketitle

\section{Introduction}

The Ising-like nature of the universal features of the one-component
fluids close to their vapor-liquid critical point is now well-established
\citet{Anisimov2000}. These universal features can be estimated by
using a renormalization approach \citet{ZinnJustin1996} of the classical-to-critical
crossover behavior \citet{Pelissetto1998} for three-dimensional (3D)
Ising-like systems with a symmetrical order parameter density. Especially,
the massive renormalization scheme \citet{Bagnuls1984a,Bagnuls1985,Bagnuls1987}
applied to the $\Phi_{d=3}^{4}\left(n=1\right)$ model ($d$ and $n$
are the dimensions of the space and order parameter density, respectively),
has been recently revisited \citet{Bagnuls2002} to provide max and
min crossover functions which include updated estimations \citet{Guida1998}
of the universal values for exponents and amplitude combinations.
Subsequently, a related paper \citet{Garrabos2006gb} has provided
a unique universal form of the mean crossover functions, which enabled
asymptotic description of the universal features valid within the
\emph{Ising-like preasymptotic domain} exactly, by eliminating the
increasing uncertainties associated with the asymptotic error-bar
propagation in the initial min and max estimations of the critical
exponents. 

These crossover functions have been used to analyze for example, the
singular behavior of the correlation length of seven pure fluids in
their homogeneous domain \citet{Garrabos2006cl} or the singular behavior
of the squared capillary length of twenty pure fluids in their non-homogeneous
domain \citet{Garrabos2007}. For both cases, it was shown that the
scale dilatation method proposed by one of us \citet{Garrabos1982,Garrabos1985,Garrabos1986,Garrabos2002,Garrabos2006qe}
gives a corresponding master singular behavior for the one-component
fluid subclass which agrees with a description by an appropriate modification
of the theoretical function \citet{Garrabos2006mcf} within the Ising-like
preasymptotic domain. As a matter of fact, such an exact behavior
of each theoretical function, approximated by a two-term Wegner-like
expansion \citet{Wegner1972} close to the non-trivial fixed point,
is an essential tool to provide better understanding of the basic
relations between the {}``measured'' asymptotic amplitudes and the
unknown (fluid-dependent) scale factors introduced by linear approximations
between bare fields and physical fields \citet{Wilson1974}. From
the analogy with the scale dilatation method which relates the master
fields to the physical fields introducing two well-defined critical
parameters of each fluid, it was then recently proposed \citet{Garrabos2006mcf}
an unambiguous modification of the mean theoretical functions to represent
the master crossover for the one-component fluid subclass (labeled
$\left\{ 1f\right\} $) in conformity with the corresponding universal
features of the Ising-like universality class (labeled $\Phi_{3}\left(1\right)$).
In the following, the theoretical functions (valid for any three-dimensional
Ising-like system) estimated in Ref. \citet{Garrabos2006gb} are called
\emph{mean} crossover functions and the related reference \citet{Garrabos2006gb}
is labeled I, while their modification (only valid for the one-component
fluids) proposed in Ref. \citet{Garrabos2006mcf} are called \emph{master}
crossover functions and the related Ref. \citet{Garrabos2006mcf}
is labeled II .

The validity range of the mean crossover functions is theoretically
funded only \emph{within} the Ising-like preasymptotic domain (i.e.
for $t\lesssim\mathcal{L}_{\text{PAD}}^{\text{Ising}}$, where $t$
is the thermal-like field along the critical isochore, i.e. for $h=0$,
where $h$ is the magnetic-like field , see II). Especially, the uniqueness
of the asymptotic scale factor, which acts as a crossover parameter
along the critical isochore, has never been clearly demonstrated for
the effective extended asymptotic domain where the mean crossover
functions fit correctly the experimental results. The main goal of
the present paper addresses to the explicit calculation of this single
crossover parameter using the mean crossover functions \emph{beyond}
the Ising-like preasymptotic domain.

We consider in detail the singular behavior of the isothermal compressibility
$\kappa_{T}\left(\Delta\tau^{*}\right)$ and the correlation length
$\xi\left(\Delta\tau^{*}\right)$ of xenon as a function of the reduced
temperature distance $\Delta\tau^{*}=\frac{T-T_{c}}{T_{c}}$, along
the critical isochore, in the homogeneous domain $T>T_{c}$, {[}$T$
($T_{c}$) is the temperature (critical temperature)]. Xenon is here
selected as a standard {}``critical fluid'' in the sense that the
present work only uses quantities which originate from theoretical
arguments, now well-understood, to link three independent dimensionless
amplitudes to the needed three scale factors \citet{background}.
As a matter of fact, the fit of susceptibility and correlation length
data obtained by Güttinger and Cannell \citet{Guttinger1981} from
their precise turbidity measurements in the temperature range $0.028\, K\leq T-T_{c}\leq29\, K$,
can then be used as representative of an {}``ideal'' result in order
to check carefully the temperature range where the value (and its
attached uncertainty) of the scale factor characteristic of the thermal
field is effectively determined. This singular behavior was first
analyzed \citet{Bagnuls1984b}, jointly with singular behaviors of
the correlation length \citet{Guttinger1981} and the heat capacity
\citet{Edwards1968}, using precisely the initial estimations \citet{Bagnuls1984a}
of the crossover functions from the massive renormalization scheme.
In addition to the experimental critical temperature $T_{c}$$\left(\simeq289.74\,\text{K}\right)$,
pressure $p_{c}\left(\simeq5.84\,\text{MPa}\right)$, and density
$\rho_{c}\left(\simeq1110\,\text{kg}\,\text{m}^{-3}\right)$, for
the first time a minimal quantity made of three Ising-like non-universal
parameters of xenon was introduced as a set of a single critical wavelength
(noted $g_{0}$) and two dimensionless scale factors (noted $\vartheta$
and $\psi$) for the thermal-like ($t$) and magnetic-like ($h$)
fields, respectively, with the analytical relations $t=\vartheta\Delta\tau^{*}$
and $h=\psi_{\rho}\Delta\tilde{\mu}$ valid for the asymptotic limits
$t\rightarrow0$ and $h\rightarrow0$ \citet{Bagnuls1984b,Wilson1974}.
$\Delta\tilde{\mu}$ is the standard notation of the dimensionless
ordering field in the fluid case \citet{Levelt1978}, which is related
to the chemical potential difference to the critical chemical potential
{[}see below Eq. (\ref{practical ordering field (16)})]. In the following,
when we refer to this initial result obtained by fitting the experimental
data with the (max) crossover functions estimated for the sixth-order,
we will use the label $\text{MR6}_{\text{max}}$ \citet{Bagnuls1984a,Bagnuls1985}.

Now, using the updated mean crossover functions given in I, the main
objective is to replace the fitting adjustment of the fluid-dependent
parameters $g_{0}$, $\vartheta$, and $\psi_{\rho}$, by their exact
asymptotical values defined in II, only introducing three master numbers
(noted $\Theta^{\left\{ 1f\right\} }$, $\mathbb{L}^{\left\{ 1f\right\} }$,
and $\Psi^{\left\{ 1f\right\} }$) which are characteristics of the
$\left\{ 1f\right\} $-subclass (see II). In that scheme, $g_{0}$-,
$\vartheta$-, and $\psi_{\rho}$-values originate from calculations
of the three fluid-dependent amplitudes $\xi_{0}^{+}$, $\Gamma^{+}$,
and $a_{\chi}^{+}$ (in standard notations \citet{Levelt1978,Privman1991}),
using the scale dilatation method where each fluid is characterized
by only four well-defined critical point coordinates. Such results
account for the master singular behavior of the $\left\{ 1f\right\} $-subclass
observed \emph{within} the Ising-like preasymtotic domain bounded
by a known limit (defined as $\mathcal{L}_{\text{PAD}}^{\left\{ 1f\right\} }=\frac{\mathcal{L}_{\text{PAD}}^{\text{Ising}}}{\Theta^{\left\{ 1f\right\} }}$
in II), in conformity with the two-scale factor-universality of the
$\Phi_{3}\left(1\right)$-universality class. As a main result explicited
below {[}see Eqs. (\ref{system Ising PAD extension (35)}) and (\ref{Xe independent scale factors (40)})],
we unambiguously probe that the temperature range $\Delta\tau_{\text{min}}^{*}<\Delta\tau^{*}<\Delta\tau_{\text{max}}^{*}$
covered by the Güttinger and Cannell's experiments is beyond the Ising-like
preasymptotic domain, i.e. $\Delta\tau_{\text{min}}^{*}\left(\simeq10^{-4}\right)>\frac{\mathcal{L}_{\text{PAD}}^{\text{Ising}}}{\vartheta}$. 

Moreover, after the initial analysis of Ref. \citet{Bagnuls1984b},
the Güttinger and Cannell's xenon data were also used \citet{Anisimov1995,Luijten2000,Hahn2001,Zhong2003,Zhong2004}
in support for discussion of several theoretical \citet{Dhom1985,Luijten1999,Luitjen1997,Luitjen1998,Muser2002}
and phenomenological \citet{Chen1990a,Chen1990b,Anisimov1992,Belyakov1992,Agayan2001}
approaches of the crossover phenomena. As an essential common result,
a single temperature-like crossover parameter seems appropriate for
a complete characterization of the classical-to-critical crossover
in pure fluids, whatever the selected crossover theory or the phenomenological
approach (for comparative analyses, see for example Ref. \citet{Anisimov2000}
and references therein). However, the theoretical understanding in
terms of a Ising-like critical crossover characterized by the asymptotic
value of the scale factor $\vartheta$ appears limited for two main
reasons: i) the fitted data are not within the Ising-like preasymptotic
domain; ii) any crossover function is only well-defined within the
Ising-like preasymptotic domain $t\lesssim\mathcal{L}_{\text{PAD}}^{\text{Ising}}$
(see I and II). Especially for the susceptibility case, such a result
cannot be used in comparative analyses which debate on the {}``correct''
shape of the temperature dependence of the effective exponent $\gamma$
defined as universal functions of a single dimensionless variable
\citet{Belyakov1992,Anisimov1995,Luitjen1997,Luijten1999,Luijten2000,Muser2002,Agayan2001,Hahn2001,Zhong2003,Zhong2004}.

Our present attention is then mainly focused on the determination
of a thermal-like scale factor, noted $\vartheta_{\mathcal{L}}$,
calculated from the application of the mean crossover function beyond
the preasymptotic domain. Our analytic procedure developed hereafter
aims to retrieve the Ising-like universal features only using well-defined
\emph{energy} and \emph{length} units, and a well-controlled number
(three) of well-defined \emph{dimensionless scale factors} given in
a form of a set $\mathbb{S}_{SF}$ as defined in II {[}see also Eqs.
(\ref{asympt Qc vs Ltabfstar set (25)}) and (\ref{(26)}) below].
We \emph{strictly} avoid adjusting the system-dependent parameters
by a minimization of fitting errors to demonstrate that $\vartheta_{\mathcal{L}}\equiv\vartheta$.
Alternatively, using both estimations of the effective exponent and
amplitude values \citet{Kouvel1964}, we calculate the local value
of $\vartheta_{\mathcal{L}}\left(\Delta\tau^{*}\right)$ to verify
the uniqueness of the $\vartheta_{\mathcal{L}}$ value and its identity
with the $\vartheta$ value when $\Delta\tau^{*}\rightarrow0$. Accordingly,
our method offers the great advantage to directly re-use previous
results obtained by fitting experimental data with an effective power
law valid in a limited temperature range \citet{Levelt1975,Levelt1976,Levelt1978}.
Therefore our attention is also focussed on the corresponding analyses
of the isothermal compressibility data obtained from $pVT$ measurements
\citet{Beattie1951,Weinberger1952,Habgood1954,Michels1954,Rabinovich1973}
covering the temperature range $0.1\,\text{K}\leq T-T_{c}\leq283.41\,\text{K}$,
light scattering measurements \citet{Giglio1969,Smith1971,Cannell1970,Swinney1973,Guttinger1980,Guttinger1981}
covering the temperature range $2.6\,\text{mK}\leq T-T_{c}\leq29\,\text{K}$,
and interferometry measurements \citet{Estler1975,Hocken1976,Sengers1978}
covering the temperature range $-0\lesssim T-T_{c}\leq29\, K$, extending
then significantly the temperature range investigated by the present
study. We simultaneously provide the effective reduced temperature
range of the extended asymptotic domain, bounded by a limit noted
$\mathcal{L}_{\text{EAD}}^{\text{Xe}}$, with $\mathcal{L}_{\text{EAD}}^{\text{Xe}}>\mathcal{L}_{\text{PAD}}^{\text{Xe}}=\frac{\mathcal{L}_{\text{PAD}}^{\text{Ising}}}{\vartheta\left(\text{Xe}\right)}$
(see II and below), where isothermal compressibility of xenon is accurately
described by the mean crossover function for susceptibility using
the single scale factor $\vartheta\left(\text{Xe}\right)$ for the
thermal-like field. Such a result gives the first complete comparison
between experimental results and crossover theories, \emph{de facto}
without any adjustable parameters in a relative temperature range
covering more than four decades, when the mean crossover functions
are appropriately modified to account for master behaviors of the
one-component fluid subclass. 

The paper is organized as follows.

In Section 2, we recall useful notations and definitions needed to
use the mean crossover functions for the correlation length and the
isothermal susceptibility in the homogeneous phase of a one-component
fluid. In Section 3, the characterization of the Ising-like preasymptotic
domain of xenon is analyzed only using the isothermal compressibility
fitting result obtained by Güttinger and Cannell, from their turbidity
measurements performed outside the Ising-like preasymptotic domain.
In Section 4, after the introduction of the three-parameter characterization
when the extension of the crossover domain remains undefined, we demonstrate
the great advantage of the mean crossover functions to provide unambiguous
scaling determination of $\vartheta_{\mathcal{L}}$ beyond the Ising-like
preasymptotic domain.

In Appendix A, we give the basic estimation of the needed amplitudes
from application of the scale dilatation method to xenon. All the
other measurement methods of the isothermal compressibility are then
considered, with a special attention to the data obtained from $pVT$
and interferometry measurements which infer a practical three point
calibration for their relative comparison to the Güttinger and Cannell's
data. We finally compare in a detailed manner all the results to the
ones calculated with the mean crossover function for susceptibility,
then justifying \emph{a posteriori} the exact values of the three
scale factors $g_{0}$, $\vartheta$, and $\psi_{\rho}$ for xenon
which are used in this paper.

\section{Mean crossover functions for correlation length and susceptibility}

\subsection{Definitions and notations }

The dimensionless mean crossover functions $F_{\ell}\left(t,h=0\right)=\frac{1}{\ell_{\text{th}}\left(t\right)}$
for the inverse correlation length, and $F_{\chi}\left(t,h=0\right)=\frac{1}{\chi_{\text{th}}\left(t\right)}$
for the inverse susceptibility, in the homogeneous phase, read as
follows (see I),\begin{equation}
\left[\ell_{\text{th}}\left(t\right)\right]^{-1}=\mathbb{Z}_{\xi}^{+}t^{\nu}{\displaystyle \prod_{i=1}^{3}\left(1+X_{\xi,i}^{+}t^{D\left(t\right)}\right)^{Y_{\ell,i}^{+}}}\label{MR lstar vs tstar (1)}\end{equation}
\begin{equation}
\left[\chi_{\text{th}}\left(t\right)\right]^{-1}=\mathbb{Z}_{\chi}^{+}t^{\gamma}{\displaystyle \prod_{i=1}^{3}\left(1+X_{\chi,i}^{+}t^{D\left(t\right)}\right)^{Y_{\chi,i}^{+}}}\label{MR khistar vs tstar (2)}\end{equation}
where $t\left(>0\right)$ is the thermal field like variable, $h\left(=0\right)$
is the magnetic field like variable. $D\left(t\right)$ is a universal
crossover function for the confluent exponents $\Delta$ and $\Delta_{\text{MF}}$
given by \begin{equation}
D\left(t\right)=\frac{\Delta_{\text{MF}}S_{2}\sqrt{t}+\Delta}{S_{2}\sqrt{t}+1}\label{Deff exponent MR universal (3)}\end{equation}
All the universal exponents $\nu$, $\gamma$, $\Delta$, $\Delta_{\text{MF}}$,
theoretical amplitudes $\mathbb{Z}_{\xi}^{+}$, $\mathbb{Z}_{\chi}^{+}$,
and theoretical parameters $S_{2}$, $X_{\xi,i}^{+}$, $Y_{\xi,i}^{+}$,
$X_{\chi,i}^{+}$, $Y_{\chi,i}^{+}$, are defined in I. They are also
reported in Table \ref{Table I} for use in the following numerical
estimations.

\begin{table*}[!]
\begin{tabular}{|c|c|c|c|c|c|c|}
\hline 
\multicolumn{1}{|c|}{a} &  exponent &  $\mathbb{Z}_{\xi}^{+}$ &  $S_{2}$ &  $i$ &  $X_{\xi,i}$ &  $Y_{\xi,i}$\tabularnewline
\hline
$\nu$ &  $0.6303875$ &  $2.121008$ &  $22.9007$ &  $1$ &  $40.0606$ &  $-0.098968$\tabularnewline
\hline
$\Delta$ &  $0.50189$ &  &  &  $2$ &  $11.9321$ &  $-0.15391$\tabularnewline
\hline
$\Delta_{\text{MF}}$ &  $0.5$ &  &  &  $3$ &  $1.90235$ &  $-0.00789505$\tabularnewline
\hline
 &  &  &  &  &  $\mathbb{Z}_{\xi}^{1,+}=$ &  $5.81623$ \tabularnewline
\hline
\end{tabular} \begin{tabular}{|c|c|c|c|c|c|c|}
\hline 
\multicolumn{1}{|c|}{b} &  exponent &  $\mathbb{Z}_{\chi}^{+}$ &  $S_{2}$ &  $i$ &  $X_{\chi,i}$ &  $Y_{\chi,i}$\tabularnewline
\hline
$\gamma$ &  $1.2395935$ &  $3.709601$ &  $22.9007$ &  $1$ &  $29.1778$ &  $-0.178403$\tabularnewline
\hline
$\Delta$ &  $0.50189$ &  &  &  $2$ &  $11.7625$ &  $-0.282241$\tabularnewline
\hline
$\Delta_{\text{MF}}$ &  $0.5$ &  &  &  $3$ &  $2.05948$ &  $-0.0185424$\tabularnewline
\hline
 &  &  &  &  &  $\mathbb{Z}_{\chi}^{1,+}=$ &  $8.56347$ \tabularnewline
\hline
\end{tabular}

\caption{Values of the universal exponents and constants of Eqs. (\ref{MR lstar vs tstar (1)})
to (\ref{MR Z1khi amplitude (7)}), for (a) the dimensionless correlation
length and (b) the dimensionless susceptibility, in the homogeneous
domain (see Ref. \citet{Garrabos2006gb}). \label{Table I}}

\end{table*}

More generally, the crossover functions estimated in I are only well-defined
for the critical line that links the Gaussian fixed point and the
non-trivial fixed point. The effects due to the second-order (and
higher) analytical contributions and the ones due to the confluent
corrections to scaling linked to critical exponents $\Delta_{2}$,
$\Delta_{3}$, ..., have been discarded. As a direct consequence,
these functions account exactly for the Ising-like universal features
only estimated in the $t$-range very close to the non-trivial fixed
point, which corresponds the Ising-like preasymptotic domain $t\leq\mathcal{L}_{\text{PAD}}^{\text{Ising}}$
defined in I {[}see also below Eq. (\ref{model Ising PAD extension (9)})]. 

\emph{Within} the Ising-like preasymptotic domain, the mean crossover
functions can be approximated by their restricted (two-term) Wegner
like expansion \citet{Wegner1972}. Then Eqs. (\ref{MR lstar vs tstar (1)})
and (\ref{MR khistar vs tstar (2)}) can be replaced by:\begin{equation}
\ell_{\text{PAD,th}}\left(t\right)=\left(\mathbb{Z}_{\xi}^{+}\right)^{-1}t^{-\nu}{\displaystyle \left[1+\mathbb{Z}_{\xi}^{1,+}t^{\Delta}\right]}\label{MR 2term elestar (4)}\end{equation}
\begin{equation}
\chi_{\text{PAD,th}}\left(t\right)=\left(\mathbb{Z}_{\chi}^{+}\right)^{-1}t^{-\gamma}{\displaystyle \left[1+\mathbb{Z}_{\chi}^{1,+}t^{\Delta}\right]}\label{MR 2term khistar (5)}\end{equation}
where the amplitudes $\mathbb{Z}_{\xi}^{1,+}$ and $\mathbb{Z}_{\chi}^{1,+}$
of the first-order term due to the lowest confluent corrections to
scaling are given by\begin{equation}
\mathbb{Z}_{\xi}^{1,+}=-{\displaystyle \sum_{i=1}^{3}}X_{\xi,i}^{+}Y_{\xi,i}^{+}\label{MR Z1ele amplitude (6)}\end{equation}
\begin{equation}
\mathbb{Z}_{\chi}^{1,+}=-{\displaystyle \sum_{i=1}^{3}}X_{\chi,i}^{+}Y_{\chi,i}^{+}\label{MR Z1khi amplitude (7)}\end{equation}
(see Table \ref{Table I}). They are related by the universal ratio
\citet{Bagnuls2002,Guida1998}\begin{equation}
\frac{\mathbb{Z}_{\xi}^{1,+}}{\mathbb{Z}_{\chi}^{1,+}}=0.67919\label{confluent Ztabksi vs khi ratio (8)}\end{equation}
The validity domain of Eqs. (\ref{MR 2term elestar (4)}) to (\ref{confluent Ztabksi vs khi ratio (8)})
corresponds to $t\leq\mathcal{L}_{\text{PAD}}^{\text{Ising}}$ where\begin{equation}
\mathcal{L}_{\text{PAD}}^{\text{Ising}}\simeq\left(\frac{0.033}{S_{2}}\right)^{2}\cong1.9\times10^{-6}\label{model Ising PAD extension (9)}\end{equation}
with $S_{2}=22.9007$ (see I and Table \ref{Table I})

Finally, any restricted two-term Wegner-like expansions can be estimated
only using the following set of three theoretical amplitudes \citet{background},\begin{equation}
\mathcal{\mathbb{S}}_{A}^{MR}\left(t\leq\mathcal{L}_{\text{PAD}}^{\text{Ising}};h=0\right)=\left\{ \mathbb{Z}_{\chi}^{1,+};\left(\mathbb{Z}_{\xi}^{+}\right)^{-1};\left(\mathbb{Z}_{\chi}^{+}\right)^{-1}\right\} \label{QcalMR PAD (10)}\end{equation}
where the subscript $A$ labels for the amplitude nature of these
characteristic parameters.

\emph{Beyond} the Ising-like preasymptotic domain, i.e. $t>\mathcal{L}_{\text{PAD}}^{\text{Ising}}$,
we also recall that the theoretical parameter $S_{2}$ acts as a convenient
sensor to estimate some orders of magnitude of $t$ which are convenient
for the analysis of the crossover. For example, the theoretical crossover
temperature $t_{\Delta}$, which corresponds to the value $D\left(t_{\Delta}\right)=\frac{\Delta+\Delta_{\text{MF}}}{2}$
of the universal confluent function of Eq. (\ref{Deff exponent MR universal (3)}),
is defined by\begin{equation}
t_{\Delta}\simeq\left(\frac{1}{S_{2}}\right)^{2}\cong1.9\times10^{-3}\label{model crossover temperature (11)}\end{equation}
For the confluent corrections, $t_{\Delta}$ characterizes the universal
crossover exchange between predominant Ising-like nature close to
the non-trivial fixed point {[}$t\ll t_{\Delta}$], to predominant
mean field-like nature close to the Gaussian fixed point {[}$t\gg t_{\Delta}$].
Indeed, $t_{\Delta}$ gives an estimation of the order of magnitude
of the $t_{e_{P},\frac{1}{2}}$-value where each effective exponent
$e_{P,e,\text{th}}\left(t\right)=-\frac{\partial Ln\left[F_{P}\left(t\right)\right]}{\partial Ln\left(t\right)}$
\citet{Kouvel1964} crosses its {}``mean'' crossover value $e_{P,\frac{1}{2}}\left(t_{e_{P},\frac{1}{2}}\right)=\frac{e_{P}+e_{P,\text{MF}}}{2}$
(see Figure 4 in I). In such a {}``critical-to-classical'' crossing
range in the homogeneous domain {[}$t>0$], $\ell_{\text{th}}\simeq20-30$
is a typical (dimensionless) order of magnitude for the theoretical
correlation length. For example, $\ell_{\text{th}}\left(t_{\Delta}\right)=28.8$,
$\ell_{\text{th}}\left(t_{\nu,\frac{1}{2}}\right)=22.2$, and $\ell_{\text{th}}\left(t_{\gamma,\frac{1}{2}}\right)=18.8$,
anticipating the following discussion of the results reported in Table
\ref{Table II} where the values of Eqs. (\ref{MR lstar vs tstar (1)})
and (\ref{MR khistar vs tstar (2)}) are calculated for eight conditions
of the effective exponents. Moreover, introducing the practical relations
$\mathcal{L}_{\text{PAD}}^{\text{Ising}}=\varpi t_{\Delta}$, with
$\varpi\cong10^{-3}$ (see I), it is easy to separate the analysis
of, either the Ising-like preasymptotic domain $t\lesssim\varpi\times t_{\Delta}$,
or the intermediate {}``Ising-like'' crossover domain $\varpi t_{\Delta}<t<t_{\Delta}$.
Especially considering the selected values of the effective exponent
$\gamma_{e,\text{th}}\left(t\right)=-\frac{\partial Ln\left[\mathcal{X}_{th}\left(t\right)\right]}{\partial Ln\left(t\right)}$
given in line 3 of Table \ref{Table II}), we note that the conditions
$\gamma_{e}\left(t_{\gamma_{\frac{1}{2}}}\right)$ (column 1) and
$\gamma_{e}\left(t_{\nu_{\frac{1}{2}}}\right)$ (column 2) are obtained
for $t>t_{\Delta}$, while, obviously, the condition $\gamma_{e}\left(t_{\Delta}\right)$
(column 3) is obtained at $t=t_{\Delta}$, where $\ell_{\text{th}}$
reach a value $\sim30$. For $\ell_{\text{th}}\gtrsim30$, or $t\lesssim t_{\Delta_{\frac{1}{2}}}$,
(columns 4 and right), we expect to observe the Ising-like universal
features of the critical phenomena which are then characterized by
the three system-dependent parameters only asymptotically well-defined
when $\ell_{\text{th}}\gtrsim1916$, or $t\lesssim\mathcal{L}_{\text{PAD}}^{\text{Ising}}$
(i.e. within the Ising-like preasymptotic domain). We note that the
effective values $\gamma_{e}=\gamma_{e,pVT}=1.16665$ \citet{Garrabos1985}),
$1.19$ \citet{Levelt1975,Levelt1978}, $1.211$ \citet{Levelt1976}
obtained from $pVT$ measurements, are precisely in this intermediate
Ising-like range $\mathcal{L}_{\text{PAD}}^{\text{Ising}}<t<t_{\Delta}$.
So that, we have underlined these three latter conditions in columns
5 to 6, respectively, showing that the value of theoretical correlation
length $\ell_{\text{th}}$ is on the range $50-220$, i.e., $2\lesssim\frac{\ell_{\text{th}}}{\ell_{\text{th}}\left(t_{\Delta}\right)}\lesssim8$.
A special attention to the finite {}``Ising-like'' temperature range
covered by $pVT$ experiments is given in Section 4.2 and Appendix
A3. Correlatively, within the Ising-like preasymptotic domain $t\lesssim\mathcal{L}_{\text{PAD}}^{\text{Ising}}$,
the theoretical values reported in lines 7 and 8 of Table \ref{Table II}
show that the condition $\ell_{\text{th}}\gtrsim2000$ (or $\frac{\ell_{\text{th}}}{\ell_{\text{th}}\left(t_{\Delta}\right)}\gtrsim70$)
is satisfied. The expected temperature-like variation of $\gamma_{e,\text{th}}\left(t\right)$,
i.e., $\gamma-\gamma_{e,\text{th}}\left(\mathcal{L}_{\text{PAD}}^{\text{Ising}}\right)\leq0.006$,
is then significantly lower than the typical uncertainty on the asymptotic
experimental value $\gamma_{\text{exp}}\simeq1.23\pm0.02$ \citet{Levelt1978}.

The eight indexations of $t$ (or $\ell_{\text{th}}$) given by these
eight conditions of Table \ref{Table II} will be used to label the
horizontal (upper) axes of the next figures with corresponding arrows
and column numbers.

\begin{table*}
\begin{tabular}{|c|c|c|c||c|c|c||c|c|}
\hline 
(a) \textbackslash{} label & $1$ & $2$ & 3 & $4$ & $5$ & $6$ & $7$ & $8$\tabularnewline
\hline 
$Condition$ & $\gamma_{\frac{1}{2}}=\frac{\gamma+\gamma_{\text{MF}}}{2}$ & $\nu_{\frac{1}{2}}=\frac{\nu+\nu_{\text{MF}}}{2}$ & $\Delta_{\frac{1}{2}}=\frac{\Delta+\Delta_{\text{MF}}}{2}$ & $\gamma_{e,pVT}$\citet{Garrabos1982} & $\gamma_{e,pVT}$\citet{Levelt1975} & $\gamma_{e,pVT}$\citet{Levelt1978} & $\mathcal{L}_{\text{PAD}}^{\text{Ising}}$ & $\Delta\tau_{\text{C1}}^{*}\left(\text{Xe}\right)$\tabularnewline
\hline
\hline 
$\gamma_{e,\text{th}}$ & $1.1198$ & $1.12776$ & $1.14081$ & $1.16665$ & $1.19$ & $1.211$ & $1.23397$ & $1.23848$\tabularnewline
\hline 
$\nu_{e,\text{th}}$ & $0.561$ & $0.5652$ & $0.572$ & $0.5862$ & $0.5994$ & $0.612$ & $0.6266$ & $0.62963$\tabularnewline
\hline 
$\Delta_{e,\text{th}}$ & $0.500768$ & $0.500834$ & $0.500945$ & $0.501176$ & $0.501396$ & $0.501601$ & $0.50183$ & $0.50188$\tabularnewline
\hline 
$t$ & $4.06\times10^{-3}$ & $3.056\times10^{-3}$ & $1.907\times10^{-3}$ & $7.033\times10^{-4}$ & $2.392\times10^{-4}$ & $6.22\times10^{-5}$ & $1.907\times10^{-6}$ & $7.272\times10^{-8}$\tabularnewline
\hline 
$\ell_{\text{th}}$ & $18.76$ & $22.0$ & $28.8$ & $51.32$ & $97.29$ & $220.1$ & $1916$ & $14929.4$\tabularnewline
\hline
$\mathbb{Z}_{\chi,e}^{+}$ & $0.73735$ & $0.70492$ & $0.65161$ & $0.54730$ & $0.45641$ & $0.37810$ & $0.29363$ & $0.27518$\tabularnewline
\hline
\hline 
(b) &  &  &  &  &  &  &  & \tabularnewline
\hline
$\Delta\tau^{*}=\frac{t}{\vartheta}$ & $0.193$ & $0.145$ & $0.0905$ & $0.0334\gtrsim\mathcal{L}_{\text{EAD}}^{\text{Xe}}$ & $0.01135$ & $0.00295$ & $0.905\times10^{-4}$ & $3.451\times10^{-6}$\tabularnewline
\hline
$\frac{\xi}{\alpha_{c}}=\frac{\ell_{tth}}{\mathbb{L}^{\left\{ 1f\right\} }}$ & $0.730$ & $0.857$ & $1.12$ & $1.997$ & $3.786$ & $8.564$ & $74.56$ & $580.94$\tabularnewline
\hline
$\Gamma_{e,\text{th}}^{+}$ & $0.099601$ & $0.098193$ & $0.095456$ & $0.088586$ & $0.080842$ & $0.072626$ & $0.061629$ & $0.058771$\tabularnewline
\hline
\end{tabular}

\caption{(a): Calculated theoretical values of the effective exponents $\gamma_{e,\text{th}}$
(line 3), $\nu_{e,\text{th}}$ (line 4), $\Delta_{e,\text{th}}$ (line
5), the thermal-like field $t$ (line 6), the correlation length $\ell_{\text{th}}$
(line 7), and the effective amplitude $\mathbb{Z}_{\chi,e}^{+}$ (line
8) of the susceptibility, using Eqs. (\ref{MR lstar vs tstar (1)})
and (\ref{MR khistar vs tstar (2)}), for the typical conditions given
in line 2, respectively; (b): Corresponding physical values of the
reduced temperature distance $\Delta\tau^{*}=\frac{t}{\vartheta}$
(line 10), the dimensionless correlation length $\frac{\xi}{\alpha_{c}}=\frac{\ell_{th}}{\mathbb{L}^{\left\{ 1f\right\} }}$
(line 11), and the effective amplitude $\Gamma_{e,\text{th}}^{+}$
(line 12) of the dimensionless compressibility, using Eqs. (\ref{gzero vs elestar (15)}),
(\ref{Ltabfstar (21)}), and (\ref{(23)}), for the xenon case, with
$\vartheta=0.21069$, $\mathbb{L}^{\left\{ 1f\right\} }=25.6988$
and $\psi_{\rho}=3.2507\,10^{-4}$ (see text).\label{Table II}}

\end{table*}

\subsection{Fluid-dependent parameters}

The introduction of the fluid-dependent parameters to fit the experimental
results using the mean crossover functions was detailed in § 3.1 of
II. We recall here the main steps to introduce definitions and notations
of the needed quantities, fixing the value $\Lambda_{qe}^{*}=1$ to
neglect the quantum effects at the microscopic length scale \citet{Garrabos2006qe}
for the xenon case.

When $t\rightarrow0$ at $h=0$, $t$ is analytically related to\begin{equation}
\Delta\tau^{*}=\frac{T-T_{c}}{T_{c}}\rightarrow0\label{dimensionless temperature distance (12)}\end{equation}
by the following linear approximation \citet{Wilson1974}\begin{equation}
t=\vartheta\Delta\tau^{*}\label{theta from tstar vs deltataustar (13)}\end{equation}
The dimensionless scale factor $\vartheta$ is a fluid-dependent parameter.
The fluid critical temperature $T_{c}$, provides the energy unit\begin{equation}
\left(\beta_{c}\right)^{-1}=k_{B}T_{c}\label{energy unit (14)}\end{equation}
which links the dimensionless free energies of the $\Phi^{4}$-model
and the selected fluid. Accordingly, the dimensionless form of the
Hamiltonian of the $\Phi^{4}$-model results of the introduction of
a finite (but arbitrary) wave number $\Lambda_{0}$, the so-called
cutoff parameter (see I), whose inverse is related to the unknown
finite short range of the microscopic molecular interaction. A convenient
method at $d=3$ consists in replacing (unknown) $\Lambda_{0}$ by
$g_{0}$ which is the adjustable critical coupling constant of the
$\Phi^{4}$ term having correct wave number dimension. The inverse
$\left(g_{0}\right)^{-1}$ acts as the physical length unit to link
the theoretical dimensionless correlation length ($\ell_{\text{th}}$)
and the physical correlation length ($\xi_{\text{exp}}$) of each
one-component fluid, through the fitting equation\begin{equation}
\ell_{\text{th}}\left(t\right)=g_{0}\xi_{\text{exp}}\left(\Delta\tau^{*}\right)\label{gzero vs elestar (15)}\end{equation}

The dimensionless ordering field for fluids is defined as \citet{Levelt1975,Levelt1976}
\begin{equation}
\Delta\tilde{\mu}=\tilde{\mu}_{\rho}-\tilde{\mu}_{\rho,c}\label{practical ordering field (16)}\end{equation}
where $\Delta\tilde{\mu}$ is written using practical dimensionless
chemical potential $\tilde{\mu}=\frac{\mu_{\rho}\rho_{c}}{p_{c}}$
\citet{Levelt1976}. $\mu_{\rho}$ ($\mu_{\rho,c}$) is the (critical)
chemical potential per mass unit, $\rho$ ($\rho_{c}$) is the (critical)
mass density, and $p$ ($p_{c}$) is the (critical) pressure. Correspondingly,
the practical dimensionless form $\Delta\tilde{\rho}$ of the order
parameter density reads as follows\begin{equation}
\Delta\tilde{\rho}=\frac{\rho}{\rho_{c}}-1\label{practical OP (17)}\end{equation}
 using the practical dimensionless form $\tilde{\rho}=\frac{\rho}{\rho_{c}}$
of the mass density \citet{Levelt1976}.

When $h\rightarrow0$ and $\Delta\tilde{\mu}\rightarrow0$ at $t=0$,
$h$ is related to $\Delta\tilde{\mu}$ as follows \begin{equation}
h=\psi_{\rho}\Delta\tilde{\mu}\label{psi from hstar vs deltamutilde (18)}\end{equation}
where $\psi_{\rho}$ is the second (fluid-dependent) scale factor.
However, the parameters $T_{c}$, $g_{0}$, $\vartheta$, and $\psi_{\rho}$,
do not provide unequivocal link between theoretical and physical thermodynamics
quantities. For example, the definition of $\tilde{\mu}_{\rho}$ introduces
a second unit $\frac{p_{c}}{\rho_{c}}\sim\left[\frac{\text{energy}}{\text{mass}}\right]$
for (specific) energy which differs from $\left(m_{\bar{p}}\beta_{c}\right)^{-1}$
by the factor $Z_{c}$ {[} $m_{\bar{p}}$ is the mass of the fluid
particle (i.e. the molecular mass)]. Correlatively, the critical mass
unit of the one-component fluid introduces a critical specific volume
$\frac{1}{\rho_{c}}$ which differs from the volume $\frac{k_{B}T_{c}}{p_{c}}=\left(\alpha_{c}\right)^{d}$
of the critical interaction cell {[}see below Eq. (\ref{1f length scale unit (19)})].
The comparison between the two volumes introduces the extensivity
of the system through the amount $\frac{m_{\bar{p}}}{Z_{c}}$ of matter
filling the volume of the critical interaction cell \citet{Garrabos1982}.
That provides alternative choice between two energy units and two
length units originating from thermodynamics. Such noticeable differences
in the system-dependent units of the dimensionless variables impose
to have careful attention when comparing the dimensionless thermodynamics
potentials. Here, the selected length unit $\alpha_{c}$ reads \begin{equation}
\alpha_{c}=\left(\frac{k_{B}T_{c}}{p_{c}}\right)^{\frac{1}{d}}\label{1f length scale unit (19)}\end{equation}
and have physical meanning in terms of the range of intermolecular
interaction in fluids. Therefore, in our notations, the superscript
star labels a dimensionless variable which uses $\left(\beta_{c}\right)^{-1}$
and $\alpha_{c}$ as energy and length units in one self-consistent
procedure to made dimensionless all the thermodynamic variables normalized
per particle \citet{Garrabos1982,Garrabos1985}, not per mass unit.
The subscript $\rho$ recalls for practical order parameter density
defined by Eq. (\ref{practical OP (17)}), and the related practical
dimensionless variables are decorated by a tilde. For example, the
experimental isothermal susceptibility $\chi_{T,\rho}$ for fluids
is defined by $\chi_{T,\rho}=\left(\frac{\partial\rho}{\partial\mu_{\rho}}\right)_{T}\sim\left[\frac{1}{\text{energy}\times\text{volume}}\right]$
when the fluid order parameter (respectively, the fluid ordering field)
is proportional to the (mass) density $\rho\sim\left[\frac{\text{mass}}{\text{volume}}\right]$
(respectively, the chemical potential per mass unit $\mu_{\rho}\sim\left[\frac{\text{energy}}{\text{mass}}\right]$).
$\chi_{T,\rho}$ is related to the isothermal compressibility $\kappa_{T}=\frac{1}{\rho}\left(\frac{\partial\rho}{\partial p}\right)_{T}\sim\left[\frac{\text{volume}}{\text{energy}}\right]$
by $\chi_{T,\rho}=\rho^{2}\kappa_{T}$. The practical fluid dimensionless
variables are $\tilde{\rho}=\frac{\rho}{\rho_{c}}$, $\tilde{\mu}=\frac{\mu_{\rho}\rho_{c}}{p_{c}}$,
while $\kappa_{T}^{*}=p_{c}\kappa_{T}$ \citet{Levelt1976}. Thus
we obtain $\tilde{\chi}_{T}\equiv\kappa_{T}^{*}$ only at $\tilde{\rho}=1$
for $\Delta\tau^{*}>0$. Such a practical interrelation between the
dimensionless variables results from the implicit use of these two
{}``thermodynamic'' length units previously defined from two distinct
volumes $v_{_{\bar{p},c}}=\frac{m_{\bar{p}}}{\rho_{c}}$ and $\left(\alpha_{c}\right)^{d}$.
On the other hand, from the theoretical scheme applied to the $\Phi_{d=3}^{4}\left(n=1\right)$-model,
after normalization of the free energies by $k_{B}T\cong k_{B}T_{c}$,
the length dimensions of the Hamiltonian quantities $r_{0}-r_{0c}\sim\left(g_{0}\right)^{2}t$,
$\left\langle \phi_{0}\right\rangle \sim\left(g_{0}\right)^{\frac{5}{2}}m$,
and $h_{0}\sim\left(g_{0}\right)^{\frac{1}{2}}h$, lead to $\chi_{0,\text{th}}=\left(g_{0}\right)^{2}\chi_{\text{th}}\sim\left[\frac{1}{\text{surface}}\right]$
(see I for notations and definitions). We can then conclude that introducing
the dimensionless scale factors $\vartheta$ and $\psi_{\rho}$ through
Eqs. (\ref{theta from tstar vs deltataustar (13)}) and (\ref{psi from hstar vs deltamutilde (18)}),
provides a subtle critical combination (not discussed here) of the
model units and the fluid units. The essential point is to guarantee
the uniqueness of the energy unit and the length unit in the description
of dimensionless singular behaviors \citet{Privman1991}. Our selected
length unit $\alpha_{c}$ {[}see Eq. (\ref{1f length scale unit (19)})]
takes thermodynamic origin, while the wave number unit $g_{0}$ {[}see
Eq. (\ref{gzero vs elestar (15)})] has theoretical interest to fit
the asymptotic singular divergence of $\xi_{\text{exp}}^{*}\left(\Delta\tau^{*}\right)=\frac{\xi_{\text{exp}}\left(\Delta\tau^{*}\right)}{\alpha_{c}}$.
Then, by exchanging  Eq. (\ref{gzero vs elestar (15)}) and the following
dimensionless form \begin{equation}
\ell_{\text{th}}\left(t\right)=\mathbb{L}^{\left\{ 1f\right\} }\xi_{\text{exp}}^{*}\left(\Delta\tau^{*}\right),\label{ksiexpstar fitting eq (20)}\end{equation}
we also introduce the supplementary scale factor \begin{equation}
\mathbb{L}^{\left\{ 1f\right\} }=\alpha_{c}g_{0}\label{Ltabfstar (21)}\end{equation}
as a dimensionless product between the two critical quantities (i.e.,
defined for $t=0;h=0$ and $\Delta\tau^{*}=0;\Delta\tilde{\mu}=0$,
respectively). 

After all, $m\rightarrow0$ and $\Delta\tilde{\rho}\rightarrow0$
are related by the equation\begin{equation}
m=\left(\mathbb{L}^{\left\{ 1f\right\} }\right)^{-d}\left(\psi_{\rho}\right)^{-1}\Delta\tilde{\rho},\label{inverse psi from mstar vs deltarhotilde (22)}\end{equation}
where $m$ is the theoretical magnetization-like order parameter.
Considering the theoretical susceptibility $\chi_{\text{th}}\left(t\right)=\left(\frac{\partial m}{\partial h}\right)_{t}$
and the experimental isothermal susceptibility $\tilde{\chi}_{T,\text{exp}}\left(\Delta\tau^{*}\right)=\left(\frac{\partial\Delta\tilde{\rho}}{\partial\Delta\tilde{\mu}}\right)_{\Delta\tau^{*}}$
for fluids, the second fitting equation is then obtained as follows\begin{equation}
\chi_{\text{th}}\left(t\right)=\left(\mathbb{L}^{\left\{ 1f\right\} }\right)^{-d}\left(\psi_{\rho}\right)^{-2}\tilde{\chi}_{T,\text{exp}}\left(\Delta\tau^{*}\right),\label{(23)}\end{equation}
with $\tilde{\chi}_{T,\text{exp}}\left(\Delta\tau^{*}\right)\equiv\kappa_{T,\text{exp}}^{*}\left(\Delta\tau^{*}\right)$
when $\Delta\tilde{\rho}=0$.

Finally, each one-component fluid is asymptotically characterized
by the set \citet{background}\begin{equation}
\mathcal{\mathbb{Q}}_{c}\left(\Delta\tau^{*}\rightarrow0\right)=\left\{ \left(\beta_{c}\right)^{-1};\alpha_{c};g_{0};\vartheta;\psi_{\rho}\right\} \label{asympt Qc vs gzero set (24)}\end{equation}
which can be rewritten in an equivalent form\begin{equation}
\mathcal{\mathbb{Q}}_{c}\left(\Delta\tau^{*}\rightarrow0\right)=\left\{ \left(\beta_{c}\right)^{-1};\alpha_{c};\mathcal{\mathbb{S}}_{SF}\right\} \label{asympt Qc vs Ltabfstar set (25)}\end{equation}
then introducing the following fluid set $\mathcal{\mathbb{S}}_{SF}$
made of three (asymptotic) dimensionless scale factors\begin{equation}
\mathcal{\mathbb{S}}_{SF}\left(\Delta\tau^{*}\rightarrow0\right)=\left\{ \vartheta;\mathbb{L}^{\left\{ 1f\right\} };\psi_{\rho}\right\} \label{(26)}\end{equation}
The subscript $SF$ recalls for the scale factor nature of the three
fluid-dependent dimensionless numbers, while the condition $\Delta\tau^{*}\rightarrow0$
indicates the asymptotic nature of the hypotheses needed by the renormalization
scheme. Within the Ising-like preasymptotic domain, $\kappa_{T}^{*}\left(\Delta\tau^{*}\right)$
and $\xi\left(\Delta\tau^{*}\right)$ can be approximated by\begin{equation}
\kappa_{T,\text{exp}}^{*}\left(\Delta\tau^{*}\right)=\Gamma^{+}\left(\Delta\tau^{*}\right)^{-\gamma}\left[1+a_{\chi}^{+}\left(\Delta\tau^{*}\right)^{\Delta}\right]\label{2term kappaTexp (27)}\end{equation}
\begin{equation}
\frac{\xi_{\text{exp}}\left(\Delta\tau^{*}\right)}{\alpha_{c}}=\frac{\xi_{0}^{+}}{\alpha_{c}}\left(\Delta\tau^{*}\right)^{-\nu}\left[1+a_{\xi}^{+}\left(\Delta\tau^{*}\right)^{\Delta}\right]\label{2term ksiexp (28)}\end{equation}
with \citet{Bagnuls2002}\begin{equation}
\frac{a_{\xi}^{+}}{a_{\chi}^{+}}=\frac{\mathbb{Z}_{\xi}^{1,+}}{\mathbb{Z}_{\chi}^{1,+}}=0.67919\label{aksiplus vs akhiplus (29)}\end{equation}
Hereabove, we have selected $\Gamma^{+}$, $\xi_{0}^{+}$ (or $\xi^{+}=\frac{\xi_{0}^{+}}{\alpha_{c}}$),
and $a_{\chi}^{+}$ as independent amplitudes to characterize each
one-component fluid. The {}``experimental\emph{''} parameter set
written as\begin{equation}
\mathcal{Q}_{c,\mathcal{L}_{\text{PAD}}^{\text{Xe}}}=\left\{ \left(\beta_{c}\right)^{-1};\alpha_{c};S_{A}\right\} \label{Ising ampl Qc vs ksiplus set (30)}\end{equation}
is then Ising-like equivalent to the set $\mathcal{\mathbb{Q}}_{c}\left(\Delta\tau^{*}\rightarrow0\right)$
of Eq. (\ref{asympt Qc vs Ltabfstar set (25)}) {[}here the subscript
$\mathcal{L}_{\text{PAD}}^{\text{Xe}}$ recalls for the fluid-dependent
temperature domain of validity]. Its dimensionless part $S_{A}$ (where
the subscript $A$ recalls for the amplitude nature of the dimensionless
numbers) is given as :\begin{equation}
S_{A}=\left\{ a_{\chi}^{+};\xi^{+};\Gamma^{+}\right\} \label{Qcalf PAD (31)}\end{equation}
which is compared with Eq. (\ref{(26)}) using the fitting equations
(\ref{(23)}) and (\ref{ksiexpstar fitting eq (20)}) {[}or (\ref{gzero vs elestar (15)})].
The successive unequivocal determinations of the scale factors, first
$\vartheta$, hence $\mathbb{L}^{\left\{ 1f\right\} }$ (or $g_{0}$),
and finally $\psi_{\rho}$, are obtained from the following hierarchy
of equations:\begin{equation}
\vartheta=\left(\frac{a_{\chi}^{+}}{\mathbb{Z}_{\chi}^{1,+}}\right)^{\frac{1}{\Delta}}=\left(\frac{a_{\xi}^{+}}{\mathbb{Z}_{\mathcal{\xi}}^{1,+}}\right)^{\frac{1}{\Delta}},\label{theta vs akhiplus (32)}\end{equation}
\begin{equation}
\mathbb{L}^{\left\{ 1f\right\} }=\left[\xi^{+}\mathbb{Z}_{\xi}^{+}\vartheta^{\nu}\right]^{-1}\; or\; g_{0}=\left[\xi_{0}^{+}\mathbb{Z}_{\xi}^{+}\vartheta^{\nu}\right]^{-1},\label{gzero vs ksizeroplus (33)}\end{equation}
\begin{equation}
\psi_{\rho}=\left[\left(\mathbb{L}^{\left\{ 1f\right\} }\right)^{-d}\Gamma^{+}\mathbb{Z}_{\chi}^{+}\vartheta^{\gamma}\right]^{\frac{1}{2}}\label{psi vs gamaplus (34)}\end{equation}
Each one-component fluid characterized by $\mathcal{\mathbb{Q}}_{c}$
of Eq. (\ref{asympt Qc vs Ltabfstar set (25)}), has the Ising-like
universal features of the $\Phi_{d=3}^{4}\left(n=1\right)$-model
in the Ising-like preasymptotic domain $\mathcal{L}_{\text{PAD}}^{\text{Xe}}$
given by\begin{equation}
\Delta\tau^{*}<\mathcal{L}_{\text{PAD}}^{\text{Xe}}\simeq\frac{1}{\vartheta}\mathcal{L}_{\text{PAD}}^{\text{Ising}}\cong\frac{1.9\times10^{-6}}{\vartheta}\label{system Ising PAD extension (35)}\end{equation}
Equation (\ref{system Ising PAD extension (35)}) demonstrates that
the knowledge of the temperature-like scale factor $\vartheta$ defines
the extension of the Ising-like preasymptotic domain of each selected
fluid, then providing an essential tool for analyzing experimental
data. Correlatively, admitting a single $\vartheta$ value whatever
the property and the thermal field range, the (fluid dependent) crossover
temperature $\Delta\tau_{\Delta}^{*}$ is given by\begin{equation}
\Delta\tau_{\Delta}^{*}\simeq\frac{1}{\vartheta}t_{\Delta}\cong\frac{1.9\times10^{-3}}{\vartheta}\label{system crossover temperature (36)}\end{equation}
and the crucial problem of how to define the temperature range $\Delta\tau^{*}\leq\mathcal{L}_{\text{PAD}}^{\text{Xe}}$
is solved. Appropriate rewriting of Eqs. (\ref{theta vs akhiplus (32)})
to (\ref{psi vs gamaplus (34)}), provides the following functional
scaling form \begin{equation}
\mathcal{\mathbb{S}}_{A}^{MR}=S_{A}\mathcal{\mathbb{F}}\left(\mathcal{\mathbb{S}}_{SF}\right)\;\text{with}\;\Delta\tau^{*}\lesssim\frac{1.9\times10^{-6}}{\vartheta}\label{QcalMR vs Qcalf (37)}\end{equation}
where $\mathcal{\mathbb{F}}\left(\mathcal{\mathbb{S}}_{SF}\right)$
are universal functions. We note the {}``theoretical'' (i.e. originating
only from the MR scheme) nature of the l.h.s. of Eq. (\ref{QcalMR vs Qcalf (37)}).

In next Section 3, we analyze the status of this expected three-scale-factor
characterization for the xenon case, first, using the values inferred
from the application of the scale dilatation method \citet{Garrabos1985},
and second, comparing the results \citet{Anisimov1995,Hahn2001} obtained
from the massive renormalization (MR) scheme (present work), the minimal
subtraction renormalization (MSR) scheme \citet{Zhong2003}, and the
crossover parametric model (CPM) \citet{Agayan2001}, with the experimental
results of Güttinger and Cannell.

\section{Xenon characterization \emph{within} the Ising-like preasymptotic
domain. }

\subsection{Hypothesized description of the Ising-like preasymptotic domain}

We hypothesize that the two terms of the asymptotic Wegner-like expansion
of $\kappa_{T,\text{exp}}^{*}\left(\Delta\tau^{*}\right)$ and{\small 
$\xi_{exp}\left(\Delta\tau^{*}\right)$} are exactly known for critical
xenon, provided by the application of the scale dilatation method
given in Appendix A. In such a situation, all the needed information
takes origin on four critical coordinates {[}see below Eq. (\ref{Xe critical coordinates (A2)})]
which localize the xenon critical point on the experimental phase
surface of equation $\Phi\left(p,v_{\bar{p}},T\right)=0$. Accordingly,
the (dimensional) values of Eqs. (\ref{energy unit (14)}) and (\ref{1f length scale unit (19)})
are the following\begin{equation}
\begin{array}{l}
T_{c}=289.733\,\text{K}\; or\;\left(\beta_{c}\right)^{-1}=4.0002\times10^{-21}\,\text{J}\\
\alpha_{c}=0.881508\,\text{nm}\end{array}\label{Xe scale units (38)}\end{equation}
leading to the dimensionless amplitude set\begin{equation}
S_{A}=\left\{ \begin{array}{l}
a_{\chi}^{+}=1.23397\\
\xi^{+}=0.209111\;\left(\xi_{0}^{+}=0.184333\,\text{nm}\right)\\
\Gamma^{+}=0.0578204\end{array}\right\} \label{Xe independent amplitudes (39)}\end{equation}
with $a_{\xi}^{+}=a_{\chi}^{+}\frac{\mathbb{Z}_{\ell}^{1,+}}{\mathbb{Z}_{\mathcal{X}}^{1,+}}=0.83810$.
Obviously, the validation of our hypotheses and the justification
of Eq. (\ref{Xe independent amplitudes (39)}) require the detailed
analysis of the isothermal compressibility data of xenon given in
Appendix A. However, we note that the essential aspects for the following
presentation are the Ising-like \emph{nature} (since the fluid characterization
originates from its critical point coordinates) and \emph{quantity}
(three) of the dimensionless amplitudes, while the quoted precision
of their numerical values is of secondary importance. Using Eqs. (\ref{theta vs akhiplus (32)})
to (\ref{psi vs gamaplus (34)}), the three dimensionless scale factors
for xenon are:\begin{equation}
\mathcal{\mathbb{S}}_{SF}=\left\{ \begin{array}{l}
\vartheta=0.021069\\
\mathbb{L}^{\left\{ 1f\right\} }=25.6936\;\left(g_{0}=29.1473\,\text{nm}^{-1}\right)\\
\psi_{\rho}=3.2507\,10^{-4}\end{array}\right\} \label{Xe independent scale factors (40)}\end{equation}
The singular behavior of $\xi^{*}\left(\Delta\tau^{*}\right)$ and
$\chi_{T}^{*}\left(\Delta\tau^{*}\right)$ of xenon can then be estimated
by using Eqs. (\ref{MR lstar vs tstar (1)}) to (\ref{Deff exponent MR universal (3)}),
(\ref{gzero vs elestar (15)}), and (\ref{(23)}), with xenon parameters
of Eq. (\ref{Xe independent amplitudes (39)}). Such an estimation
of dimensionless susceptibility (or dimensionless isothermal compressibility)
will be represented by a full black curve (with label $\text{MR}$)
in the next figures.

As a most important result already underlined, $\vartheta$ enables
estimation of the extension of the Ising-like preasymptotic domain
{[}see Eq. (\ref{system Ising PAD extension (35)})]\begin{equation}
\mathcal{L}_{\text{PAD}}^{\text{Xe}}\simeq10^{-4}\label{xenon Ising PAD extension (41)}\end{equation}
and the reduced crossover temperature {[}see Eq. (\ref{system crossover temperature (36)})]:\begin{equation}
\Delta\tau_{\Delta}^{*}\simeq10^{-1}\label{xenon crossover temperature (42)}\end{equation}
The temperature extension of the Ising-like preasymptotic domain of
xenon corresponds to the temperature range $T-T_{c}\lesssim30\,\text{mK}$
{[}from Eq. (\ref{xenon Ising PAD extension (41)})], while the Ising-like
predominant nature for crossover estimated by the mean crossover functions
cannot extend beyond $\sim T_{c}+30\,\text{K}$ {[}from Eq. (\ref{xenon crossover temperature (42)})].
More generally, the knowledge of $\vartheta$ also enables useful
estimation of any relative temperature distance and any effective
value attached to a specific condition of the mean crossover function
on the complete $\Delta\tau^{*}$ range, as reported for example in
lines 7 to 9 of Table \ref{Table II}.

\begin{table*}
\begin{tabular}{|c||c|c|c|c|c|c|c|c|}
\hline 
 & $\gamma$ & $\Delta$ & $\Gamma^{+}$ & $a_{\chi}^{+}$ & $\vartheta$ & $\mathcal{\mathbb{L}}^{\left\{ 1f\right\} }$ & $\psi_{\rho}$ & \tabularnewline
$X$ &  &  &  &  & Eq. (\ref{theta vs akhiplus (32)}) & Eq. (\ref{gzero vs ksizeroplus (33)}) & Eq. (\ref{psi vs gamaplus (34)}) & $Ref$\tabularnewline
\hline
\hline 
$\text{GC}_{\text{RG4}}$ & $1.241$ & $0.496$ & $0.0577$ & $1.29$ & $0.0230$ & $24.305$ & $ $ & \citet{Guttinger1981}\tabularnewline
(pink) &  &  &  &  &  & $\xi_{0}^{+}=0.184\,\text{nm}$ &  & \tabularnewline
\hline 
$\text{MR6}_{\text{max}}$ & $1.24194$ & $0.491$ & $0.057$ & $1.1844$ & $0.0194$ & $24.305$ & $3.25\times10^{-4}$ & \citet{Bagnuls1984b}\tabularnewline
 &  &  &  &  &  & $\xi_{0}^{+}=0.184\,\text{nm}$ &  & \tabularnewline
\hline 
$\text{MR}$ & $1.2395935$ & $0.50189$ & $0.057821$ & $1.23397$ & $0.02107$ & $25.6936$ & $3.25\times10^{-4}$ & this work\tabularnewline
 &  &  &  &  &  & $\xi_{0}^{+}=0.184333\,\text{nm}$ &  & \tabularnewline
\hline 
$\text{MSR}$ & $1.2396$ & $0.504$ & $0.0587$ & $1.11$ & $0.0171$ &  & $ $ & \citet{Hahn2001}\tabularnewline
 &  &  &  &  &  & $\xi_{0}^{+}=0.184\,\text{nm}$ &  & \tabularnewline
\hline 
$\text{CPM}$ & $1.239$ & $0.51$ & $\text{n.a.}$ & $1.08$ & $0.0162$ & $30.382$ & $ $ & \citet{Anisimov1995}\tabularnewline
 &  &  & $\left(0.058\right)^{*}$ &  &  & $\xi_{0}^{+}=0.184\,\text{nm}\,\left(\text{calculated}\right)$ &  & \tabularnewline
\hline 
$\text{LM}$ & $1.240$ & $0.508$ & $0.0594$ & $\text{n.a.}$ &  &  &  & \citet{Luijten2000}\tabularnewline
 &  &  &  & $\left(0.9\right)^{*}$ &  & $\text{n.a.}$ &  & \tabularnewline
\hline
\end{tabular}

\caption{Amplitude-exponent results (colums 3 to 6) of the fit of the isothermal
compressibility data obtained from the turbidity measurements of Güttinger
and Cannell \citet{Guttinger1981}, along the critical isochore, using
different theoretical crossover models (labels $X$, column 1) proposed
in references given in the last column. Corresponding scale factor
values (columns 7 to 9) calculated from Eqs. (\ref{theta vs akhiplus (32)})
to (\ref{psi vs gamaplus (34)}) (see text for detail); n.a.: nonavailable.
; asterisk indicates a value used in the present work. \label{Table III}}

\end{table*}

\begin{table*}
\begin{tabular}{|cc||c|c|c|c|c|c|c|c|}
\hline 
 &  & $\gamma$ & $\Delta$ & $\Gamma^{+}$ & $a_{1\chi}^{+}=a_{\chi}^{+}$ & $a_{2\chi}^{+}$ & $a_{3\chi}^{+}$ & $T_{c}$ & \tabularnewline
\# (fit) &  &  &  &  &  &  &  & $\left(16.64\text{°C}\right)$ & $Ref$\tabularnewline
\hline
\hline 
$\text{GC}_{f2}$ & pink & $1.240\pm0.002$ & $\left(0.496\right)$ & $0.0584\pm0.0009$ & $1.07\pm0.06$ & $\left(0\right)$ & $\left(0\right)$ & $\pm0.5\,\text{mK}$ & \citet{Guttinger1981}\tabularnewline
\hline
$\text{GC}_{f4}$ & pink & $1.246\pm0.002$ & $\left(0.496\right)$ & $0.0551\pm0.0012$ & $1.62\pm0.14$ & $-2.7\pm0.5$ & $3.6\pm0.8$ & $\pm0.5\,\text{mK}$ & \citet{Guttinger1981}\tabularnewline
\hline
$\text{GC}_{4}$ & pink & $\left(1.241\right)$ & $\left(0.496\right)$ & $0.0577\pm0.0001$ & $1.29\pm0.03$ & $-1.55\pm0.2$ & $1.9\pm0.5$ & $\pm0.5\,\text{mK}$ & \citet{Guttinger1981}\tabularnewline
\hline
$\text{G}_{3}$ &  & $\left(1.240\right)$ & $\left(0.5\right)$ & $0.0574$ & $1.55$ & $-2.0$ & $\left(0\right)$ &  & \citet{Garrabos1982}\tabularnewline
\hline
\end{tabular}

\caption{Lines labeled $\text{GC}_{f2}$, $\text{GC}_{f4}$, $\text{GC}_{4}$:
Amplitude-exponent values obtained by Güttinger and Cannell from three
representative fitting by Eq. (\ref{khiTexp GC (43)}) of the isothermal
compressibility data for xenon along the critical isochore; Fixed
values of the parameters are given between brackets; Line labeled
$\text{G}_{3}$: Amplitude-exponent values obtained by a three-point
calibration method proposed in Ref. \citet{Garrabos1982} and discussed
in Appendix. \label{Table IV}}

\end{table*}

Comparable values to the ones given by Eq. (\ref{Xe independent amplitudes (39)})
can be found in several published papers \emph{but without explicit
reference to the effective temperature range of the Ising-like preasymptotic
domain}.

For example, in upper part (a) of Table \ref{Table III} , we have
reported $\Gamma^{+}$ (column 3) and $a_{\chi}^{+}$ (column 5) values
obtained by using different theoretical crossover functions calculated
by several models labeled $X=\left\{ \text{GC}_{\text{RG4}},\text{MR6}_{\text{max}},\text{MR},\text{MSR},\text{CPM},\text{LM}\right\} $
(column 1) of respective Refs. \citet{Bagnuls1984b,Anisimov1995,Luijten2000,Hahn2001}
(column 9). These values result from fits of the isothermal susceptibility
data of xenon published by Güttinger and Cannell (GC) \citet{Guttinger1981},
with $\gamma$ (column 2) and $\Delta$ (column 4) values fixed to
their theoretical estimation. We note that the variations of the $\Gamma^{+}$
values are on a few percent level when the change in $\gamma$ values
affects the third digit, while the variations of the $a_{\chi}^{+}$
values cover a significative range such as $0.9\lesssim a_{\chi}^{+}\lesssim1.3$,
i.e., $a_{\chi}^{+}\simeq1.1$ with $\sim20\%$ deviation. Then, we
have also reported in Table \ref{Table III}, the calculated values
of the asymptotic scale factors $\vartheta$ (column 6), $\mathcal{\mathbb{L}}^{\left\{ 1f\right\} }$
(column 7), and $\psi_{\rho}$ (column 8), by using Eqs. (\ref{theta vs akhiplus (32)})
to (\ref{psi vs gamaplus (34)}), respectively, and an available estimation
of the leading amplitude $\xi_{0}^{+}=\alpha_{c}\xi^{+}$ of the correlation
length (see column 7). We can observe a typical uncertainty of $25\%$
in the $\vartheta$ and $\psi_{\rho}$ values. In addition, the theoretical
error-bars and differences in the estimations of exponents $\gamma$,
$\nu$, $\Delta$, and universal constants appearing in Eqs. (\ref{theta vs akhiplus (32)})
to (\ref{psi vs gamaplus (34)}), can have comparable effect on the
estimation of $\vartheta$, $\psi_{\rho}$, and $g_{0}$ than the
experimental uncertainties on the estimation of $a_{\chi}^{+}$, $\Gamma^{+}$,
and $\xi^{+}$ and the critical coordinates (such as $T_{c}$, $\rho_{c}$,
$p_{c}$, etc.).

Nevertheless, accounting for the above variations of the $\vartheta$
values has no significant effect on the order of magnitude of $\mathcal{L}_{\text{PAD}}^{\text{Xe}}\simeq10^{-4}$
estimated from Eqs. (\ref{system Ising PAD extension (35)}) and (\ref{Xe independent amplitudes (39)}).
We can then note that the apparent amplitude agreement arises in spite
of the \emph{non overlap} between the estimated temperature range
of the Ising-like preasymptotic domain {[}see Eq. (\ref{xenon Ising PAD extension (41)})]
and the temperature range $10^{-4}\lesssim\Delta\tau^{*}\lesssim10^{-1}$
{[}i.e., $30\,\text{mK}\lesssim T-T_{c}\lesssim30\,\text{K}$], covered
by the Güttinger and Cannell's measurements (see also Appendix A).
Therefore, the asymptotic amplitude evaluation provided by these fitting
results cannot be easily transformed in terms of the characteristic
scale factors for two main reasons:

i) measurements of the singular properties are made in a temperature
range which do not reach the Ising-like preasymptotic domain; 

ii) fitting of the data is made with a Wegner-like expansion, whose
validity is questionable outside the Ising-like preasymptotic domain
(see I).

Therefore, to replace with a suitable precision the independent amplitudes
of Eqs. (\ref{Xe independent amplitudes (39)}) by the independent
scale factors of Eqs. (\ref{Xe independent scale factors (40)}),
one needs to give the {}``rules'' to interpolate from fitting results
obtained in an experimental range \emph{outside} the Ising-like preasymptotic
domain, to the hypothesized ones, only valid \emph{inside} the Ising-like
preasymptotic domain.

\subsection{Present status of crossover modeling in critical xenon}

\subsubsection{Güttinger and Cannell's analysis}

\begin{figure}
\includegraphics{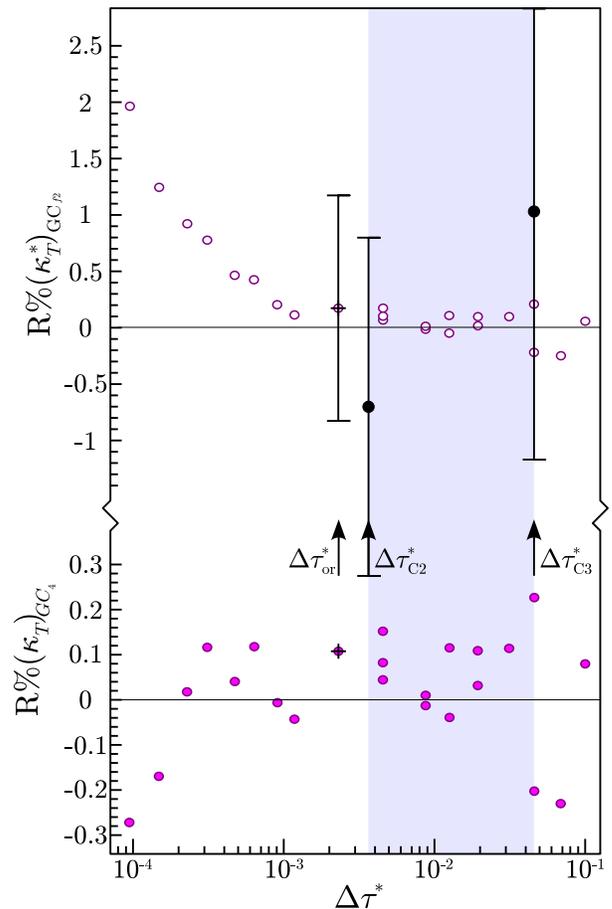}

\caption{Residuals $R\%\left(\kappa_{T}^{*}\right)_{\text{GC}_{4}}$ (expressed
in \%) between the dimensionless susceptibility data given in Table
I of Ref. \citet{Guttinger1981} and the fitting expression of Eq.
(\ref{xenon crossover temperature (42)}); a) parameters given in
line $\text{GC}_{4}$ of Table \ref{Table IV}; b) after the $T_{c}$
shift of $0.5\,\text{mK}$ to retrieve the $0.2\%$ numerical level
illustrated in Fig. 3 of Ref. \citet{Guttinger1981}; In (a) are also
illustrated the respectives deviations and related error-bars of the
three {}``calibrated'' values of the isothermal compressibility
for the three temperature distances indicated by the respective vertical
arrows labeled $\Delta\tau_{\text{or}}^{*}$ , $\Delta\tau_{\text{C2}}^{*}$
and $\Delta\tau_{\text{C3}}^{*}$ (see text and Appendix).\label{Figure1}}

\end{figure}

Due to the primary importance of Eq. (\ref{theta vs akhiplus (32)})
in a scheme where a single confluent amplitude is readily expected
independent, in the Table \ref{Table IV}, we have recalled the main
informations given from Güttinger and Cannell's experiment, which
is the common support to compare theoretical fittings. Güttinger and
Cannell observed that their susceptibility data systematically and
continuously deviate from simple power-law behavior throughout the
temperature range $9.6\times10^{-5}\lesssim\Delta\tau^{*}\lesssim10^{-1}$
with the effective exponent increasing from $\gamma_{e}=1.14$ to
$\gamma=1.246\pm0.01$. Thus the data appeared asymptotically consistent
with the theoretical estimations provided at the end of the seventies,
either using series calculation $\gamma=1.250\pm0.03$ for the Ising-model
\citet{Camp1976,Gaunt1979}, or the renormalization-group result $\gamma=1.241\pm0.002$
for the Landau-Ginzburg-Wilson Hamiltonian \citet{Baker1976,LeGuillou1977}.
To support this conclusion, their susceptibility data were fitted
to the following (four term) Wegner-like expansion\begin{equation}
\begin{array}{cc}
\kappa_{T,\text{exp}}^{*}\left(\Delta\tau^{*}\right)= & \Gamma^{+}\left(\Delta\tau^{*}\right)^{-\gamma}\left[1+a_{1\chi}^{+}\left(\Delta\tau^{*}\right)^{\Delta}+\right.\\
 & \left.a_{2\chi}^{+}\left(\Delta\tau^{*}\right)^{2\Delta}+a_{3\chi}^{+}\left(\Delta\tau^{*}\right)^{3\Delta}\right]\end{array}\label{khiTexp GC (43)}\end{equation}
with $\Delta$ fixed at $\Delta=0.496$ \citet{LeGuillou1977} which
was the theoretical value calculated at that time by the renormalization-group
approach. In initial fittings, the parameters $\Gamma^{+}$, $\gamma$,
$a_{1\chi}^{+}$, $a_{2\chi}^{+}$, $a_{3\chi}^{+}$, and $T_{c}$
were independently adjusted, by variation of the data range in temperature.
It was then shown that when the data range was narrowed to $9.6\times10^{-5}\lesssim\Delta\tau^{*}\lesssim8.8\times10^{-3}$
and the parameter $a_{2\chi}^{+}$ and $a_{3\chi}^{+}$ removed (i.e.,
with $a_{2\chi}^{+}=a_{3\chi}^{+}=0$), the fit resulted in $\gamma=1.240\pm0.002$
when $\gamma$ was freely adjusted, and the results for $\Gamma^{+}$,
$a_{1\chi}^{+}$ and $T_{c}$ are in good agreement with all the other
fits. This fit result is labeled $\text{GC}_{f2}$ and given in the
corresponding line $\text{GC}_{f2}$ of Table \ref{Table IV} where
the error bars quoted are one standard deviation allowing for the
correlation between parameters, not the experimental uncertainty which
increases for the data nearest $T_{c}$, as discussed below. In fitting
over the entire range, all the terms were required and the fit results
(labeled $\text{GC}_{f4}$) are given in line $\text{GC}_{f4}$ of
Table \ref{Table IV}. The Güttinger and Cannell's analysis to examine
the consistency with the predictions by renormalization-group calculations
was then made with $\gamma$ fixed at $1.241$. The parameters for
the fit result (labeled $\text{GC}_{4}$) for the entire range in
$\Delta\tau^{*}$ are reported in line $\text{GC}_{4}$ of Table \ref{Table IV}.

As noted by the authors, in all the fitting cases, the value of $T_{c}$
found by the fitting procedure agreed to within $0.5\,\text{mK}$
with the value observed by noting the temperature at which the meniscus
formed upon cooling in small steps. It should be also noted that a
shift of $0.5\,\text{mK}$ in $T_{c}$ amounts to a $2.5\%$ change
in fitted value at $\Delta\tau^{*}=9.6\times10^{-5}$, which was a
substantial effect in comparison to $0.2\%$ deviations reported for
all the fits (see Figure 3 in Ref. \citet{Guttinger1981}). More generally,
in the temperature range $\Delta\tau^{*}\lesssim9.1\times10^{-4}$,
the authors have accounted for the large uncertainty in the corrections
due to all the possible effects which are not at the level of about
$0.2\%$, by using a shift in $T_{c}$ in the fitting program as a
mean of compensating the systematic errors which increase for the
points nearest $T_{c}$. However, for the present discusion, it is
essential to account for the real level of the fit deviation attached
to the systematic experimental uncertainties. Therefore, we have illustrated
in Figure \ref{Figure1}, the importance of the real uncertainty using
the residuals (expressed in \%) $R\%\left(\kappa_{T}^{*}\right)_{\text{GC}_{4}}=100\left(\frac{\kappa_{T}^{*}}{\kappa_{T,\text{GC}_{4}}^{*}}-1\right)$
betwenn the raw data of the dimensionless susceptibility given in
Table 1 of Ref. \citet{Guttinger1981} and the calculated ones from
Eq. (\ref{khiTexp GC (43)}) with the parameters given in line $\text{GC}_{4}$.
The part a of Fig. \ref{Figure1}, illustrates the $2\%-3\%$ level
of the true experimental error at $\Delta\tau^{*}=9.6\times10^{-5}$,
while the part b of Fig. \ref{Figure1} illustrates similar residuals,
but after the $T_{c}$ shift of $0.5\,\text{mK}$, then reproducing
Figure 3 of Ref. \citet{Guttinger1981} to show the $0.2\%$ numerical
level of the fit deviation.

In summary, comparing the three fit results of Table \ref{Table IV},
Güttinger and Cannell have shown that the xenon susceptibility was
correctly represented by a Wegner-like expansion whose the two first
terms may be interpreted as proposed by the renormalization-group
theory with $\gamma=1.241$ and $\Delta=0.496$ \citet{LeGuillou1977}.
Despite a significative increase of experimental uncertainty as $\Delta\tau^{*}$
decreases, the uncertainty on the {}``mean'' determination of the
free leading amplitude $\Gamma^{+}$ value can be estimated of the
order of $2\%-3\%$, while the first confluent amplitude $a_{\chi}^{+}$
value was found in the range $0.6\lesssim a_{\chi}^{+}\lesssim1.45$
and seriously affected by the presence of the other two confluent
parameters and the fitted temperature range. The comparison between
$\Gamma^{+}$ and $a_{\chi}^{+}$ values given in Tables \ref{Table III}
and \ref{Table IV}, especially the low dispersion of the $\Gamma^{+}$-values
in a range $0.0570\lesssim\Gamma^{+}\lesssim0.0594$, confirms an
apparent reduction of the $a_{\chi}^{+}$-range, increasing the number
of terms of the Wegner like expansion.

\subsubsection{Crossover analyses}

\begin{figure}
\includegraphics[height=80mm,keepaspectratio]{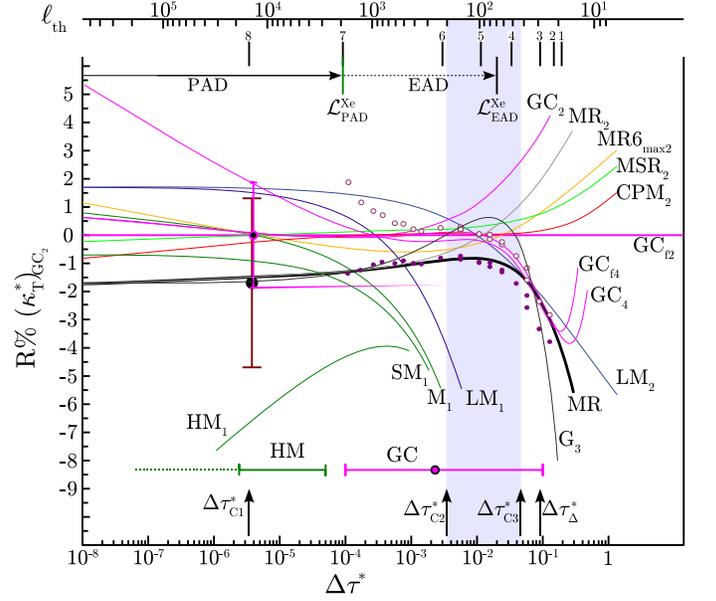}

\caption{(Color on line) Residuals $R\%\left(\kappa_{T}^{*}\right)_{\text{GC}_{f2}}$
(expressed in \%), as a function of $\Delta\tau^{*}$, for the extrapolated
description of the isothermal compressibility of xenon within the
Ising-like preasymptotic domain $\Delta\tau^{*}\lesssim\mathcal{L}_{\text{PAD}}^{\text{Xe}}$.
Each curve label $X$ refers to the line $X$ in Table \ref{Table III};
subscript indicates the number of terms of Wegner-like expansion (see
also the text); Arrows indicate the different extensions of the Ising
like preasymptotic domain (see text); horizontal (pink) segment: Güttinger
and Cannell experimental range \citet{Guttinger1981}; open circle:
temperature distance $\Delta\tau_{\text{or}}^{*}$ (see Fig. \ref{Figure1})
of the data calibration value; vertical (pink) segment: error bar
of a calibrated valueof the isothermal compressibility of xenon from
Hocken and Moldover measurements \citet{Hocken1976,Sengers1978} (see
text, Appendix A and Ref. \citet{Garrabos1982}); upper horizontal
axis: theoretical dimensionless correlation length calculated from
the mean crossover function of Eq. (\ref{MR lstar vs tstar (1)});
line indexation with labels 1 to 8: $\gamma_{e}$- or $\ell_{\text{th}}$-conditions
reported in Table \ref{Table II} (see text)\label{Figure2}}

\end{figure}

Since all the amplitude values given in Table \ref{Table III} originate
from the same experimental data, we consider that the extrapolation
at $\Delta\tau^{*}\rightarrow0$ of the two-term expansion $\kappa_{T,\text{GC}_{f2}}^{*}$
obtained with the free parameters given in line $\text{GC}_{f2}$
of Table \ref{Table III}, can act as an {}``ideal'' experimental
result obtained \emph{within} the Ising-like asymptotic domain. Making
then reference to this extrapolated experimental behavior, the relative
comparison with the predicted one using either a pure power law, or
a two-term Wegner-like expansion, or a more extended Wegner-like expansion,
or finally a complete crossover function, associated to $X=\left\{ \text{MR6}_{\text{max}},\text{MR},\text{MSR},\text{CPM},\text{LM}\right\} $
models (see Table \ref{Table III}), provides the required information
to evaluate the Ising-like characterization in the ranges $\Delta\tau^{*}<\mathcal{L}_{\text{PAD}}^{\text{Xe}}\simeq10^{-4}$
and $\Delta\tau^{*}\gtrsim\mathcal{L}_{\text{PAD}}^{\text{Xe}}\simeq10^{-4}$,
separately. We use the subscript 1, 2, etc., to refer for one-, two-,
etc., terms in the Wegner like expansion, while the abscence of this
subscript indicates a complete crossover function (except for the
$\text{MR6}_{\text{max}}$ case where the crossover function does
not account for the classical behavior close to the Gaussian fixed
point). So that, anticipating the discussion given in Appendix, we
have systematically enlighted (using a grey-blue area in the next
figures), the temperature range in between the two temperatures $T=T_{c}+1\,\text{K}$
(i.e., $\Delta\tau_{\text{C2}}^{*}=3.45137\times10^{-3}$) and $T=T_{c}+13.41\,\text{K}$
(i.e., $\Delta\tau_{\text{C3}}^{*}=4.62829\times10^{-2}$) where the
$pVT$ measurements can provide two {}``standard'' values of the
isothermal compressibility (see § A.3.b). These two temperatures are
in the experimental range covered by the optical measurements of Güttinger
and Cannell for $\Delta\tau^{*}\gtrsim\mathcal{L}_{\text{PAD}}^{\text{Xe}}\simeq10^{-4}$.

Similarly, for $\Delta\tau^{*}<\mathcal{L}_{\text{PAD}}^{\text{Xe}}\simeq10^{-4}$,
we have also indicated in the following figures, the temperature $T=T_{c}+1\,\text{mK}$
(i.e., $\Delta\tau_{\text{C1}}^{*}=3.45137\times10^{-6}$) {}``closest''
to $T_{c}$, where a carefull analysis of the uncertainty on the theoretical
estimations of the Ising-like exponent $\gamma$ provides a calibrated
value of the isothermal compressibility (see below and § A.3.a). This
value is compatible with the interferometry measurements of Hocken
and Moldover covering the temperature range $-5\, mK<T-T_{c}<15\, mK$
(i.e., $-1.5\times10^{-5}<\Delta\tau^{*}<5\times10^{-5}$) \citet{Hocken1976}.
As already shown, such a three point calibration can be used to define
a practical three-term Wegner expansion with arbitrarily {}``fixed''
values $\gamma=\gamma_{I}=1.240$ and $\Delta=\Delta_{I}=0.5$ for
the exponents, which accounts for the isothermal compressibility of
xenon in the experimental temperature range $\Delta\tau^{*}<\Delta\tau_{\text{C3}}^{*}$,
with $\pm1.5\%$ uncertainty on the {}``online'' values produced
by the distinct measurement methods around these three temperature
of calibration. The corresponding amplitude parameters of Eq. (\ref{khiTexp GC (43)})
are given in line $\text{G}_{3}$ of Table \ref{Table IV}, and each
associated fitting curve will be labeled $\text{G}_{3}$ in the next
figures.

In Figure \ref{Figure2}, we have reported the residuals (expressed
in \%), $R\%\left(\kappa_{T}^{*}\right)_{\text{GC}_{f2}}=100\left(\frac{\kappa_{T,X}^{*}}{\kappa_{T,\text{GC}_{f2}}^{*}}-1\right)$,
as a function of $\Delta\tau^{*}$. Note that the indexation of the
(lower and upper) horizontal axes of this figure are conform with
the discussion of Table \ref{Table II}, while the experimental range
of the GC measurements is given by the segment labeled GC where the
open point indicates the temperature distance (noted $\Delta\tau_{\text{or}}^{*}$
in the following) where ligth scattering data are calibrated by the
authors. The specific analysis of the residuals obtained for the two-term
expansion fitting (see all the curves labeled $X_{2}$ in Fig. \ref{Figure2})
demonstrates that the deviations in the temperature range $\mathcal{L}_{\text{PAD}}^{\text{Xe}}<\Delta\tau^{*}\lesssim10^{-3}-10^{-2}$,
induce a $\pm\left(1\%-2\%\right)$ uncertainty level in the estimation
of the $\Gamma^{+}$ value, which means that the value of the confluent
amplitude $a_{\chi}^{+}$ is {}``undetermined'' due to the correlation
between \emph{free} leading amplitude and \emph{fixed} theoretical
value of $\gamma$. 

More generally, Figure \ref{Figure2} is a well-defined tool to show
the relative importance of each term of the Wegner like expansion
as a function of each decade variation of the theoretical correlation
length (see upper axis). For example, the successive deviations at
a $1\%-2\%$ level of the curves $X_{1}$, $X_{2}$, ...., \emph{$X$,}
indicates that:

i) the pure power law can be used in the range $\ell_{\text{th}}\gtrsim\left(1-2\right)\times10^{4}$,

ii) the range $\ell_{\text{th}}\gtrsim1000-2000$ needs to add the
first confluent term and effectively corresponds to the Ising-like
preasymptotic domain,

iii) a three-term-expansion seems sufficient to cover the range $\ell_{\text{th}}\gtrsim100-200$,
and, finally,

iv) the range $\ell_{\text{th}}\gtrsim10-20$ needs to use at least
four or more terms in the Wegner-like expansion. 

To readily evaluate the relative importance of the effects due to
the $\gamma\Leftrightarrow\Gamma^{+}$ correlation, we note that the
magnitude of the difference in $\gamma$ values here considered affects
only the third digit of the theoretical estimations and cannot be
related to the realistic error-bar (one order of magnitude greater)
provided by the experimental determination of this asymptotic exponent
(see also below the related insert of Fig. \ref{Figure3}). Therefore,
in spite of the fact that measurements were performed beyond the Ising-like
preasymptotic domain, the closest extrapolated Güttinger and Cannell's
fitting to $T_{c}$ remains highly correlated to the theoretical value
of $\gamma$, especially when \emph{$T_{c}$} is fixed. This exponent-amplitude
correlation was recently used by Luijten and Meyer \citet{Luijten2000}
to re-evaluate the estimation of the leading amplitude for better
agreement with the data for $\Delta\tau^{*}<0.2$ (see curve $\text{LM}_{1}$
in Fig. \ref{Figure2}). However, such amplitude adjustment accounting
for the data in a narrowed temperature range appears equivalent to
the reverse effect of the theoretical min and max error-bar propagation
outside the Ising-like preasymptotic domain clearly shown in reference
I (or in the Fig. \ref{Figure7}b of Appendix). Such an amplitude
re-evaluation is then without gain on the Ising-like asymptotic quality
of the crossover analysis. For example, only by addition of the first
term for confluent corrections to scaling with $a_{\chi}^{+}=0.86$
(see Table \ref{Table III}, line LM), we can fit the experimental
results (with deviations at a $0.5\%$ level) on the complete experimental
range (see curve $\text{LM}_{2}$ in Fig. \ref{Figure2}). The comparison
of the curves $\text{LM}_{2}$, $\text{GC}_{f4}$, and $\text{GC}_{4}$
in Fig. \ref{Figure2} then shows that this {}``LM'' fitting result
which uses the parameters given in line $\text{LM}$ of Table \ref{Table IV},
agrees with the initial four-term fitting results given in lines $\text{GC}_{f4}$,
and $\text{GC}_{4}$ of Table \ref{Table IV}. The respective ranges
of the amplitude variations, $\Gamma^{+}=0.0584_{-0.0033}^{+0.0010}$
and $a_{\chi}^{+}=1.07_{-0.25}^{+0.55}$, illustrate the real difficulty
to account for a single specific value of $\gamma$ in the range $\gamma=1.24_{-0.001}^{+0.006}$.
More generally, as illustrated previously in Fig. \ref{Figure1},
an {}``experimental'' manner to account for each \emph{fixed} value
of $\gamma$ is to use $T_{c}$ as a \emph{free parameter} in fitting
the GC's data. But, as observed by Güttinger and Cannell from their
different fitting results, the true uncertainty on the free parameter
$\Gamma^{+}$ is then at a few percent level. Finally, we can conclude
that, due to the $2\%$ uncertainty of the GC's measurements at the
border of the Ising-like preasymptotic domain, the $1\%-2\%$ magnitude
of the residuals for extrapolated theoretical fittings at $\Delta\tau^{*}\left(\lesssim\mathcal{L}_{\text{PAD}}^{\text{Xe}}\right)\rightarrow0$,
is entirely related to the differences $\gamma_{X}-1.240$ between
respective leading exponents, while the apparent reduction of the
uncertainty in the $a_{\chi}^{+}$-value results from the mandatory
presence of (at least two) supplementary terms to correctly account
($\sim1\%$-level) for the experimental results in the temperature
range $\Delta\tau^{*}\gtrsim10^{3}>\mathcal{L}_{\text{PAD}}^{\text{Xe}}\simeq10^{-4}$.

\subsubsection{Analysis with $\gamma$ fixed}

For a fixed theoretical value of $\gamma$ \emph{with} $T_{c}$ \emph{fixed},
the leading amplitude $\Gamma^{+}$ can be obtained only using additional
measurements performed well inside the Ising-like preasymptotic domain
as defined by Eq. (\ref{model Ising PAD extension (9)}). The applicability
of Eq. (\ref{psi vs gamaplus (34)}) {[}equivalently, Eq. (\ref{gzero vs ksizeroplus (33)})
for the correlation length data] is then strictly limited within the
Ising-like preasymptotic domain, \emph{thus requiring a precision}
$<0.1\%$. In the absence of such an {}``ideal'' experimental result,
one alternative way to eliminate the {}``amplitude-exponent'' correlation
on the leading power law term, is to introduce a smallest temperature
distance to the critical temperature where a single experimental data
for the isothermal compressibility acts as a standard value \citet{Garrabos1982}.
We note $\Delta\tau_{\text{C1}}^{*}$ as this temperature distance
such that $\Delta\tau_{\text{C1}}^{*}\ll\mathcal{L}_{\text{PAD}}^{\text{Xe}}$
(see column 9, Table \ref{Table II}). For example, this method of
calibration is detailed for xenon case in Appendix A, using the interferometric
measurements of Hocken and Moldover (HM) \citet{Hocken1976} in the
range $2-3\times10^{-6}\lesssim\Delta\tau^{*}\lesssim3-5\times10^{-5}$.
In Fig. \ref{Figure2}, this experimental range is illustrated by
the segment labeled HM. The selected smallest temperature distance
is $T-T_{c}=1\,\text{mK}$, or $\Delta\tau_{\text{C1}}^{*}\left(\text{Xe}\right)=3.45\times10^{-6}$,
which effectively appears within the Ising-like preasymptotic domain,
as evidenced by the corresponding vertical dashed line in Figure \ref{Figure2}.
Unfortunatly, the required precision is not reached and here we only
discuss the relative comparison between both (HM and GC) data measurements
because there is no region of overlap between them and the two experiments
rely on completely different effects. The interferometry measurements
were performed in the very limited region of density and temperature
quite close to the critical point such that the susceptibility data
was represented by a pure power law with a highly correlated amplitude-exponent
set. For example, the curve $\text{HM}_{1}$ in Figure \ref{Figure2}
illustrates the deviation obtained using the result $\gamma=1.23$
and $\Gamma^{+}=0.062$ initialy found by Hocken and Moldover. The
Hocken and Moldover measurements, reanalyzed by Sengers and Moldover
(SM) with $\gamma=1.24$, yield $\Gamma^{+}=0.058$, leading to the
curve $\text{SM}_{1}$ in Figure \ref{Figure2}. At $\Delta\tau_{\text{C1}}^{*}\left(\text{Xe}\right)=3.45\times10^{-6}$
the deviation of $\sim6\%$ between $\text{HM}_{1}$ and $\text{SM}_{1}$
can be considered as representative of the experimental uncertainty.
These results seems to systematically deviate from fitting the raw
data (full pink points) of Güttinger and Cannell's measurements. Moreover,
accounting for the fitting results illustrated by the curves $\text{LM}_{2}$
and/or $\text{GC}_{f4}$, such a deviation appears to increase as
the $\gamma$ value increases (or as the $\Gamma^{+}$ value increases,
equivalently).

However, the Güttinger and Cannell's data are relative to the isothermal
compresibility value $\kappa_{T}\left(\Delta\tau_{\text{or}}^{*}\right)=1.9641\times10^{-5}\,\text{Pa}^{-1}$
at $\Delta\tau_{\text{or}}^{*}=2.304\times10^{-3}$ (i.e., $T_{\text{or}}=T_{c}+0.6677\,\text{K}$)
\citet{Guttinger1981}, and the typical experimental uncertainty on
the above reference value also is at a percent level. As a typical
example, the difference on the GC value $\rho_{c}\left(\text{GC}\right)=1110\,\text{kg}\,\text{m}^{-3}$
and our present value $\rho_{c}=1113.\,\text{kg}\,\text{m}^{-3}$
(see Eq. (\ref{Xe critical coordinates (A2)}) in Appendix) of the
xenon critical density, contributes for a $0.6\%$ difference on the
calibration. It is then interesting to evaluate the cumulative effect
of a $0.5\,\text{mK}$ shift of $T_{c}$ and a $1\%$ change (uncertainty
given by Güttinger and Cannell) of the reference value. The result
is represented by the open pink points in Fig. \ref{Figure2}. The
fit deviation of these {}``corrected'' points using our theoretical
function is lower than $0.5\%$, as illustrated by the curve labeled
MR in Fig. \ref{Figure2}.

Subsidiarily, Fig. \ref{Figure2} also shows that the available fitting
results of the Güttinger and Cannell's data do not lead to the same
standard value used to calibrate the mean crossover function. However,
the attached relative difference appears compatible with an estimated
error-bar of $\pm3\%$ for this standard value, which is comparable
to the fitting uncertainty attached to the leading amplitude due to
the theoretical uncertainty on the $\gamma$ estimation. For example,
considering the nine published pairs $\gamma-\Gamma^{+}$ (see Tables
\ref{Table III}, \ref{Table IV} and the interferometry result of
Refs \citet{Sengers1978}) where $1.239\leq\gamma\leq1.24194$ and
$0.056\leq\Gamma^{+}\leq0.0594$, the calculated mean value of the
dimensionless isothermal compressibility at $T=T_{c}+1\,\text{mK}$
is $\kappa_{T,mean}^{*}=3.4330\times10^{5}$, with $3.3194\times10^{5}\leq\kappa_{T,mean}^{*}\leq3.5210\times10^{5}$.
Anticipating then the introduction of the calibrated value $\kappa_{T,C1}^{*}=3.415\times10^{5}$
discussed in Appendix A, we note that the above mean value is $0.53\%$
greater than the calibrated one, while the max and min values are
$3.10\%$ greater and $2.80\%$ lower, respectively.

As an undeniable consequence of the differences $\gamma_{X}-1.240$,
(or the difference in leading amplitude, or the absence of one precise
data inside the Ising-like preasymptotic domain, equivalently), the
contribution of the highly correlated leading and first-confluent
terms propagates in any fit procedure which uses a complete Wegner-like
expansion outside the Ising-like preasymptotic domain, until a temperature
distance of the order of $\Delta\tau^{*}\simeq10^{-3}$. We can then
reasonably estimate (and precisely demonstrate below, in § 4.2 and
Fig. \ref{Figure4}) that the fitting optimization occurs at finite
temperature distance to the critical point (at least of the order
of $\Delta\tau^{*}\simeq10^{-2}$), significantly outside the Ising-like
preasymptotic domain (and such as $\Delta\tau^{*}>\Delta\tau_{\text{or}}^{*}$).
In such a situation already known as due to the low convergence of
the Wegner-like expansion, it is not easy to define with the needed
precision a \emph{single} crossover parameter which must account simultaneously
for two opposite roles:

1) it controls the asymptotic universal features of the Ising-like
singular behavior which has the restricted form of Eq. (\ref{MR 2term khistar (5)}),
but without constraint due to the lack of experimental data in the
corresponding validity range;

2) it detects the finite distance where classical-to-critical behavior
and/or non-critical behavior can occur, but may conjointly invalidate
the use of the Wegner-like expansion with unique lowest confluent
exponent.

Therefore, in the absence of the {}``ideal'' experiment within the
Ising-like preasymptotic domain, the universal form of Eq. (\ref{theta vs akhiplus (32)})
\emph{can never be directly tested}. This has three correlative consequences:

i) the numerical value of $\vartheta$ is undoubtedly related to the
effects of many correction terms to scaling (for a more detailed illustration,
see the discussion in terms of the effective exponent and Fig. 4 of
Ref. \citet{Bagnuls1985}). To account for this essential result in
the following, we note $\vartheta_{\mathcal{L}}$ as the crossover
parameter which is determined beyond the Ising-like preasymptotic
domain, to distinguish it from $\vartheta$ of Eq. (\ref{theta from tstar vs deltataustar (13)})
which is the scale factor needed to characterize the asymptotic critical
crossover within the Ising-like preasymptotic domain;

ii) the use of Eqs. (\ref{theta vs akhiplus (32)}) to (\ref{psi vs gamaplus (34)})
needs to have supplementary information such as, for example, the
unambiguous demonstration of the Ising-like uniqueness of the $\vartheta_{\mathcal{L}}$-value
(i.e. a constant $\vartheta_{\mathcal{L}}$-value independent of the
fitting domain and the fitted property);

iii) alternative equations valid beyond the Ising-like preasymptotic
domain which provide the equivalent scaling information contained
in Eqs. (\ref{theta vs akhiplus (32)}) to (\ref{psi vs gamaplus (34)})
are suitable to probe the identity $\vartheta\equiv\vartheta_{\mathcal{L}}$
and the application of the linear Eq. (\ref{theta from tstar vs deltataustar (13)})
for $\Delta\tau^{*}\rightarrow0$.

We recall that the first attempt \citet{Bagnuls1984b} to understand
the classical-to-critical crossover in xenon using the $\text{MR6}_{\text{max}}$
scheme, was made from such a global analysis of three singular properties
(correlation length, susceptibility and heat capacity) of xenon in
a rather large finite temperature range ($T-T_{c}\lesssim30\,\text{K}$,
i.e. $\Delta\tau^{*}\lesssim10^{-1}$). A rather well-defined value
of $\vartheta_{\mathcal{L}}=0.0191\pm0.0095$ was then obtained, which
was subsequently analyzed in terms of the influence of the corrections
to scaling in real systems, even outside the Ising-like preasymptotic
domain. The small difference with the present hypothesized value $\vartheta=0.021069$
{[}see Eq. (\ref{Xe independent scale factors (40)})], only reflects
the error-bar propagation of theoretical exponent and amplitude estimations
in the respective fitting of the $\text{MR6}_{\text{max}}$ or MR
discretized values, and not the true experimental uncertainty (see
a detailed discussion of the experimental uncertainty in Appendix
A). In other words, the hypothesized asymptotic value of $\vartheta$
appears independent of the Ising-like Eq. (\ref{theta vs akhiplus (32)}).
Similarly, in the other crossover modeling, the uniqueness of the
crossover temperature scale appears independent of the theoretical
approach in the sense where it is \emph{de facto} obtained at a ''single''
finite temperature distance, noted $\left\langle \Delta\tau^{*}\right\rangle _{\vartheta_{\mathcal{L}}}$.
Here \emph{single} means that, in a log-log scale, the value of the
free parameter $\vartheta_{\mathcal{L}}$ is optimized on a \emph{fine}
temperature domain (covering typically less than one decade) centered
on a large temperature distance $\left\langle \Delta\tau^{*}\right\rangle _{\vartheta_{\mathcal{L}}}$,
such that $\frac{\left\langle \Delta\tau^{*}\right\rangle _{\vartheta_{\mathcal{L}}}}{\Delta\tau_{\text{C1}}^{*}}\gg1$.
For example considering the Güttinger and Cannell's measurements in
the next Section, we show clearly that $\left\langle \Delta\tau^{*}\right\rangle _{\vartheta_{\mathcal{L}}}$
corresponds approximately to the {}``largest'' experimental value
$\left\langle \Delta\tau^{*}\right\rangle _{\vartheta_{\mathcal{L}}}\simeq0.05-0.1\lesssim\Delta\tau_{\Delta}^{*}$
of the fitted temperature range reported in Fig. \ref{Figure2}. Then
the ratio $\frac{\left\langle \Delta\tau^{*}\right\rangle _{\vartheta_{\mathcal{L}}}}{\Delta\tau_{\text{C1}}^{*}}$
is greater than $10^{4}$. This value is \emph{definitively} outside
the Ising-like preasymptotic domain and similar to the value of crossover
temperature scale $\Delta\tau_{X,M}^{*}\simeq0.23$ obtained from
fitting using CPM and MSR crossover descriptions.

The above analysis limited to the results of Figure \ref{Figure2}
extrapolated within the Ising-like preasymptotic domain, justify our
special attention in the next section to methods which account for
the calibrated contribution of the theoretical leading power law,
in order to check in a self-consistent manner the $\vartheta_{\mathcal{L}}$-determination
beyond the Ising-like preasymptotic domain.

\section{Xenon characterization \emph{beyond} the Ising-like preasymptotic
domain}

\subsection{Ising-like nature of the dimensionless scale factors}

The finite and restricted $\Delta\tau^{*}$ range where the mean crossover
functions correctly represent $\kappa_{T,\text{exp}}^{*}$ and $\xi_{\text{exp}}$
data, determines the effective extension $\mathcal{L}_{\text{EAD}}^{\text{Xe}}$
of the Ising-like extended asymptotic domain (EAD). Since the theoretical
expressions of Eqs. (\ref{MR lstar vs tstar (1)}) and (\ref{MR khistar vs tstar (2)}),
are in the form of a complete crossover, $\mathcal{L}_{\text{EAD}}^{\text{Xe}}$
may involve correction-to-scaling terms higher than the first one.
In such a situation, the following condition $\mathcal{L}_{\text{PAD}}^{\text{Xe}}<\Delta\tau^{*}\leq\mathcal{L}_{\text{EAD}}^{\text{Xe}}$
occurs. This is precisely the case for the isothermal compressibility
data of xenon obtained by Güttinger and Cannell. As a consequence,
the value of $\vartheta_{\mathcal{L}}$ introduced by the analytic
relation $t=\vartheta_{\mathcal{L}}\Delta\tau^{*}$, is related to
an {}``undefined'' domain of extension $\mathcal{L}_{\text{EAD}}^{\text{Xe}}$.
Here, {}``undefined'' means {}``beyond the Ising-like preasymptotic
domain'', so that we are not able to appreciate the effective influence
of all the numerous corrections neglected in the massive renormalization
scheme of the $\Phi^{4}$ model (see I). We must solve new correlative
difficulties concerning the effective number (which can thus be greater
than 3) and the nature (which can originate from the neglected analytical
and confluent effects in the critical massive renormalization scheme)
of the fluid-dependent parameters.

Indeed, to complete $T_{c}$ and $\alpha_{c}$ knowledge in the absence
of information concerning the true range of the Ising-like preasymptotic
domain for an actual fluid, it is proposed \citet{Bagnuls2002,Garrabos2006mcf}
to replace $\vartheta$, $\psi$, and $g_{0}$ by three new adjustable
parameters $\mathbb{X}_{0,\mathcal{L}}^{*}$, $\mathbb{L}_{0,\mathcal{L}}^{*}$,
and $\vartheta_{\mathcal{L}}$, modifying Eqs. (\ref{(23)}) and (\ref{ksiexpstar fitting eq (20)})
in the following way:\begin{equation}
\begin{array}{cl}
\frac{1}{\kappa_{T,\text{exp}}^{*}\left(\Delta\tau^{*}\right)}= & \left(\mathbb{X}_{0,\mathcal{L}}^{*}\right)^{-1}\mathbb{Z}_{\chi}^{+}\left(\Delta\tau^{*}\right)^{\gamma}\\
 & \prod_{i=1}^{K}\left(1+X_{i,\chi}^{+}t^{D\left(t\right)}\right)^{Y_{i,\chi}^{+}}\end{array}\label{khiT vs khizero fitting eq (44)}\end{equation}
\begin{equation}
\begin{array}{cl}
\frac{\alpha_{c}}{\xi_{\text{exp}}^{*}\left(\Delta\tau^{*}\right)}= & \left(\mathbb{L}_{0,\mathcal{L}}^{*}\right)^{-1}\mathbb{Z}_{\xi}^{+}\left(\Delta\tau^{*}\right)^{\nu}\\
 & \prod_{i=1}^{K}\left(1+X_{i,\xi}^{+}t^{D\left(t\right)}\right)^{Y_{i,\ell}^{+}}\end{array}\label{ksi vs elezero fitting eq (45)}\end{equation}
 with\begin{equation}
t=\vartheta_{\mathcal{L}}\Delta\tau^{*}\label{thetaLcalli vs tstar (46)}\end{equation}
The new characteristic set\begin{equation}
\mathcal{\mathbb{Q}}_{c,\mathcal{L}}=\left\{ \left(\beta_{c}\right)^{-1};\alpha_{c};\mathcal{\mathbb{S}}_{1CP,\mathcal{L}}\right\} \label{general Qc set (47)}\end{equation}
must substitute the set $\mathcal{\mathbb{Q}}_{c}\left(\Delta\tau^{*}\leq\mathcal{L}_{\text{EAD}}^{\text{Xe}}\right)$
of Eq. (\ref{asympt Qc vs Ltabfstar set (25)}), while the new dimensionless
set\begin{equation}
\mathcal{\mathbb{S}}_{1CP,\mathcal{L}}=\left\{ \vartheta_{\mathcal{L}};\mathbb{L}_{0,\mathcal{L}}^{*};\mathbb{X}_{0,\mathcal{L}}^{*}\right\} \label{(48)}\end{equation}
must substitute the set $\mathcal{\mathbb{S}}_{SF}$ of Eq. (\ref{(26)}).
The subscript $1CP,\mathcal{L}$ recalls for a single crossover parameter
to characterize the crossover behavior observed in the finite temperature
range $\mathcal{L}_{\text{PAD}}^{\text{Xe}}<\Delta\tau^{*}\lesssim\mathcal{L}_{\text{EAD}}^{\text{Xe}}$.

In comparison to Eqs. (\ref{(23)}) and (\ref{ksiexpstar fitting eq (20)}),
the noticeable modification of Eqs. (\ref{khiT vs khizero fitting eq (44)})
and (\ref{ksi vs elezero fitting eq (45)}) is the leading term in
which $\vartheta_{\mathcal{L}}$ is no longer involved in the asymptotic
scaling part of the critical behavior expressed in terms of the physical
field $\Delta\tau^{*}$. As introduced, $\mathbb{X}_{0,\mathcal{L}}^{*}$
and $\mathbb{L}_{0,\mathcal{L}}^{*}$ are prefactors for each corresponding
property, here selected as independent and characteristic of the fluid
by virtue of the two-scale-factor universality (provided that the
same length unit was used to define the dimensionless quantities \citet{Privman1991}).
Correlatively, $\vartheta_{\mathcal{L}}$ is a \emph{pure} crossover
parameter, with same value above and below $T_{c}$, which exclusively
controls the magnitude of many correction terms to scaling. In addition,
$\vartheta_{\mathcal{L}}$ can also integrate some effects of the
neglected terms linked to the supplementary confluent exponents, such
as $\Delta_{2}$ or $\Delta_{3}$, accounting for practical numerical
approximations such as $\Delta_{2}\approx2\Delta$ or $\Delta_{3}\approx3\Delta$,
or the effects of using $T$ to replace $T_{c}$ in the energy unit
and in the dimensionless form of the temperature distance to the critical
temperature. For a fluid $f$, the determination of $\vartheta_{\mathcal{L}}$
is then equivalent to the determination of $\mathcal{L}_{\text{EAD}}^{f}$.
In this general case, the physical leading amplitudes can be calculated
using:\begin{equation}
\xi_{\mathcal{L}}^{+}=\mathbb{L}_{0,\mathcal{L}}^{*}\left(\mathbb{Z}_{\xi}^{+}\right)^{-1}\label{ksiplus vs Ltabzero (49)}\end{equation}
\begin{equation}
\Gamma_{\mathcal{L}}^{+}=\mathbb{X}_{0,\mathcal{L}}^{*}\left(\mathbb{Z}_{\chi}^{+}\right)^{-1}\label{gammaplus vs khizero (50)}\end{equation}
i.e., without reference to $\vartheta_{\mathcal{L}}$. However, the
subscript $\mathcal{L}$ recalls for the determination of $\vartheta_{\mathcal{L}}$,
then correlatively, $\mathbb{L}_{0,\mathcal{L}}^{*}$ and $\mathbb{X}_{0,\mathcal{L}}^{*}$,
made beyond the Ising-like preasymptotic domain. The single (system-dependent)
first confluent amplitude can be calculated from $\vartheta_{\mathcal{L}}$
uniquely, using one independent equation among the two following equations:
\begin{equation}
a_{\chi,\mathcal{L}}^{+}=-\left(\vartheta_{\mathcal{L}}\right)^{\Delta}\mathbb{Z}_{\chi}^{1,+}\label{akhiplus vs tethalcali (51)}\end{equation}
\begin{equation}
a_{\xi,\mathcal{L}}^{+}=-\left(\vartheta_{\mathcal{L}}\right)^{\Delta}\mathbb{Z}_{\xi}^{1,+}\label{aksiplus vs tethalcali (52)}\end{equation}

As suggested in Ref. \citet{Bagnuls2002}, from similar fitting of
the correlation length and the susceptibility in the inhomogeneous
domain, the specific heat in the homogeneous and non homogeneous domains,
and the coexisting density measurements in the inhomogeneous domain,
one must verify the uniqueness of the $\vartheta_{\mathcal{L}}$ value
(along the critical isochore). Considering several properties allows
then consistent tests for the determination of the set $\mathcal{S}_{1CP,\mathcal{L}}$,
in coherence with the basic hypotheses of the renormalization group
approach at the origin of the theoretical crossover functions.

To avoid this large task, in a first approach, we consider xenon for
which we hypothetize the existence of a single scale factor $\vartheta_{\mathcal{L}}$
in the temperature range $\mathcal{L}_{\text{PAD}}^{\text{Xe}}<\Delta\tau^{*}\lesssim\mathcal{L}_{\text{EAD}}^{\text{Xe}}$.
Fitting then $\kappa_{T,\text{exp}}^{*}$ and $\xi_{\text{exp}}^{*}$
data in this temperature range with Eqs. (\ref{gzero vs elestar (15)})
and (\ref{(23)}), produces the set $\mathcal{S}_{1CP,\mathcal{L}}$,
i.e., the determination of $\vartheta_{\mathcal{L}}$. Extrapolating
the results for $\Delta\tau^{*}\rightarrow0$, we can identify the
value of the asymptotic scale factor $\vartheta$ to\begin{equation}
\vartheta\equiv\vartheta_{\mathcal{L}}\;\text{for}\;\Delta\tau^{*}\leq\mathcal{L}_{\text{PAD}}^{\text{Xe}}\label{ideal nonideal identity (53)}\end{equation}
 After that identification, we can introduce the dimensional prefactor
$\mathbb{L}_{0,\mathcal{L}}=\alpha_{c}\mathbb{L}_{0,\mathcal{L}}^{*}$
leading to define the asymptotic wave number $g_{0}$ of Eq. (\ref{gzero vs ksizeroplus (33)})
by the following parameter:\begin{equation}
g_{0}\equiv g_{0,\mathcal{L}}=\left(\mathbb{L}_{0,\mathcal{L}}\right)^{-1}\left(\vartheta_{\mathcal{L}}\right)^{-\nu}=\left[\alpha_{c}\mathbb{L}_{0,\mathcal{L}}^{*}\left(\vartheta_{\mathcal{L}}\right)^{\nu}\right]^{-1}\label{ktabnoirf (54)}\end{equation}
The two remaining asymptotic scale factors $\mathbb{L}^{\left\{ 1f\right\} }$
{[}see Eq. (\ref{Ltabfstar (21)})], and $\psi_{\rho}$ {[}see Eq.
(\ref{psi from hstar vs deltamutilde (18)})], are then obtained by
the hierarchical equations:\begin{equation}
\mathbb{L}^{\left\{ 1f\right\} }=\alpha_{c}g_{0,\mathcal{L}}=\left[\mathbb{L}_{0,\mathcal{L}}^{*}\left(\vartheta_{\mathcal{L}}\right)^{\nu}\right]^{-1}\label{nonideal gzerostar vs nonideal theta (55)}\end{equation}
\begin{equation}
\psi_{\rho,\mathcal{L}}=\left[\left(\mathbb{L}^{\left\{ 1f\right\} }\right)^{-d}\mathbb{X}_{0,\mathcal{L}}^{*}\left(\vartheta_{\mathcal{L}}\right)^{\gamma}\right]^{\frac{1}{2}}\label{nonideal psi vs nonideal theta (56)}\end{equation}
Finally, for $\Delta\tau^{*}\rightarrow0$, the scale factor set\begin{equation}
\mathcal{S}_{SF,\mathcal{L}}=\left\{ \vartheta_{\mathcal{L}};\mathbb{L}^{\left\{ 1f\right\} };\psi_{\rho,\mathcal{L}}\right\} \;\text{with}\;0<\Delta\tau^{*}<\mathcal{L}_{\text{EAD}}^{\text{Xe}}\label{(57)}\end{equation}
has the appropriate asymptotic form to compare with $\mathcal{S}_{SF}$
{[}Eq. (\ref{(26)})], except the noticeable subscript $\mathcal{L}$
which recalls the {}``non-asymptotic'' Ising-like nature of these
numbers which originates \emph{de facto} from the determination of
$\vartheta_{\mathcal{L}}$, $\mathbb{L}_{0,\mathcal{L}}^{*}$ and
$\mathbb{X}_{0,\mathcal{L}}^{*}$ made \emph{beyond} the Ising-like
preasymptotic domain.

However, when this {}``hypothetical'' xenon is characterized by
Eq. (\ref{general Qc set (47)}) {[}or Eq. (\ref{(57)})], thus Eq.
(\ref{nonideal gzerostar vs nonideal theta (55)}) is \emph{true}
and the following variable transformations\begin{eqnarray}
\xi\rightarrow\xi^{*}=\frac{\xi}{\alpha_{c}} & \rightarrow & \ell_{\text{th}}=\mathbb{L}^{\left\{ 1f\right\} }\xi^{*}\label{ele transformation (58)}\\
\Delta\tau^{*} & \rightarrow & t=\vartheta_{\mathcal{L}}\Delta\tau^{*}\label{thermal field transformation (59)}\end{eqnarray}
result in the asymptotic collapse (over the extension $\mathcal{L}_{\text{PAD}}^{\left\{ 1f\right\} }$
\citet{Garrabos2006cl,Garrabos2007}) of any physical curves of equation
$\xi\left(\Delta\tau^{*}\right)$ into the universal curve of equation
\begin{equation}
\ell_{\text{th}}\left(t\right)=\mathbb{L}^{\left\{ 1f\right\} }\xi^{*}\left(\Delta\tau^{*}\right)=g_{0,\mathcal{L}}\xi_{\text{exp}}\left(\Delta\tau^{*}\right)\label{MR elestar vs nonideal gzerostar (60)}\end{equation}

We have used this universal feature due to the scale factor nature
of $\vartheta_{\mathcal{L}}$ to obtain the xenon values of lines
8 and 9 of Table \ref{Table II} from theoretical values of lines
5 and 6, respectively. In such a numerical approach, the {}``hypothesized''
xenon values of the characteristic set $\mathcal{S}_{1CP,\mathcal{L}}$
are the following {[}see Eq. (\ref{fluid prefactors (A10)}) in Appendix]\begin{equation}
\mathcal{S}_{1CP,\mathcal{L}}=\left\{ \begin{array}{cl}
\vartheta_{\mathcal{L}} & =0.021069\\
\mathbb{L}_{0,\mathcal{L}}^{*} & =0.443526\\
\mathbb{X}_{0,\mathcal{L}}^{*} & =0.214492\end{array}\right\} \label{xenon khi-theta values (61)}\end{equation}
with\begin{equation}
\mathbb{L}_{0,\mathcal{L}}=\alpha_{c}\mathbb{L}_{0,\mathcal{L}}^{*}=0.390972\,\text{nm}\label{(62)}\end{equation}
and, obviously \begin{equation}
\mathbb{L}^{\left\{ 1f\right\} }\equiv\alpha_{c}g_{0,\mathcal{L}}=25.6936\label{1f gzerostar master value (63)}\end{equation}
More generally, our hypothesized values of Eq. (\ref{xenon khi-theta values (61)})
provide the (expected) identity $\mathcal{S}_{SF,\mathcal{L}}\equiv\mathcal{S}_{SF}$.
Accordingly, the universal scaling form of the correlation length
of xenon reads as follows\begin{equation}
\mathbb{L}^{\left\{ 1f\right\} }\xi^{*}\left(\Delta\tau^{*}\right)=\frac{\frac{\xi_{\text{exp}}\left(\Delta\tau^{*}\right)}{\left[\text{nm}\right]}}{\frac{0.0343085}{\left[\text{nm}\right]}}\equiv\ell_{\text{th}}\left(t\right)\label{universal ksi for xe (64)}\end{equation}
with\begin{equation}
t=0.021069\Delta\tau^{*}\label{universal t for xe (65)}\end{equation}
Either the {}``theoretical'' value $0.0343085\,\text{nm}$ ($=\left[g_{0,\mathcal{L}}\right]^{-1}$)
of the length unit in Eq. (\ref{universal ksi for xe (64)}), or the
{}``measured'' value $0.184333\,\text{nm}$ ($=\xi_{0,\mathcal{L}}^{+}$)
of the correlation length amplitude in Eq. (\ref{2term ksiexp (28)}),
cannot easily be related to a real microscopic length of xenon atom,
while the fact that the {}``fitting'' value $0.390972\,\text{nm}$
($=\mathbb{L}_{0,\mathcal{L}}$) of the dimensional prefactor is comparable
to the size of the xenon atom can be considered as a fortuitous result
\citet{microscopicspacing}.

As expected, more fundamental is the dimensionless nature of Eq. (\ref{1f gzerostar master value (63)})
which can then provide the needed uniqueness of the length unit for
better understanding of universality. Indeed, the above asymptotic
collapse of the correlation length implies equivalent asymptotic collapse
of any other singular thermodynamic property, by virtue of hyperscaling.
That means that, when the scale factors $\mathbb{L}^{\left\{ 1f\right\} }$
and $\vartheta_{\mathcal{L}}$ are determined in the Ising-like extended
asymptotic domain $\mathcal{L}_{\text{PAD}}^{\text{Xe}}<\Delta\tau^{*}\lesssim\mathcal{L}_{\text{EAD}}^{\text{Xe}}$,
the validity range where the universal collapse onto the theoretical
thermodynamic behavior is expected, can be alternatively discussed
in term of the value $\ell_{\text{th}}\left(t\right)>\mathbb{L}^{\left\{ 1f\right\} }$.
In other words, all the crossover functions are {}``Ising-like''
universal:

i) only over the temperature range where \emph{the crossover parameter
is unique;}

ii) only for a single dimensionless critical length $\mathbb{L}^{\left\{ 1f\right\} }$
\emph{common for all the fluids} which obey to this single parameter
crossover description.

For example, looking back on our previous analysis of the residuals
for the xenon isothermal compressibility case, by reversing Eq. (\ref{universal ksi for xe (64)}),
we can transform $\Delta\tau^{*}$ in universal values of $\ell_{\text{th}}$,
as illustrated in the upper axis of Figure \ref{Figure2}. Simultaneously,
as previously mentioned, we are also able to illustrate in this figure
(or in any figure which use a xenon $\Delta\tau^{*}$-coordinate)
all the conditions estimated in Table \ref{Table II}. In the part
of the Güttinger and Cannell's experimental range which corresponds
to the theoretical condition $\ell_{\text{th}}\gtrsim\left(2.5-3\right)\mathbb{L}^{\left\{ 1f\right\} }\simeq70$,
the universal behavior of the functional form $\chi^{*}\left(\ell_{\text{th}}\right)$
can then be analytically {}``tested'' in this Ising-like extended
asymptotic domain. Simultaneously, we will also confirm below the
uniqueness of the crossover parameter by using an alternative facet
of the universality accounted for by the mean crossover functions.
As a matter of fact, to illustrate the asymptotic Ising-like transformation
of each thermodynamic property $P^{*}\left(\Delta\tau^{*}\right)$
attached to the uniqueness of the crossover parameter, we can also
use the effective universal behavior of the local exponent $e_{P,\text{th}}\left(t\right)$
to replace the one of the correlation length $\ell_{\text{th}}\left(t\right)$,
in order to construct the {}``universal'' scaling form $F_{P}\left(e_{P,\text{th}}\right)$.
Therefore, in a second approach with the objective to illustrate this
transformation in a self-consistent manner for the susceptibility
case, we consider the universal and experimental effective amplitudes
attached to the local power laws with effective exponents, such as
introduced in Ref. \citet{Kouvel1964}.

\begin{figure*}
\includegraphics{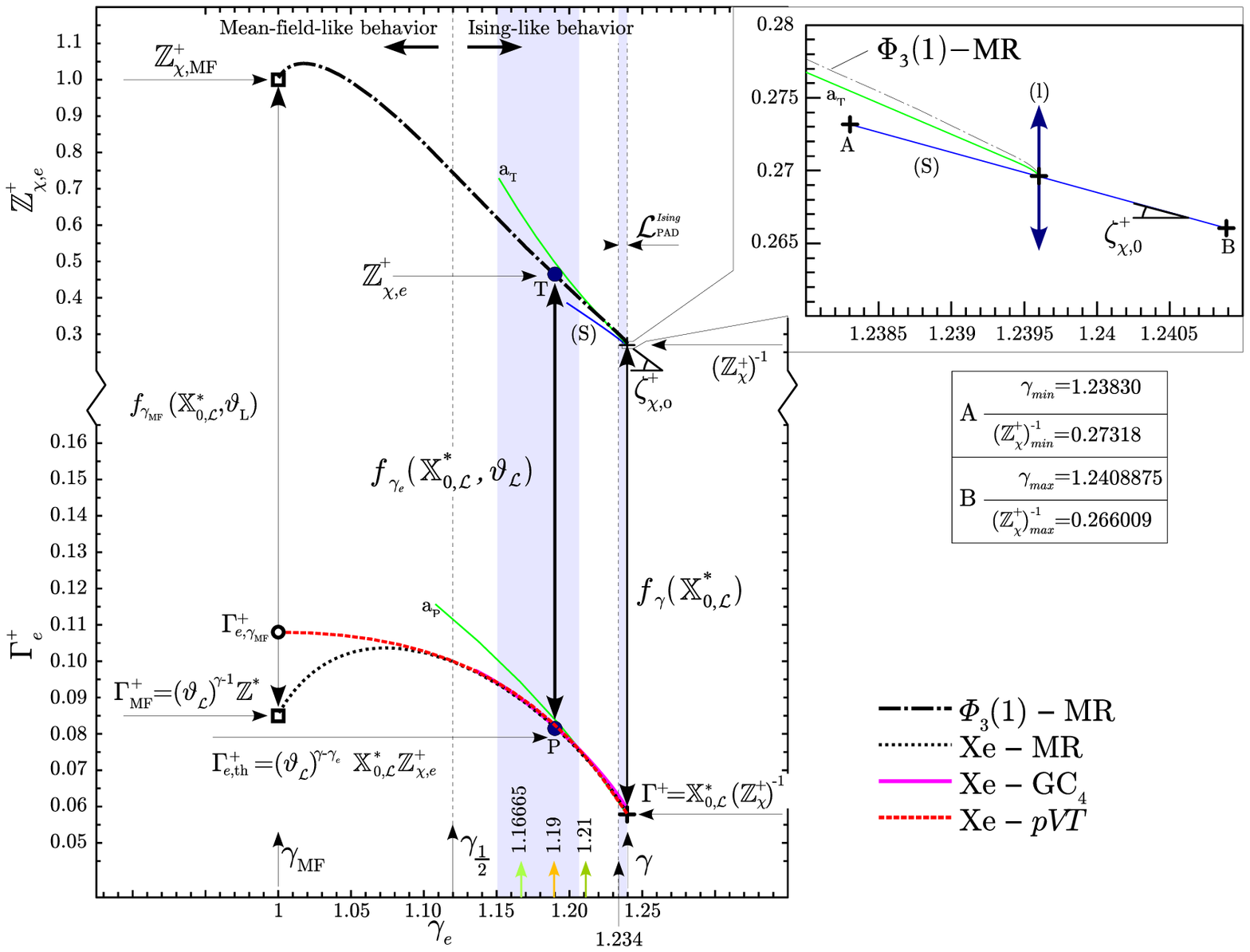}

\caption{(Color on line) Two-parameter transformations $f_{\gamma_{e}}\left(\mathbb{X}_{0,\mathcal{L}}^{*},\vartheta_{\mathcal{L}}\right)$
(see arrays), at $\gamma_{e}=\text{constant value}$, from the theoretical
universal behavior of the susceptibility (upper curve), to the xenon
effective behavior of the isothermal compressibility (lower curve)
(in dimensionless variables); $\mathbb{X}_{0,\mathcal{L}}^{*}$: adjustable
parameter for the general case of Eq. (\ref{khiT vs khizero fitting eq (44)}),
unequivocally defined from comparison in positions of the two Ising
limiting points (right crosses); $\vartheta_{\mathcal{L}}$ adjustable
parameter unequivocally defined by the two other points, either within
the Ising-like preasymptotic domain (see the insert and the text),
or outside the Ising-like preasymptotic domain (full black circles
T and P on the two curves labeled $\Phi_{3}\left(1\right)-\text{MR}$
and $\text{Xe}-\text{MR}$, respectively). Note the inversion of the
relative position between the curve labeled $\Phi_{3}\left(1\right)-\text{MR}$
and the curve labeled $a_{T}$ which is related to the asymptotic
(logarithmic) approximation of Eq. (\ref{log limit gamma ising (68)}),
beyond the Ising-like preasymptotic domain. \label{Figure3}}

\end{figure*}

\subsection{Effective exponent and amplitude beyond the Ising-like PAD}

>From $\chi_{\text{th}}\left(t\right)$ of Eq. (\ref{MR khistar vs tstar (2)}),
the related local values of the effective exponent $\gamma_{e,\text{th}}\left(t\right)$
and effective amplitude $\mathbb{Z}_{\chi,e}^{+}\left(t\right)$ are
given by the equations\begin{equation}
\gamma_{e,\text{th}}\left(t\right)=-\frac{\partial Ln\left[\chi_{\text{th}}\left(t\right)\right]}{\partial Lnt}\label{gammaeth definition (66)}\end{equation}
\begin{equation}
\mathbb{Z}_{\chi,e}^{+}\left(t\right)=\frac{\chi_{\text{th}}\left(t\right)}{t^{-\gamma_{e}}}\label{zkhipluseff MR (67)}\end{equation}
Eliminating $t$ {[}then simultaneously eliminating the scale factor
$\vartheta_{\mathcal{L}}$, since $t=\vartheta_{\mathcal{L}}\Delta\tau^{*}$],
the theoretical classical-to-critical crossover is characterized by
a {}``universal'' curve $\mathbb{Z}_{\chi,e}^{+}\left(\gamma_{e,\text{th}}\right)$
over the complete range $\gamma_{\text{MF}}\leq\gamma_{e,\text{th}}\left(t\right)\leq\gamma$
(see curve labeled $\Phi_{3}\left(1\right)$-MR in Fig. \ref{Figure3}).

Our present interest is restricted to the Ising-like range $\gamma_{e,\text{th}}\left(t\right)\geq\gamma_{\frac{1}{2}}=\frac{\gamma+\gamma_{\text{MF}}}{2}$
(see the corresponding Ising-like range defined in the upper part
of Fig. \ref{Figure3}). The theoretical Ising-like limiting point
takes universal coordinates $\left\{ \gamma;\left(\mathbb{Z}_{\chi}^{+}\right)^{-1}\right\} $
(upper cross in Fig. \ref{Figure3}). The {}``small'' extension
$\gamma-\gamma_{e,\text{th}}\lesssim\mathbb{Z}_{\chi}^{1,+}\Delta\left(\mathcal{L}_{\text{PAD}}^{\text{Ising}}\right)^{\Delta}\approx0.006$
of the Ising-like preasymptotic domain is magnified by the curve labeled
$\Phi_{3}\left(1\right)$-MR in the insert of Fig. \ref{Figure3}.
On the other hand, the curve $\text{a}_{\text{T}}$ corresponds to
the asymptotic behavior of the derivative $\left(\frac{\partial\mathbb{Z}_{\chi,e}^{+}}{\partial\gamma_{e}}\right)_{\gamma_{e}\rightarrow\gamma}$of
equation

\begin{equation}
\begin{array}{rl}
\left(\frac{\partial\mathbb{Z}_{\chi,e}^{+}}{\partial\gamma_{e}}\right)_{\gamma_{e,\text{th}}\rightarrow\gamma}= & \left(\mathbb{Z}_{\chi}^{+}\right)^{-1}\left\{ 1+\left(\frac{\gamma-\gamma_{e,\text{th}}}{\Delta\left|\mathbb{Z}_{\chi}^{1,+}\right|}\right)^{-\left(\frac{\gamma-\gamma_{e,\text{th}}}{\Delta}\right)}\right.\\
 & \left.\left(1-\log\left[\frac{\gamma-\gamma_{e,\text{th}}}{\Delta\left|\mathbb{Z}_{\chi}^{1,+}\right|}\right]\right)\left(\frac{\gamma-\gamma_{e,\text{th}}}{\Delta}\right)\right\} \end{array}\label{log limit gamma ising (68)}\end{equation}
The vertical double arrow with label$\left(1\right)$indicates the
above logarithmic divergence of $\left(\frac{\partial\mathbb{Z}_{\chi,e}^{+}}{\partial\gamma_{e}}\right)_{\gamma_{e,th}\rightarrow\gamma}$.
We note the significant difference between the curve $\text{a}_{\text{T}}$
and the curve $\left(S\right)$ which results from {}``analytic''
error-bar correlation between the Ising values of $\gamma$ and $\left(\mathbb{Z}_{\chi}^{+}\right)^{-1}$.
As a matter of fact, the curve $\left(S\right)$ corresponds to the
linearized slope $\zeta_{\chi,0}^{+}=\frac{\left(\mathbb{Z}_{\chi,\text{max}}^{+}\right)^{-1}-\left(\mathbb{Z}_{\chi,\text{min}}^{+}\right)^{-1}}{\gamma_{\text{min}}-\gamma_{\text{max}}}$
between the respective bounded coordinates of points A and B (see
inserted table in Fig. \ref{Figure3} and Ref. \citet{Bagnuls2002}
for data sources).

Since only two parameters ($\mathbb{X}_{0,\mathcal{L}}^{*}$ and $\vartheta_{\mathcal{L}}$)
are free in fitting Eq. (\ref{khiT vs khizero fitting eq (44)}),
Fig. \ref{Figure3} illustrates how the adjustable prefactor $\mathbb{X}_{0,\mathcal{L}}^{*}$
acts at the exact value of the Ising exponent, since each Ising point
of the fluid takes the coordinates $\left\{ \gamma;\Gamma^{+}=\mathbb{X}_{0,\mathcal{L}}^{*}\left(\mathbb{Z}_{\chi}^{+}\right)^{-1}\right\} $,
as represented by the lower cross in Fig. \ref{Figure3} for the critical
xenon case. The resulting prefactor $f_{\gamma}\left(\mathbb{X}_{0,\mathcal{L}}^{*}\right)\equiv\mathbb{X}_{0,\mathcal{L}}^{*}$
is schematized by the double array between two crosses in Fig. \ref{Figure3}.
Therefore, calibrating the $\Gamma^{+}$ value as suggested in previous
paragraph, fixes this prefactor value which governs the universal
collapse of the Ising-like limiting point through Eq. (\ref{gammaplus vs khizero (50)}).
$\mathbb{X}_{0,\mathcal{L}}^{*}$ acts in a equivalent manner to $\mathbb{L}^{\left\{ 1f\right\} }$
{[}see Eq. (\ref{universal ksi for xe (64)})] for the correlation
length case. In principle, the critical divergence in the initial
slope at the limiting points provides the second {}``Ising-like constraint''
which is needed to determine the asymptotic scale factor $\vartheta$.
However, our previous description of the Ising-like preasymptotic
domain and its above geometrical illustration underline the challenging
(theoretical and experimental) difficulties to provide the exact characterization
of the asymptotic scaling when a property reaches the Ising-like limiting
point along a curve of {}``universal'', but \emph{infinite}, slope
for all the physical systems.

At contrario, the description of the $\gamma_{e}$-variation in the
range $\gamma-\gamma_{e}\gtrsim0.015$ (i.e., in a temperature range
significantly beyond the Ising-like preasymptotic domain), appears
now simplified, using precisely the Güttinger and Cannell's results
of Eq. (\ref{khiTexp GC (43)}) to define the following effective
exponent by :\begin{equation}
\gamma_{e,\text{exp}}\left(\Delta\tau^{*}\right)=-\frac{\partial Ln\left[\kappa_{T,\text{exp}}^{*}\left(\Delta\tau^{*}\right)\right]}{\partial Ln\left(\Delta\tau^{*}\right)}\label{gammaeexp definition (69)}\end{equation}
and its attached effective amplitude by :\begin{equation}
\Gamma_{e}^{+}\left(\Delta\tau^{*}\right)=\frac{\kappa_{T,\text{exp}}^{*}\left(\Delta\tau^{*}\right)}{\left(\Delta\tau^{*}\right)^{-\gamma_{e,exp}}}\label{gamapluseff experimental (70)}\end{equation}

The resulting single curve $\Gamma_{e}^{+}\left(\gamma_{e,\text{exp}}\right)$
is illustrated in the lower part of Fig. \ref{Figure3} (see curve
$\text{Xe}-\text{GC}_{4}$), while the expected (two parameter) transformation
$f\left(\mathbb{X}_{0,\mathcal{L}}^{*},\vartheta_{\mathcal{L}}\right)$
able to insure the universal collapse between the theoretical curves
(labeled $\text{Xe}-\text{MR}$ and $\Phi_{3}\left(1\right)$-MR),
is schematized by a double array between two points on each curve
of well-defined \emph{finite} slope. This transformation must contain
the needed both constraints for the (point) position \emph{and} the
related (tangent) direction. Therefore, the scaling nature of the
collapse beyond the Ising-like preasymptotic domain is significantly
different in fitting procedure which either eliminates, or accounts
for, the contribution of the leading term. In the latter situation,
we can then replace the prefactor $\mathbb{X}_{0,\mathcal{L}}^{*}$
by the calibrated leading amplitude $\Gamma^{+}$, as seen above.

\begin{figure}
\includegraphics[width=80mm,keepaspectratio]{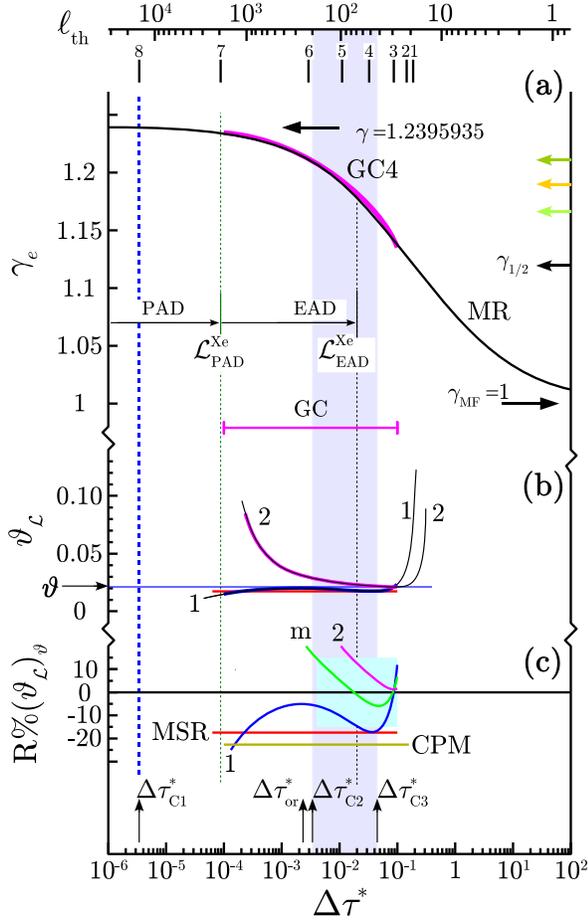}

\caption{(Color on line) Ising-like asymptotic behaviors of: (a), the effective
exponent $\gamma_{e}$ for the xenon isothermal susceptibility, (b),
the xenon crossover parameter $\vartheta_{\mathcal{L}}$, and (c),
the percent deviation between $\vartheta_{\mathcal{L}}$ and the xenon
asymptotic scale factor $\vartheta$ calculated from the scale dilatation
method, as a function of the relative temperature $\Delta\tau^{*}$
(lower horizontal axis) and the theoretical correlation length $\ell_{\text{th}}$
(upper horizontal axis); curve labeled MR: from MR crossover function
of Eq. (\ref{(23)}) and xenon parameters of Eq. (\ref{Xe independent scale factors (40)});
curve GC4: from a fit of the Güttinger and Cannell's data by Eq. (\ref{khiTexp GC (43)})
and xenon parameters in Table \ref{Table IV} line $\text{GC}_{4}$;
curve 1: from Eq. (\ref{gama effective fitting eq (71)}); curve 2:
from Eq. (\ref{theta vs amplitude ratios (74)}); Curve m: mean value
of curves 1 and 2; lines labeled CPM and MSR: with $\vartheta_{X}$-values
of Table \ref{Table III} obtained from the fitting results of the
Güttinger and Cannell's data by CPM and MSR crossover functions (see
text).\label{Figure4}}

\end{figure}

In the first case, \emph{at large temperature distance}, the fit procedure
based on Eq. (\ref{khiT vs khizero fitting eq (44)}) is mainly equivalent
to a dominant constraint in {}``direction'' given by the following
relation between the two effective exponents:\begin{equation}
\gamma_{e,\text{exp}}\left(\Delta\tau^{*}\right)\equiv\gamma_{e,\text{th}}\left[\vartheta_{\mathcal{L}}\left(\Delta\tau^{*}\right)\right]\label{gama effective fitting eq (71)}\end{equation}
We \emph{numerically} solve Eq. (\ref{gama effective fitting eq (71)}),
using the Güttinger and Cannell's fitting results given by Eq. (\ref{khiTexp GC (43)}),
then providing the $\gamma_{e}\left[\left(\Delta\tau^{*}\right)\right]$
and $\vartheta_{\mathcal{L}}\left(\Delta\tau^{*}\right)$ values as
a function of $\Delta\tau^{*}$. Both results are reported as the
curve labeled $\text{GC}_{4}$ in Fig. \ref{Figure4}a {[}$\gamma_{e}$
as a function of $\Delta\tau^{*}$], and the curve labeled 1 in Fig.
\ref{Figure4}b {[}$\vartheta_{\mathcal{L}}$ as a function of $\Delta\tau^{*}$],
respectively. 

In the second case, to account for the contribution of the leading
term needs to use the following \emph{scaling} relation between the
two effective amplitudes\begin{equation}
\Gamma_{e}^{+}=\left(\vartheta_{\mathcal{L}}\right)^{\gamma-\gamma_{e}}\mathbb{X}_{0,\mathcal{L}}^{*}\mathbb{Z}_{\chi,e}^{+}\label{gamaplus effective fitting eq (72)}\end{equation}
Now, the transformation $f\left(\mathbb{X}_{0,\mathcal{L}}^{*},\vartheta_{\mathcal{L}}\right)=\left(\vartheta_{\mathcal{L}}\right)^{\gamma-\gamma_{e}}\mathbb{X}_{0,\mathcal{L}}^{*}$
is explicit in Eq. (\ref{gamaplus effective fitting eq (72)}). Its
takes a convenient effective power law of the crossover parameter
$\vartheta_{\mathcal{L}}$, while the prefactor $\mathbb{X}_{0,\mathcal{L}}^{*}$
has (as expected above) the same value whatever the $\gamma_{e}\left(=\gamma_{e,\text{exp}}=\gamma_{e,\text{th}}\right)$
value is. The constrained {}``position and direction'' are accounted
for correctly. Therefore, using Eq. (\ref{gammaplus vs khizero (50)})
to eliminate $\mathbb{X}_{0,\mathcal{L}}^{*}$, infers the \emph{pure}
$\vartheta_{\mathcal{L}}$-dependence of equation\begin{equation}
\frac{\Gamma_{e}^{+}}{\Gamma^{+}}=\left(\vartheta_{\mathcal{L}}\right)^{\gamma-\gamma_{e}}\frac{\mathbb{Z}_{\chi,e}^{+}}{\left(\mathbb{Z}_{\chi}^{+}\right)^{-1}}\label{gamaplus effective vs gamaplus-theta (73)}\end{equation}
By appropriate combination between the Güttinger and Cannell's fitting
results and the mean crossover function for susceptibility, we \emph{numerically}
calculate the local value of the crossover parameter over the complete
experimental temperature range, using the equation\begin{equation}
\vartheta_{\mathcal{L}}=\left(\frac{1}{\mathbb{Z}_{\chi}^{+}\mathbb{Z}_{\chi,e}^{+}}\times\frac{\Gamma_{e}^{+}}{\Gamma^{+}}\right)^{\frac{1}{\gamma-\gamma_{e}}}\label{theta vs amplitude ratios (74)}\end{equation}
 It is essential to note that for each fluid for which $\Gamma^{+}$
is known, Eq. (\ref{theta vs amplitude ratios (74)}), applied in
the extended asymptotic domain $\Delta\tau^{*}\lesssim\mathcal{L}_{\text{EAD}}^{\text{Xe}}$,
takes equivalent {}``Ising-like'' meaning to Eq. (\ref{theta vs akhiplus (32)})
applied within the Ising-like preasymptotic domain $\Delta\tau^{*}\lesssim\mathcal{L}_{\text{PAD}}^{\text{Xe}}$.
In Figure \ref{Figure4}b, we have reported as a curve labeled 2,
the calculated value of $\vartheta_{\mathcal{L}}$ as a function of
$\Delta\tau^{*}$, using Eq. (\ref{theta vs amplitude ratios (74)})
with $\Gamma^{+}=0.0578204$ {[}see Eq. (\ref{Xe independent amplitudes (39)})].
The available part of these curves 1 and 2 must be restricted to the
experimental temperature range illustrated by the segment labeled
GC. 

For both cases, the $\vartheta_{\mathcal{L}}$-change as a function
of $\Delta\tau^{*}$ within the range $10^{-3}\lesssim\Delta\tau^{*}\lesssim10^{-1}$
of Fig. \ref{Figure4}b, can be approximated by our hypothetical Ising-like
asymptotic value $\vartheta=0.021069$ of Eq. (\ref{Xe independent amplitudes (39)}).
In Fig. \ref{Figure4}c the corresponding curves (labeled $1$ and
$2$) of the residuals $R\%\left(\vartheta_{\mathcal{L}}\right)_{\vartheta}=100\left(\frac{\vartheta_{\mathcal{L}}}{\vartheta}-1\right)$
(expressed in \%), together with their mean curve (labeled $m$),
are given. In the temperature range $5\times10^{-3}\lesssim\Delta\tau^{*}\lesssim10^{-1}$
of Fig. \ref{Figure4}c, the green area corresponds to the error-bar
of $\pm15\%$ for both determinations. Even at such large values of
$\Delta\tau^{*}$, the main significant result obtained from the massive
renormalization scheme, is the estimation of the temperature-like
crossover parameter in conformity with the asymptotic two-scale-factor
characterization of the fluid, especially using the true scaling Eq.
(\ref{theta vs amplitude ratios (74)}). 

Such a temperature range where a {}``measurable'' value of the exponent
difference $\gamma-\gamma_{e}$ occurs, was largely investigated in
the seventy's \citet{Levelt1975}, when the scaling approach of the
fluid universality was based on the effective {}``universal'' values
of the critical exponents - as for example $\gamma_{e}=1.211$ \citet{Levelt1976},
$\gamma_{eos}=1.19$ \citet{Levelt1975,Levelt1978}, and $\gamma_{e}=1.16665$
\citet{Garrabos1985} - involved in effective power laws and/or effective
{}``universal'' form of a rescaled equation of state. Anticipating
a more detailed discussion given in Appendix A, we can use the data
reported on Table \ref{Table V} for these $\gamma_{e}$ values to
easily demonstrate, using the corresponding arrows in Fig. \ref{Figure4}a,
the {}``Ising-like nature'' of the covered temperature range $\Delta\tau^{*}\lesssim0.05-0.1\ll\Delta\tau_{\Delta}^{*}$,
or alternatively but equivalently, the {}``Ising-like nature'' of
the covered correlation length range $\ell_{\text{th}}\gtrsim70>2.5\mathbb{L}^{\left\{ 1f\right\} }$.
That also gives interest to revisit \citet{Garrabos2006mcf} the effective
universal formulation of a parametric equation of state using the
master crossover functions to validate the universal features observed
in the well-defined Ising-like extended asymptotic domain of the fluid
subclass.

Finally we note that, when $T_{c}$ and $\alpha_{c}$ are known, the
mean crossover functions take a convenient controlled form to determine
a single crossover parameter $\vartheta_{\mathcal{L}}$ in a temperature
range beyond the Ising-like preasymptotic domain. The Ising-like nature
of this crossover parameter is then revealed by the dimensionless
value of a single characteristic length. However, to define the {}``minimal''
set of three characteristic parameters such as $\left\{ \vartheta_{\mathcal{L}};\mathbb{X}_{0,\mathcal{L}}^{*};\mathbb{L}_{0,\mathcal{L}}^{*}\right\} $,
or, alternatively but equivalently, $\left\{ \vartheta;\psi_{\rho};g_{0}\right\} $,
and $\left\{ a_{\mathcal{\chi}}^{+};\Gamma^{+};\xi_{0}^{+}\right\} $,
one needs a {}``data calibration'' from measurements performed within
the Ising-like preasymptotic domain. The practical interest of this
data calibration is given in the Appendix A, without reference to
the estimated precision of the experimental method able to provide
such measurements in this closest temperature range.

In the absence of explicit thermodynamic definition for the prefactors
$\mathbb{L}_{0,\mathcal{L}}^{*}$, $\mathbb{X}_{0,\mathcal{L}}^{*}$
and the crossover parameter $\vartheta_{\mathcal{L}}$ {[}or for the
scale factors $g_{0}$, $\vartheta$, and $\psi_{\rho}$], the remaining
difficulty is to compare between distinct fluids which show differences
in their fluid-dependent amplitudes $\xi_{0}^{+}$, $\Gamma^{+}$,
and $a_{\mathcal{\chi}}^{+}$. This difficulty can be solved by application
of the scale dilatation method when the localization of their liquid-vapor
critical point is known, as shown in Appendix A for the xenon case.

\section{Conclusion}

Using xenon as a standard critical fluid, and the mean crossover function
for susceptibility in the homogeneous phase as an illustrative example,
we have estimated the values of the fluid-dependent parameters which
are compatible with the universal features predicted by the massive
renormalization scheme. A special mention for the three dimensionless
parameter characterization \emph{within} the Ising-like preasymptotic
domain was given, in spite of the large theoretical and experimental
uncertainties at such {}``closest'' temperature distance to the
critical temperature. Using the abundant literature now available
from several fittings of Güttinger and Cannell's data, we have demonstrated
the great advantage of the mean crossover functions to provide an
unambiguous determination of one fluid-dependent crossover parameter
outside the Ising-like preasymptotic domain. Specifically, we have
clearly shown that the value of this crossover parameter is entirely
governed by fitting the data at {}``largest'' distance to the critical
point, leading for example to an apparent reduction of the uncertainty
in the determination of the amplitude of the first confluent correction-to-scaling
term. Finally, the magnitude of the resulting deviations and the range
of temperature where it is to be observed are exactly accounted for.

However, in the absence of controlled information about the Ising-like
preasymptotic domain description, only the similar use of the complete
set of mean crossover functions applied to several properties is able
to demonstrate that the crossover parameter conserves its {}``Ising-like''
nature, even outside the Ising-like preasymptotic domain. Alternatively,
it was recently proposed in II and herafter justified in the Appendix
A for the xenon case, an appropriate modification of the mean crossover
functions which only uses three master (i.e. constant) parameters.
The modified crossover functions represent the master (i.e. unique)
singular behaviors of the one component fluid subclass in a well-defined
Ising-like extended asymptotic domain. In such a situation, the real
extension and amplitude of the singular behavior of the fluid properties
can be estimated for any one-component fluid for which the vapor-liquid
critical point is localized in the $pVT$ phase surface. 

\textbf{Acknowledgments}

The authors are indebted to C. Bervillier for valuable discussion
and constructive comments during this work.

\appendix

\section{Isothermal compressibility of xenon}

In the first part of this Appendix we calculate the characteristic
parameters of xenon involved in Eqs. (\ref{Xe scale units (38)})
and (\ref{Xe independent amplitudes (39)}), by using the scale dilatation
method \citet{Garrabos1985,Garrabos1986}. The needed information
is then given by the set $Q_{c,a_{\bar{p}}}^{min}$ made of four critical
coordinates of the xenon critical point. Indeed, for any one-component
fluid, $Q_{c,a_{\bar{p}}}^{min}$ reads as follows (II) \begin{equation}
Q_{c,a_{\bar{p}}}^{min}=\left\{ T_{c};p_{c};v_{\bar{p},c};\gamma_{c}^{'}\right\} \label{Qcmin (A1)}\end{equation}
where $v_{\bar{p},c}=\frac{m_{\bar{p}}}{\rho_{c}}$ is the critical
molecular volume and $\gamma_{c}^{'}$ is the common critical direction
at the critical temperature of the critical isochoric line and the
saturation pressure curve in the $p;T$ diagram.

The second part shows that the xenon parameters given by Eq. (\ref{Xe independent scale factors (40)})
can be used as entry data for the Eqs. (\ref{Deff exponent MR universal (3)})
and (\ref{(23)}) to represent accurately the singular behavior of
the xenon isothermal compressibility in the temperature range such
as $T-T_{c}\lesssim5-10\,\text{K}$.

The last part discusses the data calibration and the uniqueness of
the Ising-like crossover parameter within the extended asymptotic
domain.

\subsection{Xenon data sources from application of the scale dilatation method}

\subsubsection{Xenon critical coordinates}

The selected critical coordinates of xenon are:\begin{equation}
\begin{array}{ccl}
T_{c} & = & 289.733\,\text{K};\\
p_{c} & = & 5.84\,\text{MPa};\\
\rho_{c} & = & 1113\,\text{kg}\,\text{m}^{-3}\;\text{or}\; v_{\bar{p},c}=0.19596\,\text{nm}^{3};\\
\gamma_{c}^{'} & = & 0.1192\,\text{MPa}\,\text{K}^{-1};\\
\text{with}\, m_{\bar{p}} & = & 2.181\times10^{-25}\,\text{kg}.\end{array}\label{Xe critical coordinates (A2)}\end{equation}
They result from the combined analysis \citet{Garrabos1982,Garrabos1985}
of $pVT$ measurements of Habgood and Schneider \citet{Habgood1954},
and coexisting density measurements of Weinberger and Schneider \citet{Weinberger1952},
Cornfeld and Carr \citet{Cornfeld1972}, Thoen and Garland \citet{Thoen1974},
and Balzarini and coworkers \citet{Balzarini1983,Narger1990}.

The xenon critical temperature and pressure were fixed to the values
recently provided by Gillis et al \citet{Gillis2004} which are compatible
with $T_{c}=289.740\pm0.003\,\text{K}$ and $p_{c}=5.8400\pm0.0005\,\text{MPa}$
obtained from Schneider et al's measurements.

The $\rho_{c}$ value has an uncertainty of $\pm5\,\text{kg}\,\text{m}^{-3}$
($\sim0.5\%)$, which accounts for the $\rho_{c}$ values of Schneider
et al's ($\rho_{c}=1110\pm2\,\text{kg}\,\text{m}^{-3}$) \citet{Weinberger1952,Habgood1954},
Cornfeld and Carr's ($\rho_{c}=1111.2_{-3.4}^{+1.9}\,\text{kg}\,\text{m}^{-3}$
for three different estimations) \citet{Cornfeld1972}, Baidakov et
al's ($\rho_{c}=1112.8\pm n.a.\,\text{kg}\,\text{m}^{-3}$) \citet{Baidakov1988},
and Balzarini et al's ($\rho_{c}=1099\pm n.a.\,\text{kg}\,\text{m}^{-3}$
\citet{Balzarini1983}, $\rho_{c}=1116.0\pm1.7\,\text{kg}\,\text{m}^{-3}$
and $\rho_{c}=1114.0\pm1.7\,\text{kg}\,\text{m}^{-3}$ \citet{Narger1990}).

The value $\gamma_{c}^{'}=0.1192\pm0.0005\,\text{MPa}\,\text{K}^{-1}$
($\sim0.5\%)$ was estimated by one of us \citet{Garrabos1982} by
graphical analysis of the $pVT$ measurements of Habgood and Schneider,
which agrees to other literature values $\gamma_{c}^{'}=0.1192\,\text{MPa}\,\text{K}^{-1}$
\citet{Cannell1970}, $\gamma_{c}^{'}=0.1196\,\text{MPa}\,\text{K}^{-1}$
\citet{Swinney1973}, and $\gamma_{c}^{'}=0.120\,\text{MPa}\,\text{K}^{-1}$
\citet{Baidakov1988}. The $\gamma_{c}^{'}$ value used in Ref. \citet{Gillis2004}
differs by $0.19\%$.

More generally, we note the remarkable agreement with the critical
set defined by Gillis et al \citet{Gillis2004} in their recent analysis
of the sound attenuation (in the frequency range $100<f\left(\text{Hz}\right)<7500$)
within thermoacoustic layers between solid surfaces and xenon at critical
density.

\subsubsection{Physical and master amplitudes from the scale dilatation method}

>From Eq. (\ref{Xe critical coordinates (A2)}), the critical values
{[}see Eq. (\ref{Xe scale units (38)})] of the energy $\left(\beta_{c}\right)^{-1}$
and length $\alpha_{c}$ units of xenon are the following: \begin{equation}
\begin{array}{rl}
\left(\beta_{c}\right)^{-1}=k_{B}T_{c}= & 4.0003\times10^{-21}\,\text{J}\\
\alpha_{c}=\left(\frac{k_{B}T_{c}}{p_{c}}\right)^{\frac{1}{d}}= & 0.881508\,\text{nm}\end{array}\label{xenon scale units (A3)}\end{equation}
while the values of two xenon scale factors $Y_{c}$ and $Z_{c}$
are the following:\begin{equation}
\begin{array}{rl}
Z_{c}=\frac{p_{c}m_{\bar{p}}}{\rho_{c}k_{B}T_{c}}= & 0.28601\\
Y_{c}=\gamma_{c}^{'}\frac{T_{c}}{p_{c}}-1= & 4.91373\end{array}\label{xenon scale factors (A4)}\end{equation}
We have calculated the values {[}see Eq. (\ref{Xe independent amplitudes (39)})]
of the xenon amplitudes $\Gamma^{+}$, $\xi_{0}^{+}$, and $a_{\chi}^{+}$,
(or $a_{\xi}^{+}$) by using the following relations :\begin{equation}
\begin{array}{rl}
a_{\chi}^{+}= & \mathcal{Z}_{\chi}^{1,+}\left[\left(Y_{c}\right)^{\Delta}\right]=1.23399\\
\xi^{+}= & \mathcal{Z}_{\xi}^{+}\left[\left(Y_{c}\right)^{-\nu}\right]=0.209111\\
\Gamma^{+}= & \mathcal{Z}_{\chi}^{+}\left[\left(Z_{c}\right)^{-1}\left(Y_{c}\right)^{-\gamma}\right]=0.0578204\\
(\text{with}\,\,\alpha_{c}\xi^{+}= & 0.184333\,\text{nm})\end{array}\label{fluid amplitude eqs (A5)}\end{equation}
where the respective values of the master amplitudes $\mathcal{Z}_{\xi}^{+}$,
$\mathcal{Z}_{\chi}^{+}$, and $\mathcal{Z}_{\chi}^{1,+}$ (or $\mathcal{Z}_{\xi}^{1,+}$)
are \citet{Garrabos2006mcf} \begin{equation}
\begin{array}{rl}
\mathcal{Z}_{\xi}^{+}= & 0.570481\\
\mathcal{Z}_{\chi}^{+}= & 0.119\\
\mathcal{Z}_{\chi}^{1,+}= & 0.555\end{array}\label{master amplitude values (A6)}\end{equation}
Hereabove, universal features within the Ising-like preasymptotic
domain are correctly accounted for by the following equations\begin{equation}
\begin{array}{rl}
\mathcal{Z}_{\xi}^{1,+}= & 0.37695\\
a_{\xi}^{+}= & \mathcal{Z}_{\xi}^{1,+}\left[\left(Y_{c}\right)^{\Delta}\right]=0.83812\\
\text{with}\;\frac{\mathcal{Z}_{\xi}^{1,+}}{\mathcal{Z}_{\chi}^{1,+}}= & \frac{a_{\xi}^{+}}{a_{\chi}^{+}}=0.67919\end{array}\label{(A7)}\end{equation}
The amplitude value $\xi_{0}^{+}=0.184333\,\text{nm}$ estimated using
the scale dilatation method, compares favorably with the one $\xi_{0}^{+}=0.1866\pm0.001\,\text{nm}$
recently used by Gillis et al \citet{Gillis2004} to analyze sound
attenuation within thermoacoustic layers between solid surfaces and
xenon at critical density. Moreover, this amplitude value is also
in agreement with the following ones obtained from analyses of (static
and dynamic) light scattering measurements: (i) $\xi_{0}^{+}=0.2\,\text{nm}$,
with $\nu=0.63$, in the temperature range $22\,\text{mK}\leq T-T_{c}\leq3.3\,\text{K}$
{[}i.e., $8\times10^{-5}\lesssim\Delta\tau^{*}\lesssim10^{-2}$] (see
Refs. \citet{Giglio1969,Smith1971,Swinney1973}); (ii) $\xi_{0}^{+}=0.1934\,\text{nm}$,
with $\nu=0.62$, in the temperature range $2.6\,\text{mK}\leq T-T_{c}\leq10\,\text{K}$
{[}i.e., $9\times10^{-6}\lesssim\Delta\tau^{*}\lesssim3.4\times10^{-2}$]
from Ref. \citet{Guttinger1980}; (iii) $\xi_{0}^{+}=0.184\pm0.009\,\text{nm}$
and one confluent correction term (with $a_{\xi}^{+}=0.55$ and $\Delta=0.5$),
in the temperature range $28\,\text{mK}\leq T-T_{c}\leq3.65\,\text{K}$
{[}i.e., $9.6\times10^{-5}\lesssim\Delta\tau^{*}\lesssim1.26\times10^{-2}$]
from Ref. \citet{Guttinger1981}.

Using the above xenon parameters in the theoretical estimation of
the correlation length, we also underline that the agreement with
the experimental measurements extends to the range $\frac{\xi}{\alpha_{c}}\gtrsim3$,
i.e., in a temperature range $\Delta\tau^{*}\lesssim L_{EAD}^{Xe}\simeq\left(2-3\right)\times10^{-2}$
which extends largely beyond the Ising-like preasymptotic domain (see
Ref. \citet{Garrabos2006cl}). For example, at the calibration temperature
$T=T_{c}+1\,\text{K}$ ($\Delta\tau_{\text{C2}}^{*}=3.45137\times10^{-3}$,
see below), our calculated value of the correlation length is $\xi=68.6323\,\textrm{\AA}$
(i.e., $\frac{\xi}{\alpha_{c}}\simeq7.8$), while the experimental
values are $\xi=71.14\,\textrm{\AA}$ \citet{Smith1971}, $\xi=64.66\,\textrm{\AA}$
\citet{Guttinger1980}, and $\xi=67.56\,\textrm{\AA}$ \citet{Guttinger1981}.
Especially considering the light scattering measurements of the isothermal
susceptibility and the turbidity of xenon reported by Güttinger and
Cannell \citet{Guttinger1981}, we note that the turbidity data are
fitted in the Orstein-Zernike approximation within a $1\%$ precision
in the temperature range $28\,\text{mK}\leq T-T_{c}\leq2.54\,\text{K}$
{[}i.e., $9.6\times10^{-5}\lesssim\Delta\tau^{*}\lesssim8.76\times10^{-3}$]
, using the present theoretical estimation of the correlation length
and isothermal compressibility. In particular, we estimate the reference
value $\tau\left(T_{\text{or}}=T_{c}+0.6677\,\text{K}\right)=4.1067\,\text{m}^{-1}$
of the turbidity, in excellent agreement with the Güttinger and Cannell's
one $\tau\left(T_{\text{or}}\right)=4.1\,\text{m}^{-1}$ \citet{Guttinger1981},
without any adjustable parameter.

\subsubsection{Mean and master forms of a theoretical crossover function}

The master modifications of the mean crossover functions of Eqs. (\ref{MR lstar vs tstar (1)})
and (\ref{MR khistar vs tstar (2)}) use the following values of three
master (i.e., constant) factors \citet{Garrabos2006mcf} \begin{equation}
\begin{array}{l}
\Theta^{\left\{ 1f\right\} }=4.288\times10^{-3}\\
\mathbb{L}^{\left\{ 1f\right\} }=25.6936\\
\Psi^{\left\{ 1f\right\} }=1.73847\times10^{-4}\end{array}\label{master scale factor values (A8)}\end{equation}
The two scale factors $\vartheta$ and $\psi_{\rho}$ {[}see Eq. (\ref{Xe independent scale factors (40)})]
needed by the massive renormalization scheme are related to $Y_{c}$
and $Z_{c}$ through the equations\begin{equation}
\begin{array}{ccl}
\vartheta= & Y_{c}\Theta^{\left\{ 1f\right\} } & =0.021069\\
\psi_{\rho}= & \left(Z_{c}\right)^{-\frac{1}{2}}\Psi^{\left\{ 1f\right\} } & =3.2507\times10^{-4}\end{array}\label{fluid theta psi scale factors (A9)}\end{equation}
In addition, the values of Eq. (\ref{xenon khi-theta values (61)})
for the metric prefactors $\mathbb{L}_{0,\mathcal{L}}^{*}$ and $\mathbb{X}_{0,\mathcal{L}}^{*}$
are calculated by using the following equations\begin{equation}
\begin{array}{rcl}
\mathbb{L}_{0,\mathcal{L}}^{*}= & \mathcal{Z}_{\xi}^{\pm}\mathbb{Z}_{\xi}^{\pm}\left(Y_{c}\right)^{-\nu}\\
= & \frac{\left(Y_{c}\right)^{-\nu}}{\left[\mathbb{L}^{\left\{ 1f\right\} }\times\left(\Theta^{\left\{ 1f\right\} }\right)^{\nu}\right]} & =0.443526\\
\mathbb{X}_{0,\mathcal{L}}^{*}= & \mathcal{Z}_{\varkappa}^{\pm}\mathbb{Z}_{\chi}^{\pm}\left(Z_{c}\right)^{-1}\left(Y_{c}\right)^{-\gamma}\\
= & \frac{\left(Z_{c}\right)^{-1}\left(Y_{c}\right)^{-\gamma}}{\left[\left(\mathbb{L}^{\left\{ 1f\right\} }\right)^{-d}\left(\Psi^{\left\{ 1f\right\} }\right)^{-2}\left(\Theta^{\left\{ 1f\right\} }\right)^{\gamma}\right]} & =0.214493\end{array}\label{fluid prefactors (A10)}\end{equation}
These two independent prefactors are two characteristic parameters
of xenon, which permit to calculate all the other xenon prefactors
of the modified crossover functions, in conformity with the two-scale-factor
universality.

\subsection{Theoretical representation of the isothermal compressibility in the
temperature range $T-T_{c}\lesssim5-10\,\text{K}$}

\begin{figure*}
\includegraphics{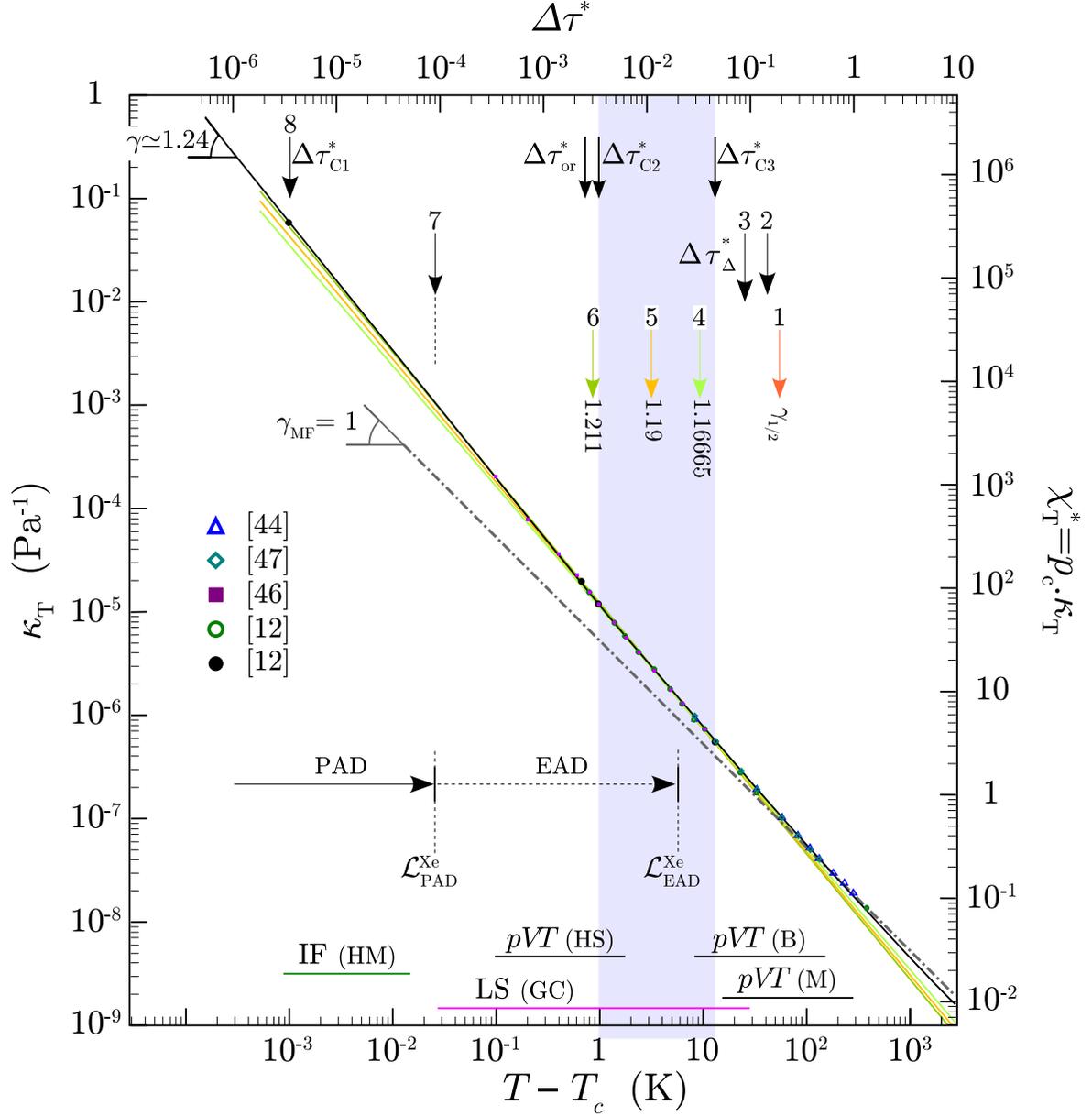}

\caption{Isothermal compressibility of xenon along the critical isochore; Only
the points corresponding to the $pVT$ data given in Table \ref{Table V}
are reported; Inserted labels are common for all the figures (see
text).\label{Figure5}}

\end{figure*}

We can immediately compare all the published $\kappa_{T}\left(\Delta T\right)=\frac{1}{\rho}\left(\frac{\partial\rho}{\partial p}\right)_{T}$
data for xenon at $\rho=\rho_{c}$, to the estimated values by using
the mean crossover function of Eq. (\ref{MR khistar vs tstar (2)})
and the xenon parameter set of Eq. (\ref{Xe independent amplitudes (39)}).
A complete view of the results is shown in Fig. \ref{Figure5} which
covers about six decades on the temperature distance (the data sources
are given below; see Table \ref{Table V}). The related numerical
resolution of this picture is such that the size of each data point
(including the calibration data points defined below) corresponds
to a relative error-bar of $\pm5\%$, while the thickness of the theoretical
curve (label MR) accounts for one of $\pm3\%$. As expected from our
previous analysis \citet{Garrabos2006cl}, no difference larger than
$\pm5\%$ is visible between the curve and the experimental data points
on the extended asymptotic domain (labeled EAD) such as $T-T_{c}\lesssim5\,\text{K}$,
i.e., $\Delta\tau^{*}\lesssim\mathcal{L}_{\text{EAD}}^{\text{Xe}}=\frac{\mathcal{L}_{\text{EAD}}^{\left\{ 1f\right\} }}{Y_{c}}\approx2\times10^{-2}$
(see also below the Fig. \ref{Figure7} and the related discussion).
In addition, the thickness of each (colored) line, having characteristic
slope $\gamma_{e,pVT}$, amplitude $\Gamma_{e,pVT}^{+}$, and color
indexation given in Table \ref{Table V}, represents $\pm1.5\%$ relative
error-bar.

\begin{table*}
\begin{tabular}{|c||c|c|c|c|c|c|c|c|c|c|c|c|c|c|}
\hline 
$T-T_{c}$ & $283.41$ & $233.41$ & $183.41$ & $133.41$ & $108.41$ & $83.41$ & $58.41$ & $33.41$ & $23.41$ & $13.41$ & $10.41$ & $8.41$ & $6.41$ & Ref.\tabularnewline
\hline
\hline 
$10^{6}\kappa_{T}$ & $0.01883$ & $0.02348$ & $0.02912$ & $0.04078$ & $0.05095$ & $0.06772$ & $0.10074$ & $0.18695$ &  &  &  & $0.8961$ &  & \citet{Beattie1951}\tabularnewline
\hline
 &  &  &  & $0.04145$ & $0.05185$ & $0.06904$ & $0.10227$ & $0.19268$ & $0.29065$ & $0.56344$ &  & $1.0099$ &  & \citet{Michels1954}\tabularnewline
\hline
 &  &  &  &  &  &  &  &  &  & $0.54791$ & $0.73594$ &  & $1.3020$ & \citet{Garrabos1982}\tabularnewline
\hline
 &  &  &  &  &  &  &  &  &  & $0.555$ &  &  &  & \citet{Garrabos1982}\tabularnewline
 &  &  &  &  &  &  &  &  &  & $\left(\pm0.008\right)$ &  &  &  & \tabularnewline
\hline
\hline 
$T-T_{c}$ & $4.91$ & $3.41$ & $2.41$ & $1.80$ & $1.4$ & $1.0$ & $0.6677^{*}$ & $0.8$ & $0.6$ & $0.4$ & $0.2$ & $0.1$ & $0.001$ & \tabularnewline
\hline
\hline 
$10^{6}\kappa_{T}$ &  &  &  & $5.7371$ & $7.8543$ & $11.858$ &  &  & $22.701$ & $36.640$ & $79.11$ & $204.86$ &  & \citet{Habgood1954}\tabularnewline
\hline
 & $1.7835$ & $2.7224$ & $4.1139$ &  &  &  &  & $15.432$ &  &  &  &  &  & \citet{Garrabos1982}\tabularnewline
\hline 
 &  &  &  &  &  & $11.95$ & $19.64^{*}$ &  &  &  &  &  & $58562$ & \citet{Garrabos1982}\tabularnewline
 &  &  &  &  &  & $\left(\pm0.15\right)$ & $\left(\pm1\%\right)$ &  &  &  &  &  & $\left(\pm800\right)$ & \tabularnewline
\hline
\end{tabular}

\caption{Values of the isothermal compressibility $\kappa_{T}$ (expressed
in $\text{Pa}^{-1}$) of xenon obtained from the $pVT$ measurements
along the critical isochore $\rho_{c}=1113\,\text{kg}\,\text{m}^{-3}$,
as a function of $T-T_{c}$ (expressed in $\text{K}$); The lower
part corresponds to data obtained within the Ising-like extended asymptotic
domain; Asterisk indicates the reference values used by Güttinger
and Cannell to calibrate their light scattering measurements of the
isothermal compressibility; (see text).\label{Table V}}

\end{table*}

\subsubsection{Data sources for xenon isothermal compressibility}

As shown in the lower part of Figure \ref{Figure5}, the complete
temperature range $1\,\text{mK}\lesssim\Delta T\lesssim400\,\text{K}$
(i.e. $3\times10^{-6}\lesssim\Delta\tau^{*}\lesssim1$) is covered
by successive overlap between $\kappa_{T}$ data which originate from
three distinct experimental methods: $pVT$ measurements with labels
$pVT$ (B) \citet{Beattie1951}, (M) \citet{Michels1954}, and (HS)
\citet{Habgood1954}; light diffusion or turbidity measurements with
label LS (GC)\citet{Guttinger1981}, and Franhauffer interferometry
measurements with label IF (HM) \citet{Estler1975,Hocken1976}.

Within the main part of the temperature distance to $T_{c}$ {[}typically
$0.1\,\text{K}\lesssim\Delta T\lesssim400\,\text{K}$, (i.e., $3\times10^{-3}\lesssim\Delta\tau^{*}\lesssim1$)],
$\kappa_{T}$ is obtained from $pVT$ measurements \citet{Beattie1951,Michels1954,Habgood1954},
generally using graphical \citet{Garrabos1982} or numerical fitting
\citet{Levelt1975} of the $\left[p\left(\rho\right)\right]_{T}$
isotherms to obtain the slope $\left(\frac{\partial\rho}{\partial p}\right)_{T}$
at the selected critical density. We have also reported in Table \ref{Table V}
the $\kappa_{T}$-values obtained from $pVT$ measurements which were
used in our analysis. A noticeable specific situation occurs for xenon
where the high precision of the $pVT$ measurements of Habgood and
Schneider \citet{Habgood1954}, can be used to check carefully the
overlap {[}in the temperature range $0.2\,\text{K}\lesssim\Delta T\lesssim1.8\,\text{K}$,
(i.e. $6\times10^{-4}\lesssim\Delta\tau^{*}\lesssim6\times10^{-3}$)]
with the optical methods.

In the intermediate temperature range {[}typically $30\,\text{mK}\lesssim\Delta T\lesssim15\,\text{K}$,
(i.e., $10^{-4}\lesssim\Delta\tau^{*}\lesssim5\times10^{-2}$)], the
analyses of the light (diffusion or transmission) intensity measurements
\citet{Guttinger1981}, as a function of $\Delta T=T-T_{c}$, provide
interrelated data of the correlation length $\xi$ and isothermal
susceptibility $\chi_{\rho,T}$. $\kappa_{T}$ is related to $\chi_{\rho,T}$
by $\chi_{\rho,T}=\left(\frac{\partial\rho}{\partial\mu_{\rho}}\right)_{T}=\rho^{2}\kappa_{T}$.

In the closest temperature range {[}$1\,\text{mK}\lesssim\Delta T\lesssim10\,\text{K}$,
(i.e., $10^{-4}\lesssim\Delta\tau^{*}\lesssim3\times10^{-2}$)], the
fluid density profile versus the cell height subjected to the gravitational
field generates Franhauffer interferograms \citet{Estler1975,Hocken1976}
which are related to the isothermal compressibility \citet{Sengers1978,Hocken1976}.
However, as already noted by Levelt-Sengers et al \citet{Levelt1976},
the published results in Ref. \citet{Estler1975} needs to be reconsidered
before a quantitative comparison with other ones in overlaping temperature
ranges.

In order to focus our attention in the restricted temperature range
$0.3\,\text{K}\lesssim\Delta T\lesssim5\,\text{K}$, around the central
value $T\simeq T_{c}+1.25\, K$ (see below next §), we have also used
the isothermal compressibility data obtained from the dynamic ligth
scattering data of Cannell and Benedek \citet{Cannell1970}, and from
the static ligth scattering data of Smith et al \citet{Smith1971},
as complementary data sources.

A realistic estimation of the experimental uncertainty is of the order
of $10\%$ when comparison is made between the data obtained from
the different methods (the optical method being of high relative precision
but lower absolute precision; see for example Fig. \ref{Figure6}
below).

\subsubsection{Effective fitting results}

\begin{table*}
\begin{tabular}{|c||c|c|c|c|c|c|c|c|c|}
\hline 
 & $\gamma_{e,pVT}$ & $\Gamma_{e,pVT}^{+}$ & Ref. & $\Delta\tau_{\text{min}}^{*}$ & $\Delta\tau_{\text{max}}^{*}$ & $\left\langle \Delta\tau_{pVT}^{*}\right\rangle $ & $\Delta\tau_{\text{th}}^{*}$ & $R\%\left(\Gamma_{e,pVT}^{+}\right)_{\Gamma_{e,\text{th}}^{+}}$ & $\vartheta_{\mathcal{L}}$\tabularnewline
 &  &  &  &  &  &  & (Table \ref{Table II})  &  & Eq. (\ref{theta vs amplitude ratios (74)})\tabularnewline
\hline
\hline 
$\#6$ & $1.211\left(\pm0.02\right)$ & $0.0743\left(\pm0.015\right)$ & \citet{Levelt1976} & $6.9\times10^{-4}$ & $6.2\times10^{-3}$ & $2.07\times10^{-3}$ & $2.95\times10^{-3}$ & $2.3$ & $0.04676$\tabularnewline
\hline 
$\#5$ & $1.190$ & $0.0793$ & \citet{Levelt1975} & $2.8\times10^{-3}$ & $4.6\times10^{-2}$ & $1.13\times10^{-2}$ & $1.135\times10^{-2}$ & $-1.9$ & $0.01429$\tabularnewline
\hline 
$\#4$ & $1.1665$ & $0.089$ & \citet{Garrabos1986} & $6.2\times10^{-3}$ & $8.1\times10^{-2}$ & $2.24\times10^{-2}$ & $3.338\times10^{-2}$ & $0.1$ & $0.02618$\tabularnewline
\hline 
$\#\gamma_{\text{MF}}$ & $1$ & $0.11$ & \citet{Garrabos1982} & $0.5$ & $1$ & $7.1\times10^{-1}$ & $\left(\infty\right)$ & $30.$ & $\text{n.a.}$\tabularnewline
\hline
\end{tabular}

\caption{Lines with labels $\#6,\,5,\,4$ (corresponding to the columns with
same labels in Table \ref{Table II}): Effective exponent $\gamma_{e,pVT}$
(column 2) and amplitude $\Gamma_{e,pVT}^{+}$ (column 3) of a power
law description of Eq. (\ref{kapaT pVT power law (A11)}), for $\kappa_{T}^{*}$
obtained from xenon $pVT$ measurements in the temperature range and
(geometrical) mean temperature $\left\langle \Delta\tau_{pVT}^{*}\right\rangle =\sqrt{\Delta\tau_{\text{min}}^{*}\Delta\tau_{\text{max}}^{*}}$
(column 6); see references given in column 4; Calculated values of
the local temperature distance $\Delta\tau_{\text{th}}^{*}$ (column
7) are from Table \ref{Table II} line10; Residuals $R\%\left(\Gamma_{e,pVT}^{+}\right)_{\Gamma_{e,\text{th}}^{+}}$
(column 8), expressed in \%, between experimental ($\Gamma_{e,pVT}^{+}$)
xenon amplitude and calculated ( $\Gamma_{e,\text{th}}^{+}$ ) xenon
amplitude from Table \ref{Table II} line 12; Calculated values of
xenon crossover parameter $\vartheta_{\mathcal{L}}$ (column 10) from
Eq. (\ref{theta vs amplitude ratios (74)}); Line with label $\#\gamma_{\text{MF}}$
: Equivalent results for $\gamma_{e}=\gamma_{\text{MF}}=1$ (see text);
n.a.: non available.\label{Table VI}}

\end{table*}

To complete our analysis of the Güttinger and Cannell's measurements,
we have also made a comparison with the published fitting results,
obtained from $pVT$ data, using the following effective power law
with an adjustable \emph{non-Ising} exponent, \begin{equation}
\kappa_{T}^{*}=\Gamma_{e,pVT}^{+}\left(\Delta\tau^{*}\right)^{-\gamma_{e,pVT}}\label{kapaT pVT power law (A11)}\end{equation}
The $\gamma_{e,pVT}$ and $\Gamma_{e,pVT}^{+}$ values are reported
in Table \ref{Table VI}, in addition with the related finite extension
$\Delta\tau_{\text{min}}^{*}-\Delta\tau_{\text{max}}^{*}$ of the
experimental temperature range. The labels $\#6,\,5,\,4$, of the
lines of Table \ref{Table VI} correspond to the respectives labels
$\#6,\,5,\,4$, of the columns of Table \ref{Table II}. The typical
error-bar values of these fitting results are given in line $\#6$.
At each restricted temperature range, we have attached the central
value $\left\langle \Delta\tau_{e,pVT}^{*}\right\rangle =\sqrt{\Delta\tau_{\text{min}}^{*}\Delta\tau_{\text{max}}^{*}}$
(in log scale). So that, in the next Fig. \ref{Figure7}, each effective
power law result is illustrated as a form of a (colored) point-segment
(with $\gamma_{e,pVT}$-value as a label) covering the range $\Delta\tau_{\text{min}}^{*}\leq\Delta\tau^{*}\leq\Delta\tau_{\text{max}}^{*}$,
around a central (full) point fixed at $\left\langle \Delta\tau_{pVT}^{*}\right\rangle $
.

In this Table \ref{Table VI}, results \citet{Garrabos1982} given
in line labeled $\#\gamma_{\text{MF}}$ correspond to a fit of the
$pVT$ data of Beattie et al \citet{Beattie1951}, where the mean-field
value $\gamma_{e}=\gamma_{\text{MF}}=1$ of the effective exponent
can be observed at large temperature distance from $T_{c}$ (typically
$\Delta\tau^{*}>0.5$, see also Ref. \citet{Bagnuls1984b}).

\begin{table*}
\begin{tabular}{|r||c|c|c|c|c|c|c|c|c|c|}
\hline 
 & $\gamma_{e}$ & $\Gamma_{e}^{+}$ & Ref. & $\begin{array}{cc}
\Delta\tau_{\text{min}}^{*}\\
\Delta\tau_{\text{max}}^{*}\end{array}$ & $\left\langle \Delta\tau_{\gamma_{e}}^{*}\right\rangle $ & $\Delta\tau_{\text{th}}^{*}$ & $\Gamma_{e,\text{th}}^{+}$ & $R\%\left(\Gamma_{e}^{+}\right)_{\Gamma_{e,\text{th}}^{+}}$ & $\mathbb{Z}_{\chi,e}^{+}$ & $\vartheta_{\mathcal{L}}$\tabularnewline
 &  &  &  &  &  &  &  &  &  & Eq. (\ref{theta vs amplitude ratios (74)})\tabularnewline
\hline
\hline 
$\text{GC}_{e}$ & $1.205879$ & $0.07551466$ & t.w. & $\begin{array}{rc}
9.115 & \times10^{-4}\\
1.95 & \times10^{-2}\end{array}$ & $4.215\times10^{-3}$ & $4.347\times10^{-3}$ & $0.0747481$ & $1.03$ & $0.396926$ & $0.0285147$\tabularnewline
\hline 
$pVT$ & $1.204519$ & $0.07590869$ & t.w. & $\begin{array}{rc}
3.451 & \times10^{-4}\\
3.593 & \times10^{-2}\end{array}$ & $3.521\times10^{-3}$ & $4.778\times10^{-3}$ & $0.075298$ & $0.81$ & $0.40195$ & $0.0265262$\tabularnewline
\hline 
$\text{CB}$(a) & $1.21$ & $0.076845$ & \citet{Cannell1970} & $\begin{array}{rc}
6.9 & \times10^{-4}\\
6.9 & \times10^{-2}\end{array}$ & $6.9\times10^{-3}$ &  &  &  &  & \tabularnewline
(b) & $1.201586$ & $0.0780485$ & t.w. & $\begin{array}{rc}
3.451 & \times10^{-4}\\
2.5886 & \times10^{-2}\end{array}$ & $2.989\times10^{-3}$ &  &  &  &  & \tabularnewline
(c) & $1.201586$ & $0.0768207$ & t.w. &  &  & $5.804\times10^{-3}$ & $0.076466$ & $0.46$ & $0.41283$ & $0.0238014$\tabularnewline
\hline 
$\text{SGB}$(a) & $1.21\pm0.03$ & $0.06742$ & \citet{Smith1971} & $\begin{array}{rc}
1.553 & \times10^{-4}\\
1.76 & \times10^{-2}\end{array}$ & $1.653\times10^{-3}$ &  &  &  &  & \tabularnewline
(b) & $1.204381$ & $0.0679096$ & t.w. & $\begin{array}{rc}
1.5877 & \times10^{-4}\\
1.723 & \times10^{-2}\end{array}$ & $1.653\times10^{-3}$ &  &  &  &  & \tabularnewline
(c) & $1.204381$ & $0.0755227$ &  t.w. &  &  & $4.823\times10^{-3}$ & $0.075354$ & $0.22$ & $0.40246$ & $0.0224563$\tabularnewline
\hline
\end{tabular}

\caption{Fitting results using an effective power law equation $\kappa_{T}^{*}=\Gamma_{e}^{+}\left(\Delta\tau^{*}\right)^{-\gamma_{e}}$
in a restricted temperature range $\Delta\tau_{\text{min}}^{*}\leq\Delta\tau^{*}\leq\Delta\tau_{\text{max}}^{*}$
(see text); The calculated local values of the temperature distance
$\Delta\tau_{\text{th}}^{*}$ (column 7), the xenon amplitude $\Gamma_{e,\text{th}}^{+}$
(column 8), the universal amplitude $\mathbb{Z}_{\chi,e}^{+}$ (column
10), and the xenon crossover parameter $\vartheta_{\mathcal{L}}$
(column 11), are for each condition $\gamma_{e,\text{th}}=\gamma_{e}$
; t.w.: this work.\label{Table VII}}

\end{table*}

\begin{figure}
\includegraphics[width=80mm,height=80mm,keepaspectratio]{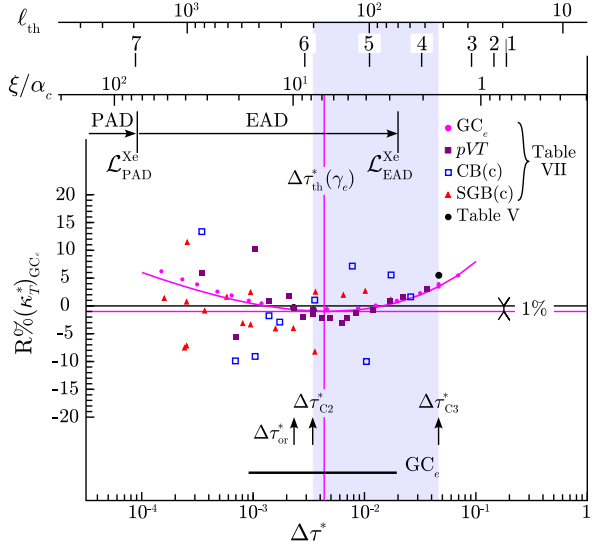}

\caption{Residuals $\%R\left(\kappa_{T}^{*}\right)_{\text{GC}_{e}}$ (expressed
in \%) of the $\kappa_{T}^{*}$ data from the effective power law
$\kappa_{T,\text{GC}_{e}}^{*}=0.07551466\left(\Delta\tau^{*}\right)^{-1.205879}$
(see text); The horizontal and vertical (pink) lines permit to define
the local values of the effective amplitude ($\Gamma_{e,\gamma_{e}}^{+}=0.99\times0.07551466$)
and the temperature distance {[}$\Delta\tau_{th}^{*}\left(\gamma_{e}\right)=4.347\times10^{-3}$,
see Table \ref{Table VII} ] where the line of slope $\gamma_{e}=1.205879$
and equation $\kappa_{T}^{*}=\Gamma_{e,\gamma_{e}}^{+}\left(\Delta\tau^{*}\right)^{-\gamma_{e}}$,
is tangent to the experimental singular behavior measured by Güttinger
and Cannell, at the contact point of coordinates $\left\{ \Delta\tau_{th}^{*}\left(\gamma_{e}\right);\,0.99\kappa_{T,\text{GC}_{e}}^{*}\left[\Delta\tau_{th}^{*}\left(\gamma_{e}\right)\right]\right\} $
(see also Table \ref{Table VII} and text); Horizontal segment labeled
$\text{GC}_{e}$: restricted temperature range of the fit by the effective
power law $\kappa_{T,\text{GC}_{e}}^{*}=0.07551466\left(\Delta\tau^{*}\right)^{-1.205879}$;
(colored) symbols: see inserted legend and Tables \ref{Table V} and
\ref{Table VII}; All the other inserted labels are identical to the
ones previously used. \label{Figure6}}

\end{figure}

A similar introduction of the effective power law to analyse the results
of light scattering experiments gives access to a quantitative comparison
between the $pVT$ and light scattering measurements of the isothermal
compressibility. As a matter of fact, Güttinger and Cannell have claimed
that the correction to scaling terms are important by demonstrating
that the susceptibility deviates systematically from a simple power
law behavior with the effective exponent value $\gamma_{e}=1.206$.
More precisely, Fig. 2 of Ref. \citet{Guttinger1981} shows that $\gamma_{e}\simeq1.206$
is the slope of the tangent line to the rough experimental behavior
at $\Delta\tau_{\gamma_{e}=1.206}^{*}\simeq4.5\times10^{-3}$, i.e.,
the temperature distance which corresponds to the minimum of the deviation
curve in this Fig. 2. We have used this result to renew a fit of twelve
compressibility data measured in the restricted temperature range
$9.115\times10^{-4}\leq\Delta\tau^{*}\leq1.95\times10^{-2}$ (i.e.,
$0.26\,\text{K}\leq\Delta T\leq6.90\,\text{K}$), i.e., a temperature
range with a central value $\left\langle \Delta\tau_{e,\text{GC}}^{*}\right\rangle =4.215\times10^{-3}$,
very close to the $\Delta\tau_{\gamma_{e}=1.206}^{*}$ one. As expected,
our fitting values $\gamma_{e}=1.205879$ and $\Gamma_{e}^{+}=0.07551466$
(reported in line $\#\text{GC}_{e}$ of Table \ref{Table VII}), are
in excellent agreement with the Güttinger and Cannell's ones $\gamma_{e}=1.206$
and $\Gamma_{e}^{+}=0.075135$. The latter amplitude value was calculated
applying a $-4\%$ correction (corresponding to the fit deviation
observed in this restricted temperature range) to the effective amplitude
of the published original fit. This resulting power law behavior may
be seen from Fig. \ref{Figure6} which gives the \% residual between
the dimensionless isothermal compressibility and the fitting equation
$\kappa_{T,\text{GC}_{e}}^{*}=0.07551466\left(\Delta\tau^{*}\right)^{-1.205879}$.
For us, the most important consequence of the high \emph{relative}
precision of the Güttinger and Cannell's measurements is the demonstration
that a well-defined local value $\gamma_{e}\simeq1.20588$ of the
effective exponent can be measured at a well-defined local value ($T-T_{c}\simeq1.260\, K$)
of the temperature distance to $T_{c}$. Accordingly, by expressing
now the condition $\gamma_{e,\text{th}}=1.205879$ in complement to
the similar eight conditions given in Table \ref{Table II} for the
theoretical crossover function of Eq. (\ref{MR khistar vs tstar (2)}),
such a local value of the effective exponent must be observed at $\Delta\tau_{th}^{*}\left(\gamma_{e,th}=1.2058\right)=4.347\times10^{-3}$,
while the calculated local values of the effective theoretical and
xenon amplitudes are $\mathbb{Z}_{\chi,e}^{+}=0.396926$ and $\Gamma_{e,th}^{+}\left(\gamma_{e,th}=1.2058\right)=0.0747481$.
We will exploit the noticeable agreement between the experimental
and theoretical local value of the effective power law behavior in
the last § of the Appendix. We note that the calculated value of the
correlation length is then $\frac{\xi\left(\gamma_{e}\simeq1.20588\right)}{\alpha_{c}}\simeq6.764$,
i.e, within our expected range for the Ising-like extended asymptotic
domain previously defined by the condition $\ell^{*}=\frac{\xi}{\alpha_{c}}\gtrsim3$.

Previously, in a first approach independent of any one-parameter crossover
theory, it is necessary to control that the compressibility data obtained
by any other measurement method covering a similar restricted temperature
range, are satisfying this effective power law behavior with comparable
values of the related effective amplitudes. This is the object of
the results given in lines $\#pVT,\, CB,\, SGB$ of Table \ref{Table VII}
(see also Fig. \ref{Figure6}), where we have used:

i) line $\#pVT$, the $pVT$ data of Table \ref{Table V} covering
the temperature range $0.1\,\text{K}\leq T-T_{c}\leq10.41\,\text{K}$
(i.e., $3.451\times10^{-4}\leq\Delta\tau^{*}\leq3.593\times10^{-2}$);

ii) lines $\#CB$, the dynamic ligth scattering data of Cannell and
Benedek \citet{Cannell1970} covering the temperature range $0.2\,\text{K}\leq T-T_{c}\leq20\,\text{K}$,
i.e., $6.9\times10^{-4}\leq\Delta\tau^{*}\leq6.9\times10^{-2}$, and

iii) lines $\#SGB$, the static ligth scattering data of Smith et
al \citet{Smith1971} covering the temperature range $0.045\,\text{K}\leq T-T_{c}\leq5.1\,\text{K}$,
i.e., $1.553\times10^{-4}\leq\Delta\tau^{*}\leq1.76\times10^{-2}$
.

In the $pVT$ case, the fitting values of the exponent-amplitude pair
confirm the power law behavior observed from Güttinger and Cannell's
measurements of high relative precision. On the other hand, we have
reported the initial fitting results {[}lines $\#CB,\, SGB$, (a)]
of Refs. \citet{Cannell1970,Smith1971}. They are only used to illustrate
quantitatively the effects due to a large uncertainty on measurements
of the geometrical factors in light scattering experiments\citet{Smith1971},
or to an indirect estimation of the magnitude of the compressibility
from an elaborate and complex analysis of the Brillouin spectra of
xenon \citet{Cannell1970}. For example, the numerical values found
from Ref. \citet{Smith1971} with $\gamma_{e}=1.21$ can be multiplied
by a factor $\frac{1.63}{1.43}=\frac{0.076845}{0.06742}\simeq1.14$
to match those reported in Ref. \citet{Cannell1970}. However, in
spite of the importance ($\simeq\left(10-20\right)\%$) of these effects,
we have confirmed with our fitting results {[}lines $\#CB,\, SGB$,
(b)] that the value of the effective exponent is well in the range
$\gamma_{e}\simeq1.20-1.21$ for the restricted temperature range
selected here. In addition, the large uncertainty on the experimental
value of the effective amplitude can be decreased by using our calibrated
value of the isothermal compressibility at $T=T_{c}+1\, K$. For example,
after calibration {[}lines $\#CB,\, SGB$, (c)], the effective amplitude
for Cannell and Benedek's data was lowered by $1.573\%$, while the
one for Smith et al's data was increased by $11,21\%$, in agreement
with around $\simeq14\%$ initial deviation between these two data
set.

Finally, as reported in column 9 of Table \ref{Table VII}, the percent
deviation $\%R\left(\Gamma_{e}^{+}\right)_{\Gamma_{e,\text{th}}^{+}}=100\left(\frac{\Gamma_{e}^{+}}{\Gamma_{e,\text{th}}^{+}\left(\gamma_{e,\text{th}}=\gamma_{e}\right)}-1\right)$
between the experimental and theoretical effective amplitudes for
the four experimental values of the effective exponent $\gamma_{e}$
is on the $1\%$-level. In Fig. \ref{Figure6}, we have also reported
the residuals $\%R\left(\kappa_{T}^{*}\right)_{\text{GC}_{e}}=100\left(\frac{\kappa_{T,}^{*}}{\kappa_{T,\text{GC}_{e}}^{*}}-1\right)$
(expressed in \%) between each experimental data $\kappa_{T}^{*}$
and the calculated one $\kappa_{T,\text{GC}_{e}}^{*}$ using the effective
power law $\kappa_{T,\text{GC}_{e}}^{*}=0.07551466\left(\Delta\tau^{*}\right)^{-1.205879}$
as reference (see line $\#\text{GC}_{e}$ in Table \ref{Table VII}).
We note the data agreement at the same percent level that the precision
on the calibrated values of the isothermal compressibility, which
leads to a conclusion that the {}``experimental'' values of the
effective amplitude $\Gamma_{e}^{+}$ are estimated with an uncertainty
of $1\%$. 

Therefore, in a second approach focussed on the validity test of the
one-parameter crossover modelling in pure fluids, which will be discussed
below in § A.3, all these fitting results with effective values of
the exponents observed in a small restricted temperature range at
large temperature distance to $T_{c}$ are appropriate:

i) to check the calibration of the leading asymptotic amplitude $\Gamma^{+}$
(with $\gamma$ fixed);

ii) to verify the uniqueness of the scale factor $\vartheta$ by estimating
the related local values of $\vartheta_{\mathcal{L}}$, and then,

iii) to control the master values of the leading amplitude $\mathcal{Z}_{\chi}^{+}=0.119$
{[}related to $\Gamma^{+}$, see Eq. (\ref{fluid amplitude eqs (A5)})]
and the confluent amplitude $\mathcal{Z}_{\chi}^{1,+}=0.555$, {[}related
to $\vartheta$, see Eq. (\ref{fluid amplitude eqs (A5)})]. 

\begin{figure}
\includegraphics[width=1\columnwidth,keepaspectratio]{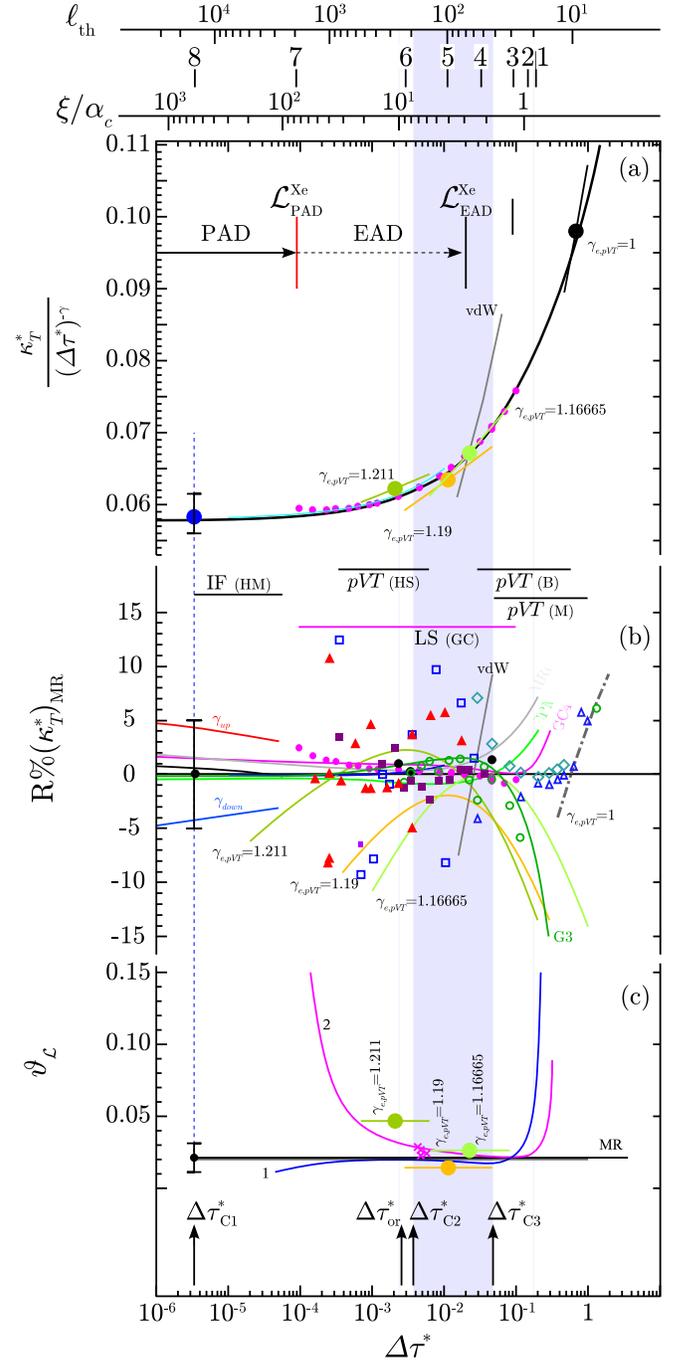}

\caption{Fitting results of xenon isothermal compressibility $\kappa_{T}^{*}$,
as a function of $\Delta\tau^{*}$(lower horizontal axis) or correlation
lengths (upper horizontal axes), by using Eq. (\ref{(23)}) with xenon
parameters of Eq. (\ref{Xe independent scale factors (40)}); (a)
$\frac{\kappa_{T}^{*}}{\left(\Delta\tau^{*}\right)^{-\gamma}}$ (full
black line: calculated confluent correction from the crossover function);
(b) Residuals $R\%\left(\kappa_{T}^{*}\right)_{\text{MR}}$ (expressed
in \%), for all the experimental data and fitting results selected
in this work from reference to the theoretical estimation by the crossover
function; (c) Same as Fig. \ref{Figure4}(b), calculated $\vartheta_{\mathcal{L}}$
{[}line MR: using the scale dilatation method; line 1: from Eq. (\ref{gama effective fitting eq (71)});
line 2: from Eq. (\ref{theta vs amplitude ratios (74)}); colored
point-segments: from Table \ref{Table VI}, column 10; crosses: from
Table \ref{Table VII}, column 11]; All the inserted labels and symbols
are identical to the ones of the previous figures (see text).\label{Figure7}}

\end{figure}

\subsubsection{Data comparison from reference to the master crossover function}

Data comparisons are magnified in the three parts (a), (b), and (c),
of Figure \ref{Figure7} (in lin-log scale), either as a function
of $\Delta\tau^{*}$ in the lower horizontal axis, or as a function
of the theoretical ($\ell_{\text{th}}$) and master ($\ell^{*}=\frac{\xi}{\alpha_{c}}$)
correlation lengths \citet{Garrabos2006cl} in the upper horizontal
axis. The extensions of the Ising-like preasymptotic and extended
asymptotic domains, the selected $\gamma_{e}$-conditions of Table
\ref{Table II}, and each experimental temperature range, are illustrated
as in Fig. \ref{Figure2}.

The asymptotic behavior of the quantity $\frac{\kappa_{T}^{*}}{\left(\Delta\tau^{*}\right)^{-\gamma}}$
estimated from the crossover function {[}(black) curve MR] is given
in Figure \ref{Figure7}a. Such a presentation magnifies the role
of the theoretical $\gamma$-value on the determination of the leading
amplitude $\Gamma^{+}$close to $T_{c}$ (given by an horizontal asymptotic
limit for $\Delta\tau^{*}\rightarrow0$). Correlatively, at large
values of $\Delta\tau^{*}$, the increasing contribution of the confluent
corrections is well demonstrated by the decreasing $\gamma_{e,pVT}$-values,
here given by the direction of each point-segment closely tangent
to the MR curve (see Table \ref{Table VI}). Only the published experimental
data points {[}(pink) small circles] and fitting result {[}GC4 (pink)
line] of Güttinger and Cannell's measurements are reported in this
Fig. \ref{Figure7}(a).

The residual $R\%\left(\kappa_{T}^{*}\right)_{\text{MR}}=100\left[\frac{\kappa_{T}^{*}}{\kappa_{T,\text{MR}}^{*}}-1\right]$,
expressed in \%, for all the selected data, are reported in part (b).
The accurate ($\sim\pm3\%$) theoretical representation of $pVT$
data of Habgood and Schneider and light scattering data of Güttinger
and Cannell confirms that the extended asymptotic domain is well such
that $\Delta\tau^{*}\lesssim\mathcal{L}_{\text{EAD}}^{\text{Xe}}\sim0.02-0.03$,
which well corresponds to the range $\ell^{*}=\frac{\xi}{\alpha_{c}}\gtrsim2-3$
of accurate theoretical representation of the master singular behavior
of the correlation length, as previously observed.

In Figs \ref{Figure7}(a) and (b), a significant difference occurs
in the case of the actual mean field like behavior represented by
the segment $\gamma_{pVT}=\gamma_{\text{MF}}=1$ (see line 5 of Table
\ref{Table II}). In the temperature range $90\,\text{K}\lesssim\Delta T\lesssim300\,\text{K}$,
(i.e., $3\times10^{-1}\lesssim\Delta\tau^{*}\lesssim1$), the effective
classical-to-critical crossover for xenon is not accounted for by
the mean crossover function. In addition, each part (a) and (b) also
contains the relative position of a grey segment (labeled $\text{vdW}$)
which corresponds to the mean-field-like equation $\kappa_{T}^{*}\Delta\tau^{*}=\frac{1}{6}$,
where xenon is assimilated to a van der Waals fluid (i.e. with $\gamma_{e,\text{vdW}}=1$
and $\Gamma_{e,\text{vdW}}^{+}=\frac{1}{6}$). That clearly illustrates
the failure of the van der Waals equation of state close to the liquid-gas
critical point.

Figure \ref{Figure7}(c) is similar to Figure \ref{Figure4}(b). Now
we have added the point-segment representation of the values of $\vartheta_{\mathcal{L},pVT}$
calculated using Eq. (\ref{theta vs amplitude ratios (74)}) and the
fitting results of the $pVT$ measurements with Eq. (\ref{kapaT pVT power law (A11)}),
as reported in Table \ref{Table V}. We will return below to the discussion
of these $pVT$ results which confirm that the xenon crossover is
characterized by a unique value {[}Eq. (\ref{xenon khi-theta values (61)})]
of the scale factor $\vartheta\left(\text{Xe}\right)$ over the temperature
range where $\frac{\xi}{\alpha_{c}}\gtrsim2-3$.

\subsection{Data calibration and related discussion to the uniqueness of the
crossover parameter }

As already evidenced in the seventy's \citet{Levelt1975}, fitting
the experimental singular behavior using Wegner-like expansion with
a limited number of terms generates large uncertainty in the amplitude
determination, due to the low convergence of this expansion. Examining
more carefully the isothermal compressibility data obtained from interferometry
\citet{Hocken1976}, light scattering \citet{Guttinger1981}, and
$pVT$ \citet{Beattie1951,Habgood1954,Michels1954} measurements,
it was noted by one of us \citet{Garrabos1982}, that a restricted
three-term Wegner-like expansion, with fixed exponent values $\gamma_{I}=1.24\pm0.01$
and $\Delta_{I}=\frac{1}{2}\pm0.05$, can provide satisfactory representation
of the $\kappa_{T}$ singular behavior in the range $T-T_{c}\lesssim13\,\text{K}$
(see for example the curve G3 in Figure \ref{Figure7}b) which corresponds
to Eq. (\ref{khiTexp GC (43)}) with $\Gamma^{+}=0.0577$, $a_{1\chi}^{+}=a_{\chi}^{+}=1.55$,
$a_{2\chi}^{+}=-1.3$, and $a_{3\chi}^{+}=0$; see Table \ref{Table IV},
last line). It was conjointly shown that $\Gamma^{+}$, $a_{\chi}^{+}$,
and $a_{2\chi}^{+}$ can also be calculated using three calibrated
values at three selected temperature distances (in logarithmic scale,
see below). The correlative main results were the estimation of min
($a_{\chi,\text{min}}^{+}\simeq0.6$) and max ($a_{\chi,\text{max}}^{+}\simeq1.8$)
values of $a_{\chi}^{+}$ and the evaluation of the correlated error-bars
which can then be controlled by a careful analysis of the residuals.

We recall the main advantages of such a three points calibration approach,
first by examining the estimation of the leading amplitude and its
attached uncertainty for the closest point to the critical point.

\subsubsection{Calibration of the leading amplitude the closest to the critical
point}

At $T-T_{c}=1\,\text{mK}$ ($\Delta\tau_{\text{C1}}^{*}=3.45137\times10^{-6}$),
which corresponds to the {}``lowest'' temperature distance accessible
by interferometry experiments of Hocken and Moldover (HM) \citet{Hocken1976},
the standard dimensionless value of the isothermal compressibility
was defined such that $\kappa_{T,1}^{*}\left(\Delta\tau_{\text{C1}}^{*}\right)=\Gamma_{I}^{+}\left(\Delta\tau_{\text{C1}}^{*}\right)^{-\gamma_{I}}=\Gamma_{\text{MR6}_{\text{max}}}^{+}\left(\Delta\tau_{\text{C1}}^{*}\right)^{-\gamma_{\text{MR6}_{\text{max}}}}=\left(3.415\pm0.035\right)\times10^{5}$
with $\left\{ \gamma_{I}=1.240;\,\Gamma_{I}^{+}=0.0574\pm0.0006\right\} $
and $\left\{ \gamma_{\text{MR6}_{\text{max}}}=1.24194;\,\Gamma_{\text{MR6}_{\text{max}}}^{+}=0.0563\pm0.0006\right\} $.
These initial values of the leading amplitude $\Gamma^{+}$ were re-evaluated
accounting for the small difference with $\gamma=1.2395935$. Similar
re-evaluation was conjointly made for the leading amplitudes $\xi^{+}$
(of the dimensionless correlation length $\xi^{*}$) and $B$ (of
the dimensionless order parameter density $\Delta\rho_{LV}^{*}$)
due to the small differences in $\nu$ and $\beta$ values. The updated
values of the corresponding leading amplitudes (such as $\Gamma^{+}\left(\text{Xe}\right)=0.05782\pm0.0006$)
were then used to optimize the new central values of the corresponding
master amplitudes (such as $\mathcal{Z}_{\chi}^{+}=0.119$ reported
here), using Eqs. (\ref{master amplitude values (A6)}), corresponding
values of the critical parameters {[}see Eq. (\ref{Xe critical coordinates (A2)})],
and the updated value of the universal amplitude combination $\left(\xi^{+}\right)^{-d}\frac{\Gamma^{+}}{B^{2}}=\left(\mathcal{Z}_{\xi}^{+}\right)^{-d}\frac{\mathcal{Z}_{\chi}^{+}}{\mathcal{Z}_{M}^{2}}=\left(\mathbb{Z}_{\xi}^{+}\right)^{d}\frac{1}{\mathbb{Z}_{\chi}^{+}\mathbb{Z}_{M}^{2}}=\left(R_{\xi}^{+}\right)^{-d}R_{C}^{+}=2.92638$
(with $R_{\xi}^{+}=0.2696967$ and $R_{C}^{+}=0.057406$ \citet{Bagnuls2002}).
Finally, selecting $\mathcal{Z}_{\chi}^{+}=0.119$ and $\mathcal{Z}_{M}=0.468$
\citet{Bamplitude} as {}``central'' values for two independent
master amplitudes, any other (master and physical) leading amplitude
can be calculated with the numerical precison of the massive renormalization
scheme. For example, the number of digits in the quoted value $\mathcal{Z}_{\xi}^{+}=0.570481$
is similar to the numerical precision of $\mathbb{Z}_{\xi}^{+}=2.121008$
(see Table \ref{Table I}a), and more generally, data of Eqs. (\ref{master amplitude values (A6)})
to (\ref{kapaT pVT power law (A11)}) are obtained according to this
scheme.

Considering now the $\gamma$ differences between the theoretical
estimations, we can admit that the mean value $\gamma=1.2395935$
(see Table \ref{Table I}b) is on the range $\gamma_{\text{low}}\simeq1.235\leq\gamma\leq\gamma_{\text{high}}\simeq1.242$.
The max value $\gamma_{\text{high}}\simeq1.242$ was used in our initial
fitting of the GC data (see line 2 of Table \ref{Table III}), while
the min value $\gamma_{\text{low}}\simeq1.235$ was recently estimated
from the minimal subtraction scheme \citet{Zhong2003}. The correlative
effect on the $\Gamma^{+}$ value is of the order of $\pm3$\%, as
shown by the curves (labeled $\gamma_{\text{high}}$ and $\gamma_{\text{low}}$
in Fig. \ref{Figure7}c) related to the power law $\kappa_{\text{high,low}}^{*}=\Gamma_{\text{high,low}}^{+}\left(\Delta\tau^{*}\right)^{-\gamma_{\text{up,down}}}$,
with $\Gamma_{\text{high}}^{+}=0.056$ and $\Gamma_{\text{low}}^{+}=0.062$,
respectively. This theoretical uncertainty level appears then comparable
to the experimental one ($\sim$$\pm5$\%), when the $\gamma$ and
$\Gamma^{+}$ values are obtained from interferometry measurements
close to the critical point. For example, the initial {}``free''
values are $\gamma=1.23$ and $\Gamma^{+}=0.062\pm0.006$ \citet{Hocken1976},
while a subsequent analysis made for fixed $\gamma=1.24$ gives $\Gamma^{+}=0.058\pm0.002$
\citet{Sengers1978}. As a result, we note that the calibrated value
of $\kappa_{T,1}^{*}\left(\Delta\tau_{\text{C1}}^{*}\right)=\left(3.415\pm0.035\right)\times10^{5}$
remains well representative of the theoretical analyses of the interferometry
measurements at this closest temperature distance to $T_{c}$. However,
we also recall that the uncertainty associated to the experimental
determination of the critical temperature (roughly estimated of the
order of $\pm0.5\,\text{mK}$) cannot be accounted for in the above
analyses of the interferometry measurements.

To partly conclude, when the exponent $\gamma$ is fixed to its \emph{Ising-like}
theoretical value $\gamma=1.2395935$, the above standard value of
the isothermal compressibility at $T-T_{c}=1\,\text{mK}$ is a realistic
asymptotic constraint to fix the {}``central'' value $\Gamma^{+}\left(\text{Xe}\right)=0.05782(\pm0.0006)$
of the leading amplitude, before to analyse the isothermal compressibility
data over larger temperature distances to the (known) critical temperature
with the objective to estimate the contribution of the confluent corrections
to scaling in the xenon case.

\subsubsection{Calibration of the first amplitude of the critical confluent correction
to scaling beyond the Ising-like preasymptotic domain}

As indicated above, by using two other points of calibration properly
selected to cover the temperature range of optical and $pVT$ measurements,
provides an analytical estimation of $a_{\chi}^{+}\equiv a_{1\chi}^{+}$
and $a_{2\chi}^{+}$ in Eq. (\ref{khiTexp GC (43)}) {[}with $\gamma$,
$\Delta$ fixed, and $a_{3\chi}^{+}=0$]. Indeed, from the calibrated
values $\left\{ \Delta\tau_{\text{C2}}^{*};\kappa_{T,2}^{*}\right\} $
and $\left\{ \Delta\tau_{\text{C3}}^{*};\kappa_{T,3}^{*}\right\} $,
we can define the pairs $\left\{ \Delta\tau_{2}^{*}=\frac{\Delta\tau_{\text{C3}}^{*}}{R_{32}};Y_{2}=\frac{\kappa_{T,2}^{*}}{\Gamma^{+}\left(\Delta\tau_{\text{C2}}^{*}\right)^{-\gamma}}-1\right\} $
and $\left\{ \Delta\tau_{\text{C3}}^{*};Y_{3}=\frac{\kappa_{T,3}^{*}}{\Gamma^{+}\left(\Delta\tau_{C3}^{*}\right)^{-\gamma}}-1\right\} $.
We obtain\begin{equation}
a_{\chi}^{+}\equiv a_{1\chi}^{+}=\frac{Y_{2}\left(R_{32}\right)^{2\Delta}-Y_{3}}{\Delta\tau_{\text{C3}}^{*}\left[\left(R_{32}\right)^{\Delta}-1\right]}\label{akhiplus cal}\end{equation}
 and\begin{equation}
a_{2\chi}^{+}=\frac{Y_{3}-Y_{2}\left(R_{32}\right)^{2\Delta}}{\left(\Delta\tau_{\text{C2}}^{*}\Delta\tau_{\text{C3}}^{*}\right)^{\Delta}\left[\left(R_{32}\right)^{\Delta}-1\right]}\label{a2khiplus cal}\end{equation}
Equations (\ref{akhiplus cal}) and (\ref{a2khiplus cal}) have convenient
analytic forms to check the influence of the selected values for the
exponents and calibrated points \citet{Garrabos1982}. For example,
we can choose the calibrated points at $T_{2}-T_{c}=1\,\text{K}$
($\Delta\tau_{\text{C2}}^{*}=3.45137\times10^{-3}$) with $\kappa_{T,2}=\left(1.195\pm0.012\right)\times10^{-5}\,\text{Pa}^{-1}$
($\kappa_{T,2}^{*}=69.8\pm0.7$), and $T_{3}-T_{c}=13.41\,\text{K}$
($\Delta\tau_{\text{C3}}^{*}=4.62829\times10^{-2}$) with $\kappa_{T,3}=\left(5.55\pm0.08\right)\times10^{-7}\,\text{Pa}^{-1}$
($\kappa_{T,3}^{*}=3.24\pm0.04$) (see Table \ref{Table V}). Using
the updated theoretical values $\gamma=1.2395935$, $\Delta=0.50189$
(see Table \ref{Table I}b), $\Gamma^{+}\left(\text{Xe}\right)=0.0578204$,
and the two calibrated pairs $\left\{ R_{32}=13.41;Y_{2}=0.071051\right\} $
and $\left\{ \Delta\tau_{\text{C3}}^{*}=4.62829\,10^{-2};Y_{3}=0.242478\right\} $,
Eq. (\ref{akhiplus cal}) gives $a_{\chi}^{+}\equiv a_{1\chi}^{+}=1.25557$,
while Eq. (\ref{a2khiplus cal}) gives $a_{2\chi}^{+}=-0.569934$.
The small difference with the value $a_{\chi}^{+}=1.22961$ calculated
using Eq. (\ref{Xe independent amplitudes (39)}), is due to the similar
{}``constrained'' adjustment of the first amplitude $a_{M}^{+}$
of the order parameter density (not reported here), which maintains
the universal value $\frac{a_{M}^{+}}{a_{\chi}^{+}}=0.9$ of the related
amplitude ratio \citet{Garrabos2006gb}. In such an optimized result
from two properties, the best central values of the master confluent
amplitudes are $\mathcal{Z}_{\chi}^{1,+}=0.555$ and $\mathcal{Z}_{\mathcal{M}}^{1}=0.4995$,
respectively. However, as previously indicated, a true error-bar of
the order of $\pm35\%$ (at least) can be attached to these confluent
amplitudes in the absence of data calibration closer to $T_{c}$.

\subsubsection{The uniqueness of the crossover parameter and the effective extension
of the critical domain beyond the Ising-like preasymptotic domain}

Since the asymptotic amplitude $\Gamma^{+}$ is now fixed, the effective
exponent-amplitude pair $\gamma_{e,pVT};\,\Gamma_{e,pVT}^{+}$ reported
in columns 2 and 3 of Table \ref{Table VI}, can be used to calculate
the value of a crossover parameter $\vartheta_{\mathcal{L},pVT}$
at each {}``local'' temperature distance $\left\langle \Delta\tau_{pVT}^{*}\right\rangle $,
using the scale transformation of Eq. (\ref{gamaplus effective vs gamaplus-theta (73)}),
and the theoretical values of $\mathbb{Z}_{\chi}^{+}$ and $\mathbb{Z}_{\chi,e}^{+}\left(\gamma_{e,pVT}\right)$
given in Tables \ref{Table I} and \ref{Table II}. The corresponding
values of $\left.\vartheta_{\mathcal{L},pVT}\right|_{\gamma_{pVT}=cte}$
are given in last column of Table \ref{Table VI}. Each $pVT$ result
for $\left.\vartheta_{\mathcal{L},pVT}\right|_{\gamma_{pVT}}$ as
a function of $\Delta\tau^{*}$, has then been illustrated in Fig.
\ref{Figure7} (c) as a point-segment form. The {}``local'' value
$\left\langle \Delta\tau_{pVT}^{*}\right\rangle $ is close to the
calculated one $\Delta\tau_{\text{th}}^{*}$ (see columns 6 and 7
in Table \ref{Table VI}). In addition, the \%-residuals $R_{\Gamma_{e,\text{th}}^{+}}=100\left(\frac{\Gamma_{e,}^{+}\left(\gamma_{e,pVT}\right)}{\Gamma_{e,\text{th}}^{+}\left(\gamma_{e,pVT}\right)}-1\right)$
between the effective amplitudes compare favorably with the estimated
experimental precision (see column 8 in Table \ref{Table VI}). Therefore,
at large temperature distance from $T_{c}$, we obtain a significant
confirmation that the values of the effective crossover parameter
are close to the one of the asymptotic scale factor $\vartheta=0.021069$,
and can then be considered as independent of $\Delta\tau^{*}$ in
the Ising-like extended asymptotic domain.

Using in a similar manner the fitting results given in Table \ref{Table VII},
we have obtained the corresponding values of $\vartheta_{\mathcal{L}}$
given in column 11. However, we have now a better control of the related
uncertainty, thanks to the high precision of the ligth scattering
experiment of Güttinger and Cannell. As a matter of fact, admitting
in a first approach that the values of $\Gamma^{+}$, $\left\langle \Delta\tau_{\gamma_{e}}^{*}\right\rangle $,
$\mathbb{Z}_{\chi}^{+}$ and $\mathbb{Z}_{\chi,e}^{+}\left(\gamma_{e}\right)$
are known with zero uncertainty, while $\Gamma_{e}^{+}=\Gamma_{e,\text{th}}^{+}\left(1+\delta\Gamma\right)$,
with $\delta\Gamma\sim1-2\%$, it is easy to show from Eq. (\ref{theta vs amplitude ratios (74)})
and $\vartheta=\left(\frac{1}{\mathbb{Z}_{\chi}^{+}\mathbb{Z}_{\chi,e}^{+}}\times\frac{\Gamma_{e,\text{th}}^{+}}{\Gamma^{+}}\right)^{\frac{1}{\gamma-\gamma_{e}}}$
that\begin{equation}
\vartheta=\frac{\theta_{\mathcal{L}}}{\left(1+\delta\Gamma\right)^{\frac{1}{\gamma-\gamma_{e}}}}\simeq\frac{\theta_{\mathcal{L}}}{\left(1+\frac{\delta\Gamma}{\gamma-\gamma_{e}}\right)}\label{(A14)}\end{equation}
Looking then to the percent deviation reported in Fig. \ref{Figure6},
we can observe that the true tangent (pink) line of slope $\gamma_{e}=1.20588$
has an amplitude $\simeq1\%$ lower than the amplitude of the effective
power law $\kappa_{T,\text{GC}_{e}}^{*}=0.07551466\left(\Delta\tau^{*}\right)^{-1.205879}$
used as a reference (see also Table \ref{Table VII} column 9, where
the calculated value of the residual is $-1\%$ in line $\#\text{GC}_{e}$)
. From Eq. (\ref{(A14)}) with $\delta\Gamma=0.01$ and $\vartheta_{\mathcal{L}}\simeq0.02851$
(see line $\#\text{GC}_{e}$, colum 11, Table \ref{Table VII}), we
obtain $\vartheta\simeq0.021988$ which is in excellent agreement
($+4.38\%$) with our initial estimation $\vartheta\simeq0.21069$
from the scale dilatation method. As previously underlined, the precise
description by a local exponent value defining the slope of the tangent
line to the singular behavior of the isothermal compressibility of
xenon at a well-defined temperature distance to $T_{c}$, is one of
the major points of interest of the Güttinger and Cannell's results
to validate the one-parameter crossover modelling predicted by the
massive renormalization scheme. In a similar manner, using Eq. (\ref{(A14)})
with $\delta\Gamma$ and $\vartheta_{\mathcal{L}}$ given in lines
$\#pVT$, CB(c), SGB(c) of Table \ref{Table VII}, we obtain $\vartheta\simeq0.02160,\,0.021232,\,0.021136$,
and the corresponding deviations$+2.52\%,\,0.77\%,\,0.32\%$), respectively
for the three other fitting results reported in Table \ref{Table VII}.

Obviously, similar effect is produced by the uncertainty level attached
to the determination of the leading amplitude $\Gamma^{+}$, justifying
oncemore its independent estimation from a {}``standard'' value
of the isothermal compressibility at a temperature distance well-inside
the Ising-like preasymptotic domain.

Henceforth, the importance of the scaling form of Eq. (\ref{theta vs amplitude ratios (74)})
in the determination of a unique asymptotic value for the scale factor
$\vartheta$ is clearly established. Moreover, Eq. (\ref{theta vs amplitude ratios (74)})
is valid in the range $\frac{\xi}{\alpha_{c}}\gtrsim2.5-3$, or $\Delta\tau^{*}\lesssim\mathcal{L}_{\text{EAD}}^{\text{Xe}}\simeq0.02$
in xenon case.

One complementary remark can be formulated.

Outside the Ising-like extended asymptotic domain, i.e., typically
for $\Delta\tau^{*}\geq10^{-1}$ in xenon case, it is well-established
that the Ising-like universality is not valid. For example, increasing
the temperature distance to $T_{c}$ on the $pVT$ data analyzes,
we are able to observe the continuous decreasing behavior of $\gamma_{e,pVT}$
until a value close to mean-field value $\gamma_{e,pVT}\approx\gamma_{\text{MF}}=1$
when $\Delta\tau^{*}\lessapprox1$ (see Table \ref{Table VI}). That
unambiguously discriminates a sharp domain, i.e. typically $0.3\leq\Delta\tau^{*}\leq0.5$,
where $\gamma_{e,pVT}$ crosses the {}``mean'' crossover value $\gamma_{\frac{1}{2}}=\frac{\gamma+\gamma_{\text{MF}}}{2}\approx1.12$
(as initially reported in Ref. \citet{Bagnuls1984b}). For xenon,
this classical-to-critical crossover {}``crossing'' is expected
close to $\frac{\xi}{\alpha_{c}}\lesssim1,$ that means that the correlation
length is of the same order of magnitude or lower than the short-range
molecular interaction. In Table \ref{Table VI}, the non-defined value
$\left.\vartheta_{\mathcal{L},pVT}\right|_{\gamma_{pVT}=1}\left(\sim\infty\right)$,
corroborates that the mean-field behavior predicted by the theoretical
crossover function with $\vartheta=0.0210$, is not compatible with
the $pVT$ experimental result $\Gamma_{e,\gamma_{\text{MF}}}^{+}\simeq0.11$.
Such a typical limit of the mean-field-like range (see upper part
of Figure \ref{Figure3}) is well-illustrated by the corresponding
transformation between the opened squares represented in Figure \ref{Figure3},
which cannot account for the {}``experimental'' location of the
opened circle when $\gamma_{e}=\gamma_{\text{MF}}=1$. 

Finally, Figure \ref{Figure7} (c) confirms that the effective extended
critical domain of xenon, corresponding to the condition $\frac{\xi}{\alpha_{c}}\gtrsim3$
discussed in a detailed manner in Ref. \citet{Garrabos2006cl}, is
well characterized by a single crossover parameter whose value at
the largest temperature range is comparable to the one of the asymptotic
scale factor estimated from the scale dilatation method. We can conclude
that singular behavior of any property in this extended critical domain
of xenon can be calculated in conformity with the universal features
predicted by the massive renormalization scheme, only using the required
four critical coordinates to define the position and tangent surface
of its actual liquid-gas critical point on the $p,v_{\bar{p}},T$
phase surface, as expected by one of us two decades ago.


\begin{thebibliography}{10}
\bibitem{background}In this approach of scaling, we use the correlation
length and the susceptibility as independent properties in terms of
the two-scale-factor universality. We can then ignore the universal
features related to the additive universal constant of the heat capacity.
Moreover, in a pure fluid case, the critical background constant is
mixed with the non-universal contribution of the regular background
terms.
\bibitem{microscopicspacing}For example, the microscopic spacing
order in xenon fluid can be driven by the Lennard Jones like diameter
$\sigma^{LJ}$ of xenon atom, which typically is $\sigma^{\text{LJ}}\left(\text{Xe}\right)\simeq0.4\,\text{nm}$,
i.e. comparable in order of magnitude to the estimated value $\mathbb{L}_{0,\mathcal{L}}\left(\text{Xe}\right)\simeq0.391\,\text{nm}$.
However, this typical size of xenon atom is one order of magnitude
greater than $\left[g_{0,\mathcal{L}}\left(\text{Xe}\right)\right]^{-1}\simeq34.3\,10^{-3}\,\text{nm}$
and also significantly greater than the critical thermal wavelength
$\Lambda_{T,c}\left(\text{Xe}\right)\simeq8.95\,10^{-3}\,\text{nm}$.
\end{thebibliography}
\end{document}